\documentclass[]{aa}

\usepackage{graphicx, times, txfonts, float, color, bm}
\usepackage{rotating}
\usepackage{longtable,lscape}

\newcommand{\less}{\raisebox{-1.1mm}{$\stackrel{<}{\sim}$}}
\newcommand{\more}{\raisebox{-1.1mm}{$\stackrel{>}{\sim}$}}
\newcommand{\msol}{\mbox{M$_{\odot}$}}

\newcommand{\msolyr}{{M$_{\odot}$}\,yr$^{-1}$}

\newcommand{\lsol}{\mbox{L$_{\odot}$}}

\begin{document}

\title{
The VMC Survey - XXXVII. Pulsation periods of dust enshrouded AGB stars in the Magellanic Clouds
\thanks{
Tables~\ref{App-Sample}, \ref{App-Periods2}, \ref{App-Periods1}, and \ref{Tab-Nanni} are available in electronic form at the CDS via
anonymous ftp to cdsarc.u-strasbg.fr (130.79.128.5) or via
http://cdsweb.u-strasbg.fr/cgi-bin/qcat?J/A+A/.
Figures~\ref{Fig-LCs} and \ref{Fig-SEDs} are available in the on-line edition of A\&A.
}
\thanks{
Based on observations made with VISTA at ESO under programme ID~179.B-2003.
}
}

\author{
M.~A.~T.~Groenewegen
\inst{1}
\and
A.~Nanni
\inst{2}
\and
M.~-R.~L.~Cioni
\inst{3}
\and
L.~Girardi\inst{4}
\and 
R.~de~Grijs\inst{5,6,7} 
\and
V.~D.~Ivanov\inst{8}
\and
M.~Marconi\inst{9}
\and
M.-I.~Moretti\inst{9}
\and
J.~M.~Oliveira\inst{10}
\and
M.~G.~Petr-Gotzens\inst{8,11}
\and
V.~Ripepi\inst{9}
\and
J.~Th.~van Loon\inst{10}
}

\institute{
Koninklijke Sterrenwacht van Belgi\"e, Ringlaan 3, B--1180 Brussels, Belgium \\ \email{martin.groenewegen@oma.be}
\and
Aix Marseille Universit\'e, CNRS, CNES, LAM, 38, rue Fr\'ed\'eric Joliot-Curie, F-13388 Marseille, Cedex 13 France
\and
Leibniz Institut f\"{u}r Astrophysik Potsdam, An der Sternwarte 16, D-14482 Potsdam, Germany
\and
Osservatorio Astronomico di Padova-INAF, Vicolo dell'Osservatorio 5, I-35122 Padova, Italy
\and
Department of Physics and Astronomy, Macquarie University, Balaclava Road, Sydney, NSW 2109, Australia
\and
Research Centre for Astronomy, Astrophysics and Astrophotonics, Macquarie University, Balaclava Road, Sydney, NSW 2109, Australia
\and
International Space Science Institute--Beijing, 1 Nanertiao, Zhongguancun, Hai Dian District, Beijing 100190, China
\and
European Southern Observatory, Karl-Schwarzschild-Str. 2, D--85748 Garching bei M\"unchen, Germany
\and
INAF -- Osservatorio Astronomico di Capodimonte, via Moiariello 16, I--80131, Naples, Italy  
\and
Lennard-Jones Laboratories, School of Chemical and Physical Sciences, Keele University, ST5 5BG, UK
\and
Universit\"ats-Sternwarte, Ludwig-Maximilians-Universit\"at M\"unchen, Scheinerstr 1, D-81679 M\"unchen, Germany
}

\date{received: 2019, accepted: 2020}

\offprints{Martin Groenewegen}

\authorrunning{Groenewegen et al.}
\titlerunning{Pulsation periods of enshrouded AGB stars in the Magellanic Clouds}

\abstract {
Variability is a key property of stars on the asymptotic giant branch (AGB). Their pulsation period is related to the luminosity and
mass-loss rate of the star. The long-period variables (LPVs) and Mira variables are the most prominent of all types of variability
of evolved stars. The reddest, most obscured AGB stars are too faint in the optical and have eluded large variability surveys.
}
{Obtain a sample of LPVs by analysing $K$-band light-curves (LCs) of a large number of sources in the direction
of the Magellanic Clouds with the colours expected for red AGB stars ($(J-K) > 3$~mag or equivalent in other colour combinations).
}
{Selection criteria are derived based on colour-colour and colour-magnitude diagrams from the combination of VISTA Magellanic Cloud (VMC) survey,
\it Spitzer \rm IRAC and AllWISE data.
After eliminating LPVs with known periods shorter than 450 days, a sample of 1299 candidate obscured AGB stars is selected.
$K$-band LCs are constructed combining the epoch photometry available in the VMC survey with literature data,
analysed for variability and fitted with a single period sine curve to derive mean magnitudes, amplitudes and periods.
A subset of 254 stars are either new variables, or known variables where the period we find is better determined than
the literature value, or longer than 1000 days.
The spectral energy distributions (SEDs) of these stars are fitted to a large number of templates.
For this purpose the SEDs and \it Spitzer \rm IRS spectra of some non-AGB stars (Be stars, H\sc ii \rm regions and
young stellar objects [YSOs]) are also fitted to have templates of the most likely contaminants
in the sample. 
}
{A sample of 217 likely LPVs is found.
Thirty-four stars have periods longer than 1000 days although some of them have alternative shorter periods.
The longest period of a known Mira in the Magellanic Clouds from Optical Gravitational
Lensing Experiment (OGLE) data (with $P= 1810$~d) is derived to have a period of 2075~d based
on its infrared LC.
Two stars are found to have longer periods, but both have lower luminosities and smaller pulsation amplitudes than expected for Miras.
Mass-loss rates and luminosities are estimated from the template fitting.
Period-luminosity relations are presented for C- and O-rich Miras that appear to be extensions of relations derived in the literature for
shorter periods. The fit for the C-stars is particularly well defined (with 182 objects) and reads
$M_{\rm bol} = (-2.27 \pm 0.20) \cdot \log P +  (1.45 \pm 0.54)$~mag with an rms of 0.41~mag.
Thirty-four stars show pulsation properties typical of Miras while the SEDs indicate they are not.
Overall, the results of the LC fitting are presented for over 200 stars that are associated to YSOs.
}
{}


\maketitle

\section{Introduction}

At the end of their lives essentially all low- and intermediate mass stars (roughly $\sim 0.9$ to $\sim 10$~\msol\ on the
main sequence) will go through the (super)-asymptotic giant branch (AGB) phase. They end up as $\sim 0.55-1.4$~\msol\
white dwarfs which means that a large fraction of the initial mass of the star is lost to the interstellar medium.
An important characteristic of AGB stars is that they pulsate, classically divided into stars with small amplitudes
(the semi-regular variables, SRVs) and the large amplitude Mira variables. Its now common to use the term long period variable (LPV)
irrespective of pulsation amplitude for a pulsating star on the AGB. 
Pulsation-induced shock waves and radiation pressure on dust is the most promising mechanism to explain wind driving,
especially regarding the more evolved AGB stars with low effective temperatures, large pulsation amplitudes,
and high mass-loss rates (MLRs; see the review by \citealt{HO18}).

It was known from observations in the Large Magellanic Cloud (LMC) that Miras follow a period-luminosity ($PL$) relation \citep{GE81,Feast1989},
but the breakthrough came with the advent of micro-lensing surveys. \citet{Wood1999} and \citet{Wood2000} showed that
late-type stars in the LMC followed 5 sequences (labelled A, B, C, D and E).
Subsequent work demonstrated that sequence B consists actually of two sequences (B and C$^{\prime}$) and that there exists an
additional sequence (A$^{\prime}$), including possible substructure \citep{Ita04,Soszynski07}.
Sequence D consists of long secondary periods (LSPs), while sequence E is due to binary stars.
A theoretical interpretation using linear, radial, non-adiabatic pulsation models and a population synthesis model show
that sequences A$^{\prime}$, A, B + C$^{\prime}$, and C, correspond to third, second, first overtone, and fundamental mode pulsation,
respectively \citep{Trabucchi2017}. The LPVs with the largest amplitudes, the classical Miras, dominate sequence C.

The Mira $PL$-relation has been used as a distance indicator, traditionally inside the Milky Way galaxy
in the pre-{\it Gaia} era (e.g. \citealt{Gr02,Ramstedt14}). The method has gained recent interest in the context of the distance ladder
as an independent check on Cepheid or tip of the red giant branch distances, for example recent works on the mega-maser galaxy
NGC 4258 \citep{Huang18} or the SN host galaxy NGC 1559 \citep{Huang19}.
As Miras are as bright as Cepheids in the NIR they will be important secondary calibrators in the upcoming era of
ground-based extremely large telescopes and the {\it James Webb Space Telescope}.

The currently largest survey of LPVs in the Magellanic Clouds (MCs) is the Optical Gravitational
Lensing Experiment OGLE-III catalogue, which contains
1667 Mira stars, 11~128 SRVs and 79~200 OGLE Small Amplitude Red Giants (OSARGs) in the LMC \citep{SoszynskiLPVLMC}, and
 352 Mira stars,   2222 SRVs and 16~810 OSARGs in the Sarge Magellanic Cloud (SMC) \citep{SoszynskiLPVSMC}.

However, OGLE and other current surveys in the optical domain will miss the reddest, most obscured AGB stars. At the very
end of the AGB the (dust) MLR may become so large that the object becomes very faint, beyond
the OGLE $I$-band detection limit of about 21~mag. The dust grains in the circumstellar envelope (CSE) scatter and
absorb the emission in the optical to re-emit it in the near- and mid-infrared (NIR, MIR).

These stars are known to exist in the LMC. They were typically selected as having {\it Infrared Astronomical Satellite} (IRAS) colours
similar to obscured AGB stars in our Galaxy.
\cite{Wood1992} give periods for 9 sources in the LMC. Periods range between 930 and 1390~d, with peak-to-peak amplitudes
in $K$ between  0.3 and 2.1 mag, and $(J-K)$ colours that range from 2.1 to 5.7 mag (an additional 7 LMC and 3 SMC sources
were monitored but showed small amplitudes and no periods were given).
\citet{Wood1998} monitored 12 sources and 9 turned out to be LPVs with periods in the range 530 to 1295~d.
Amplitudes in $K$ are between 0.85 and 1.9 mag, and colours range from  $(J-K)=$ 3.9 to 6.1 mag (with two not detected in $J$).
Similarly, \citet{Whitelock2003} give periods for two dozen stars which have an IRAS identifier.
Periods range from 540 to 1390 days, $K$-band amplitudes are between 0.52 and 1.8 mag, and $(J-K)$ colours range from 1.9 to 4.0 mag
(with several not detected in $J$).
The faintest stars in these works have (mean light) magnitudes in the range $K= 12-13$~mag.

The ESO public VISTA Magellanic Cloud (VMC) survey is a photometric survey in three filters, $Y$, $J$,
and $K_{\rm s}$ performed at the VISTA telescope using the VIRCAM camera. The latter provides a spatial
resolution of 0.34\arcsec\ per pixel and a non-contiguous field-of-view of 1.65~deg$^2$ sampled by 16 detectors.
To homogeneously cover the field-of-view it is necessary to fill these gaps using a six-point mosaic.
This is the unit area of VISTA surveys called a tile and covers uniformly with at least two exposures per pixel
an area of 1.5~deg$^2$ \citep{Emerson10}.

The VMC survey covers an area of approximately 170~deg$^2$ (110 tiles) of the MC system and
includes stars as faint as 22 mag in $K_{\rm s}$ (5$\sigma$, Vega mag), see \citet{Cioni11} for description of the survey.
The large areal coverage and the survey depth are the main strengths of the VMC survey compared to earlier works in the IR.
Unlike the VMC papers on classical variables
(Type-II and Classical Cepheids, RR Lyrae, e.g. \citealt{Ripepi12, Ripepi15, Ripepi16, Ripepi17, Muraveva2017})
the expected periods for AGB pulsators are much longer than the baseline of the VMC observations.
Therefore the VMC data is combined with $K$-band data from the literature when available to increase the time span.

In the present paper $K$-band light-curves (LCs) are inspected for a large number of candidate dust-obscured AGB stars.
The sample is selected based on colour-magnitude and colour-colour criteria derived from the properties
of known AGB stars, using both VMC and other infrared data.

The paper is organised as follows.
Section~\ref{S-sample} discusses how the sample of potential LPVs is selected based on different photometric catalogues,
and the creation of the $K$-band LCs.
Section~\ref{S-modelling} discusses how the LCs are analysed.
Section~\ref{S-analysis} discusses the results. The sample of potential LPVs is further reduced by considering the shape of the
spectral energy distributions (SEDs).

Preliminary results of this work were shown or used in \citet{Groenewegen2016}, \citet{Cioni2017} and \citet{GS18} (hereafter GS18)
but the results on the LC parameters are superseded by this paper.

\section{Sample selection}
\label{S-sample}

In this section the sample of candidate obscured AGB stars is generated.
It is based on three different photometric catalogues.
The first selection uses only VMC data\footnote{All VMC data refers to observations obtained until the end of September 2016,
  including complete observations of the Small Magellanic Cloud.},
but requires a detection in $J$ and $K_{\rm s}$ for reliability (sample~1).
The second selection combines the VMC $K_{\rm s}$-band data with {\it WISE} data from the AllWISE catalogue \citep{Cutri_Allwise} (sample~2).
The third selection starts with data exclusively from the Surveying the Agents of a Galaxy's Evolution (SAGE) survey of the MCs, and
then looks for a possible counterpart in the VMC data (sample~3).
During this selection procedure only the mean magnitudes as given by the
VISTA Science Archive (VSA; \citealt{Cross12}) are used, and the quality flags indicating warnings and
possible issues (the {\it ppErrBits}) are not considered.
Only in the second step is the time-resolved VMC $K_{\rm s}$-band data used and combined with literature data to produce a LC.

\subsection{Sample 1: VMC data only}

\begin{figure}
   \centering

\begin{minipage}{0.24\textwidth}
\resizebox{\hsize}{!}{\includegraphics{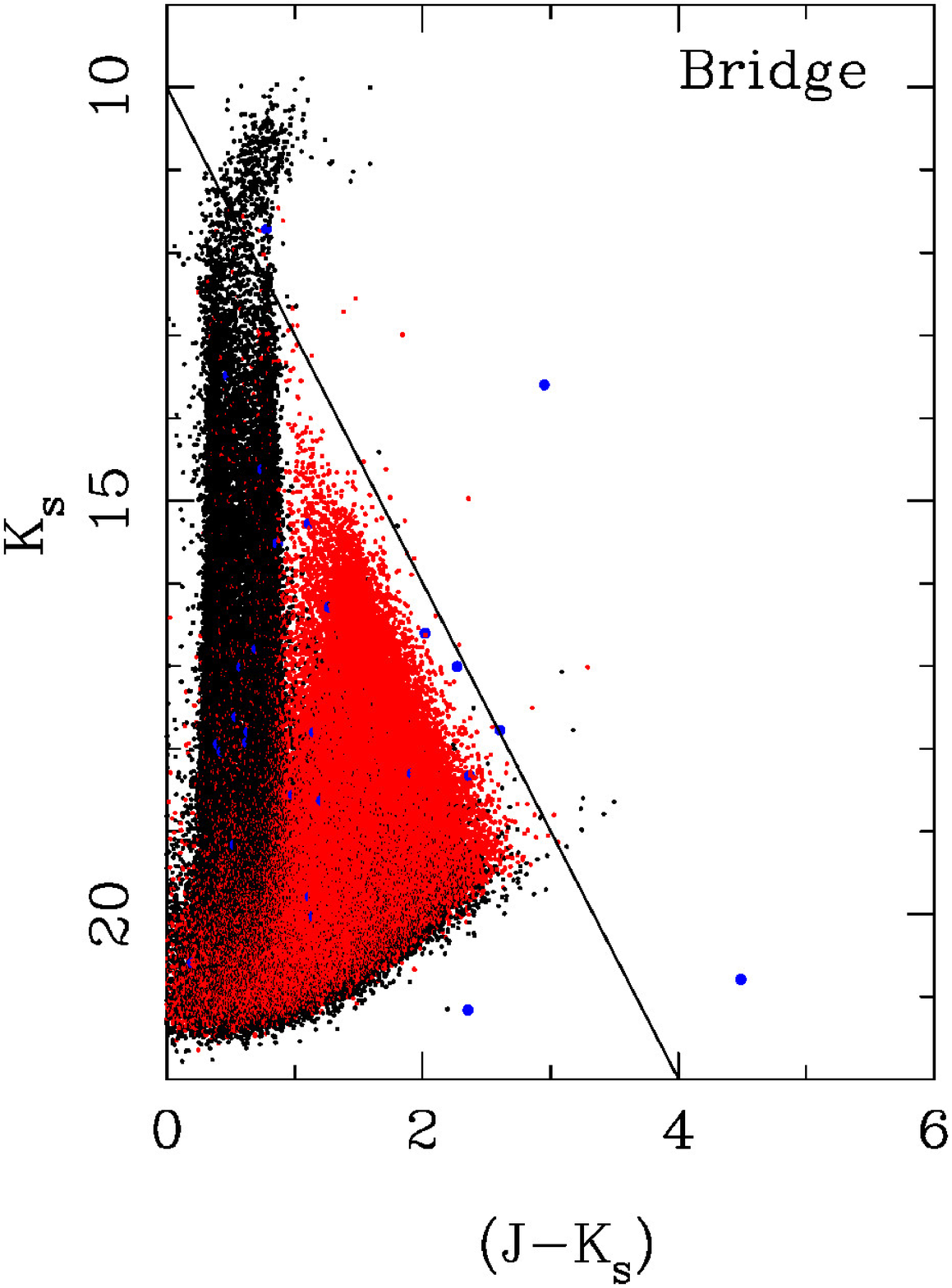}}
\end{minipage}
\begin{minipage}{0.24\textwidth}
\resizebox{\hsize}{!}{\includegraphics{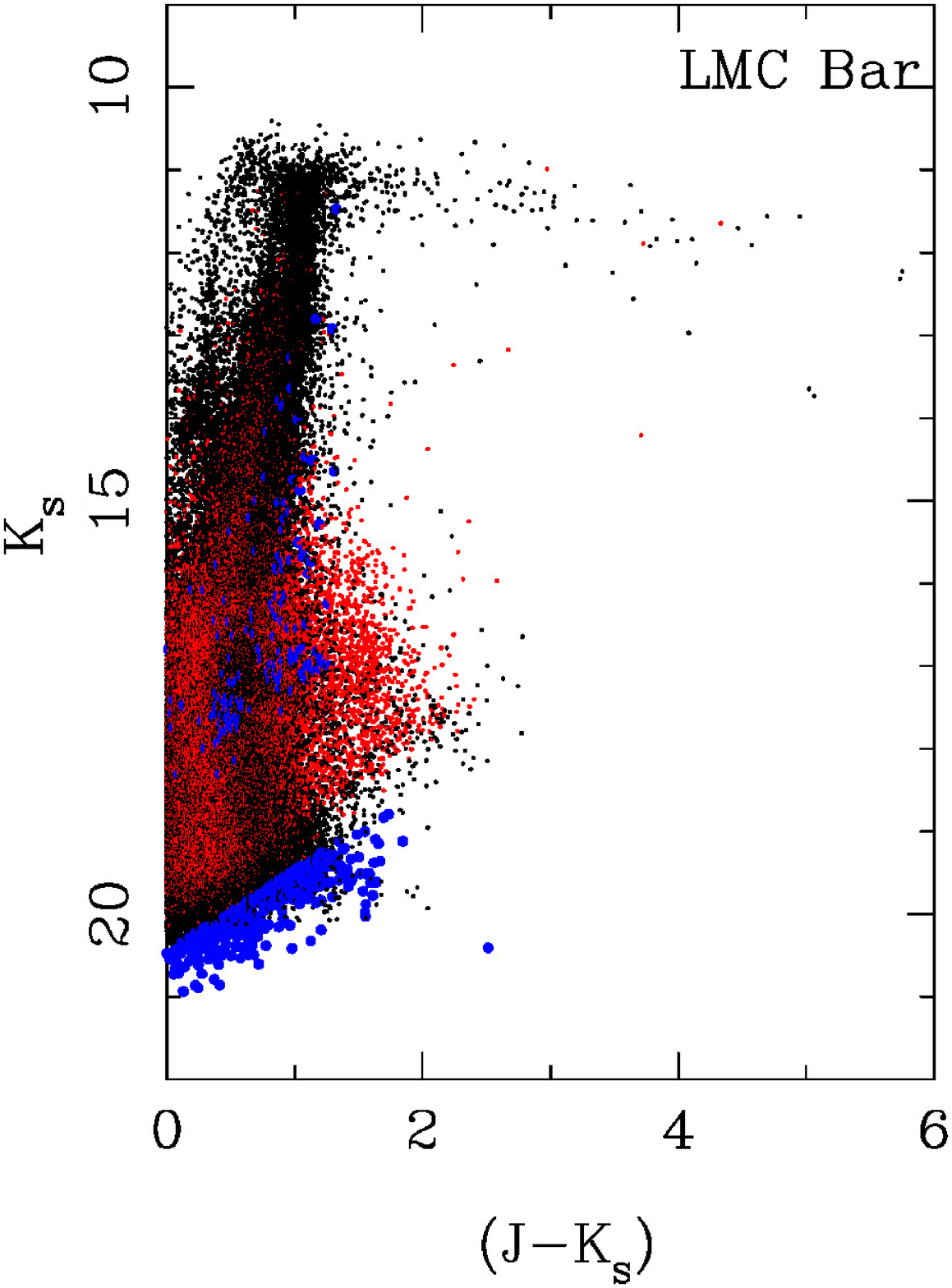}}
\end{minipage}

\caption[]{
VMC ($K_{\rm s}$, $(J-K_{\rm s})$) CMD in a region in the Magellanic Bridge  (left-hand panel;
a strip at $42\degr <$ R.A. $< 52\degr$ and  $-72.5\degr <$  Dec. $< -73.8\degr$)
and in the LMC bar (right-hand panel; a box defined
as $80\degr <$ R.A. $< 82\degr$, and  $-69\degr <$  Dec. $< -70\degr$).
Objects that are (probable) stars are indicated in black, and (probable) galaxies (extended objects) are indicated in red.
Objects classified differently are indicated in blue.
The line, $K= 3 \cdot (J-K) + 10$~mag, indicates the approximate boundary of galaxies.
}
\label{Fig-VMC-BG-BAR}
\end{figure}

Figure~\ref{Fig-VMC-BG-BAR} shows the VMC ($K_{\rm s}$, $(J-K_{\rm s})$) colour-magnitude diagram (CMD) for a region in the Bridge between the
two Clouds, and a region located in the bar of the LMC. The region in the Bridge is used to delineate the locus in colour space occupied
by extended objects, most of which are background galaxies but that can also result from
blends (colour coded in red)\footnote{Indicated by the {\it mergedClass} parameter which is $-9$ for a saturated source,
  $-3$ for a probable galaxy, $-2$ for a probable star, $-1$ for a star, $0$ for noise,
and $+1$ for a galaxy.}, while the region in the bar shows the well-known sequence
of AGB stars redder than $(J-K_{\rm s}) \sim 1.5$~mag at relatively bright magnitudes.
At the brightest level ($K_{\rm s} ~\less~ 10.5$~mag) one notices the effect of saturation.
This is not expected to influence the selection of obscured AGB stars for which the distribution in the CMD bends to faint magnitudes.
This is shown more clearly in Fig. 2.

\begin{figure}
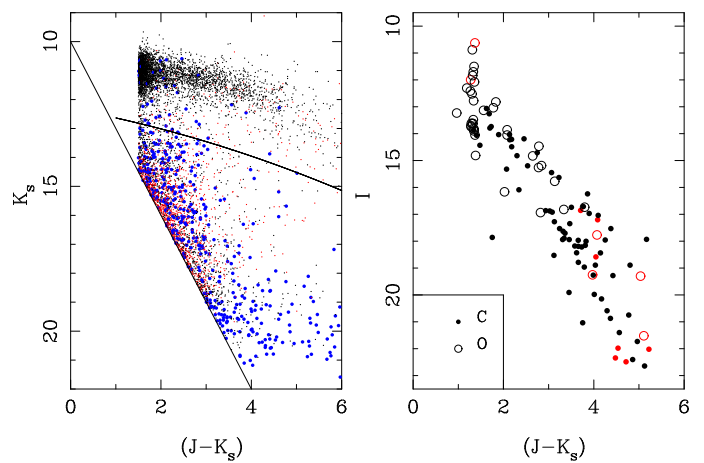

   \centering

\begin{minipage}{0.24\textwidth}
\resizebox{\hsize}{!}{\includegraphics{TRGBCMD_VMC.ps}}
\end{minipage}
\begin{minipage}{0.24\textwidth}
\resizebox{\hsize}{!}{\includegraphics{hrd_I_JK_P450_MC.ps}}
\end{minipage}

\caption[]{
Left-hand panel: All VMC stars with a detection in $J$ and $K_{\rm s}$, redder than $(J-K_{\rm s}) = 1.5$~mag, and above
the line $K_{\rm s}= 3 \cdot (J-K_{\rm s}) + 10$~mag.
The second-order polynomial is a fit to the red sequence of AGB stars, shifted downward by 2 magnitudes for comparison purposes (see text).
Objects that are (probable) stars are indicated in black, and (probable) galaxies are indicated in red.
Objects classified differently are indicated in blue.
Right-hand panel: Synthetic photometry from GS18 for 29 SMC and 133 LMC O- (open circles) and C-rich (filled circles) AGB stars
with pulsation periods longer than 450 days.
The objects marked in red have pulsation periods derived from $K/K_{\rm s}$ or {\it WISE}-band data, the black points
from optical data (OGLE, EROS, MACHO).
}
\label{Fig-IJK}
\end{figure}

In a subsequent step all (about 7000) stars in VMC with a detection in $J$ and $K_{\rm s}$, $(J-K_{\rm s}) > 1.5$~mag,
and $K_{\rm s} < 3 \cdot (J-K_{\rm s}) + 10$~mag were selected to represent candidate obscured AGB stars.
Their CMD distribution is shown in the left-hand panel of Fig.~\ref{Fig-IJK}, and shows that there is still an important contribution by
background galaxies and other extended objects.
This CMD also clearly exhibits the sequence of increasing mass loss along the AGB.
A fit is made to this sequence, $K_{\rm s}= 10.318 +0.287 \cdot (J-K_{\rm s}) + 0.0306 \cdot (J-K_{\rm s})^2$~mag, and this relation 
shifted by 2 magnitudes to fainter magnitudes to delineate the approximate lower boundary of this sequence, is shown as well.

Another consideration in selecting the stars is that the primary interest is in studying the reddest pulsating AGB stars,
objects that are too faint for the current optical surveys like OGLE-III and OGLE-IV, that have a limiting magnitude
in the $I$-band around 20-21~mag.

GS18 fitted the SEDs and spectra taken with the {\it Spitzer} Infrared Spectrograph (IRS, \citealt{Houck2004}) spectra of
almost 400 AGB and red supergiant (RSG) stars in the MCs, and
determined periods for a large fraction of them. The periods were derived by analysing existing optical data
(from the OGLE, Massive Compact Halo Objects (MACHO), and Exp\'erience de Recherche d’Objets Sombres (EROS) surveys), or data in the $K$ and $K_{\rm s}$ bands (including VMC data) and in the {\it WISE} W2 band for the reddest objects.

The right-hand panel of Fig.~\ref{Fig-IJK} shows the $I$, $(J-K_{\rm s})$ CMD for about 150 O- and C-rich AGB stars in the MCs
with a pulsation period longer than 450 days, based on
synthetic photometry, that is, the simulated magnitudes taking into account the filter transmission curves and photometric zero points
in many filters using the best-fitting model to the SEDs in GS18.
Marked in red are stars where the pulsation period is based on $K/K_{\rm s}$ or W2 data, the other periods are derived from
optical data (mainly OGLE).
The diagram shows that the $I$-band limit of 20~mag is reached for stars that have a $(J-K_{\rm s})$ colour of $\sim 4$~mag.

To have sufficient overlap with the optical surveys, and to be on the conservative side, a limit of $(J-K_{\rm s}) = 3$~mag is chosen.
Sample~1 consists of the 857 stars in the VMC survey that have a detection in $J$ and $K_{\rm s}$, $(J-K_{\rm s}) > 3$~mag and
$K_{\rm s} < 12.318 +0.287 \cdot (J-K_{\rm s}) + 0.0306 \cdot (J-K_{\rm s})^2$~mag.

\subsection{Sample 2: VMC and {\it WISE} data combined}

While sample~1 has the advantage that it uses exclusively data from the VMC survey, the reddest AGB stars, by their
very nature, will  become  $J$-band dropouts for large enough MLRs. The aim of constructing samples~2 and 3 is to combine
VMC $K_{\rm s}$-band data with data at longer wavelengths, in particular {\it WISE} (sample~2), or IRAC (sample~3).
The advantage of {\it WISE} is that it is an all-sky survey, covering the complete area observed by the VMC survey.

The AllWISE catalogue \citep{Cutri_Allwise} is available as a table in the VSA and the queries were therefore performed within the VSA.
A search radius of 1\arcsec\ was used and an error on the $W1$ magnitude of less than 0.2~mag was imposed (equivalent to {\it w1snr} $>$5).

\begin{figure*}[h]
   \centering

\begin{minipage}{0.25\textwidth}
\resizebox{\hsize}{!}{\includegraphics{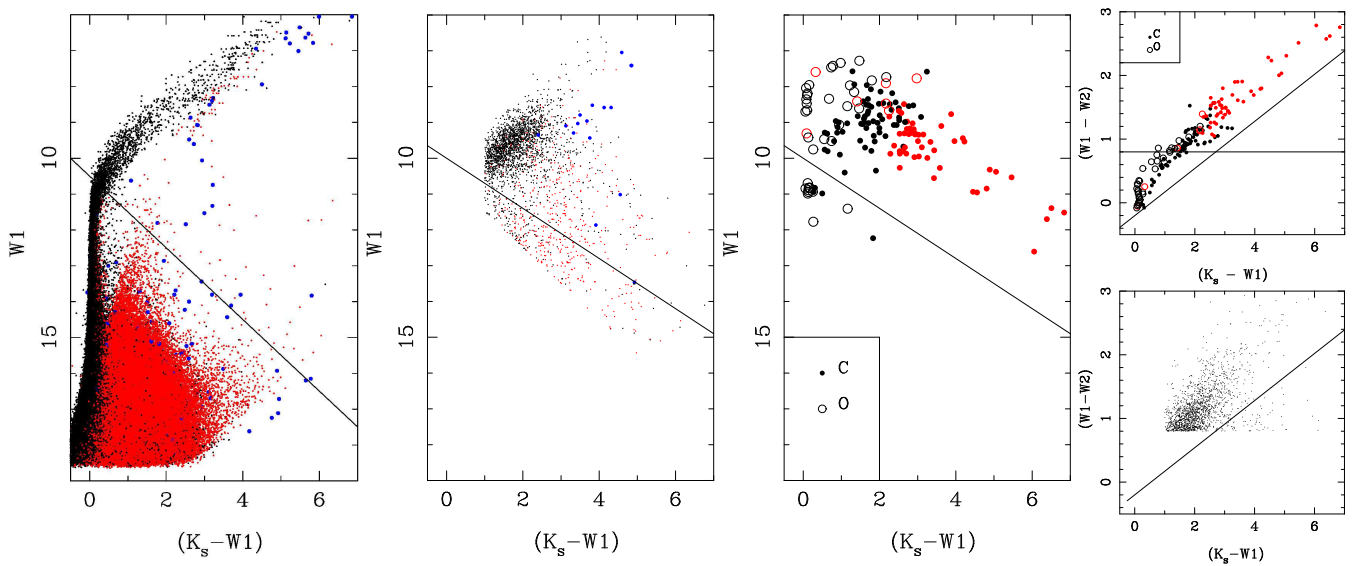}}
\end{minipage}
\begin{minipage}{0.25\textwidth}
\resizebox{\hsize}{!}{\includegraphics{TRGBCMD_WISE_All_initial.ps}}
\end{minipage}
\begin{minipage}{0.25\textwidth}
\resizebox{\hsize}{!}{\includegraphics{cmd_W1_KW1.ps}}
\end{minipage}
\begin{minipage}{0.21\textwidth}
  \begin{minipage}{0.91\textwidth}
    \resizebox{\hsize}{!}{\includegraphics{ccd_W1W2_KW1.ps}}
  \end{minipage}

  \begin{minipage}{0.91\textwidth}
    \resizebox{\hsize}{!}{\includegraphics{TRGBCCD_WISE_All_initial.ps}}
  \end{minipage}

\end{minipage}

\caption[]{
WISE-VMC ($W1, K_{\rm s}-W1$) CMD and $(W1-W2)-(K_{\rm s}-W1)$ CCDs.
In the left-most panel the CMD in a region between the SMC and LMC (a strip at $30\degr <$ R.A. $< 60\degr$) is presented.
The line $W1 = 1.0 \cdot (K-W1) +10.5$~mag indicates the upper boundary of the region predominantly occupied by galaxies.
In the second panel from the left all WISE-VMC sources are shown which are brighter
than the boundary line in the left-hand panel, $(K_{\rm s}-W1) > 1$~mag and $(W1-W2) > 0.5$~mag.
Objects that are (probable) stars are indicated in black, and (probable) galaxies (extended objects) are indicated in red, while
objects classified differently are indicated in blue in these two panels.
The provenance of the line in the second panel is explained in the next panel.
The corresponding CCD is shown in the bottom half of the right-most panel.
The third panel from the left shows a synthetic CMD based on GS18 for 29 SMC and 133 LMC O- (open circles) and C-rich (filled circles) AGB stars
with  pulsation periods longer than 450 days.
The objects marked in red have pulsation periods derived from $K/K_{\rm s}$ or {\it WISE}-band data, the black points
from optical data (OGLE, EROS, MACHO).
The line $W1= 0.7 \cdot (K_{\rm s}-W1) +10.0$~mag is the empirical lower boundary to the reddest AGB stars, and is
also shown in the second panel from the left.
The corresponding CCD diagram is shown in the top-half of the right-most panel
with the lines $(W1-W2) = 0.8$~mag and $(W1-W2) = 0.37 \cdot (K_{\rm s}-W1) -0.2$~mag, which indicates
the empirical lower boundary to the reddest AGB stars, and is also shown in the panel below it.
}
\label{Fig-WISE}
\end{figure*}

The left-hand panel of Fig.~\ref{Fig-WISE} shows the ($W1, K_{\rm s}-W1$) CMD for a region in the Bridge, with the
same colour coding for background galaxies and stars as in Fig.~\ref{Fig-VMC-BG-BAR}. The line $W1 = 1.0 \cdot (K_{\rm s}-W1) +10.5$ indicates
the approximate boundary beyond which (at bright magnitudes and red colours) the number of galaxies significantly decreases.
In the second panel all sources are shown which have $(K_{\rm s}-W1) > 1$~mag, $(W1-W2) > 0.5$~mag, and are brighter than this boundary.
The corresponding colour-colour diagram (CCD) is shown in the bottom half of the right-most panel.

The third panel from the left shows a synthetic CMD based on the AGB stars studied by GS18. 
The line $W1 = 0.7 \cdot (K_{\rm s}-W1) +10.0$ is the empirical conservative lower boundary to the reddest AGB stars, and is
also shown in the second panel from the left.
The corresponding CCD is shown in the top-half of the right-most panel
with the lines $(W1-W2) = 0.8$~mag and $(W1-W2) = 0.37 \cdot (K_{\rm s}-W1) -0.2$~mag.

The final selection is based on a detection in $K_{\rm s}$, a signal-to-noise ratio (SNR) in the $W1$ band $>5$, $(W1-W2) > 0.8$~mag,
$W1 < 0.7 \cdot (K_{\rm s}-W1) +10.0$~mag, and $(W1-W2) > 0.37 \cdot (K_{\rm s}-W1) -0.2$~mag for a total of 1317 sources.

\subsection{Sample 3: SAGE data first}

SAGE \citep{Meixner2006}) and SAGE-SMC \citep{Gordon2011}
are surveys of the LMC and SMC with the {\it Spitzer} telescope. The advantage of {\it Spitzer} is that it has a smaller
point-spread-function than {\it WISE}, and confusion is less of an issue. On the other hand the SAGE observations do not cover
entirely the area sampled by the VMC survey.
The combined LMC and SMC source catalogue in the [3.6], [4.5], [5.8], and [8.0] bands is available
electronically\footnote{ViZieR catalogue II/305/archive}.

\begin{figure}
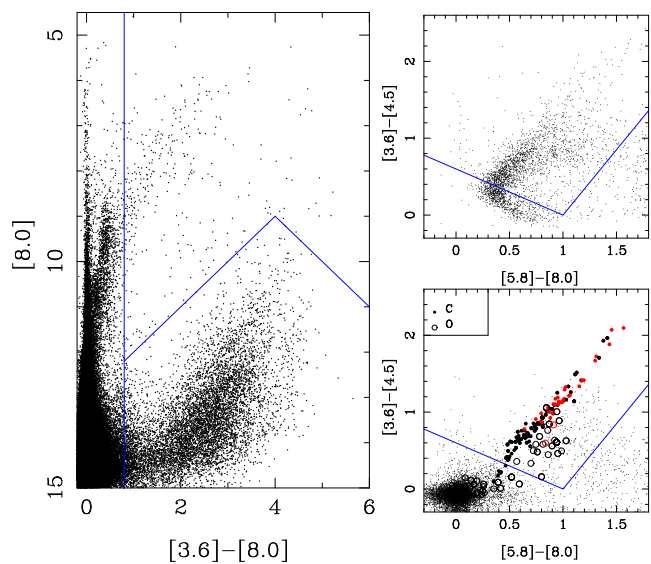

   \centering

\begin{minipage}{0.26\textwidth}
\resizebox{\hsize}{!}{\includegraphics{cmd_every10th.ps}}
\end{minipage}
\begin{minipage}{0.21\textwidth}

  \begin{minipage}{0.91\textwidth}
    \resizebox{\hsize}{!}{\includegraphics{ccd_linesCMD.ps}}
  \end{minipage}

  \begin{minipage}{0.91\textwidth}
    \resizebox{\hsize}{!}{\includegraphics{sirtf_ccd3.ps}}
  \end{minipage}

\end{minipage}

\caption[]{
Left-hand panel: the ([8.0], $[3.6]-[8.0]$) CMD of LMC and SMC objects in the SAGE survey. Only every 10th star is plotted.
The line at $[3.6]-[8.0] = 0.8$~mag indicates the blue limit of red AGB stars, and the two other lines indicate the limits
of the region occupied mainly by background galaxies (see text).
The top right-hand panel shows the ($[3.6]-[4.5]$, $[5.8]-[8.0]$) CCD of all sources that are located in the upper right of the
CMD delineated by these lines.
The bottom right-hand panel shows the same diagram with
synthetic CCD based on GS18 for 29 SMC and 133 LMC O- (open circles) and C-rich (filled circles) AGB stars
with  pulsation periods longer than 450 days.
The objects marked in red have pulsation periods derived from $K/K_{\rm s}$ or {\it WISE}-band data,
the black points from optical data (OGLE, EROS, MACHO).
The background sources in this panel are randomly picked from the S$^3$MC survey \citep{Bolatto07}.
}
\label{Fig-SAGE}
\end{figure}

Figure~\ref{Fig-SAGE} shows in the left panel the ([8.0], $[3.6]-[8.0]$) CMD of all sources with detections in all four bands.
The region occupied by background galaxies is clearly visible, as well as a tail of bright red AGB stars.

\citet{Boyer2011} used this diagram to select very red AGB stars, and we follow their work in selecting stars
with $[3.6]-[8.0] >$ 0.8~mag. Our elimination of background galaxies is less conservative than theirs, and we select objects above the line
$[8.0] < 13.0- ([3.6]-[8.0])$ (for $[3.6]-[8.0] <$ 4~mag), and
$[8.0] <  9.0+ ([3.6]-[8.0] -4.0)$~mag for redder objects.

The ($[3.6]-[4.5]$, $[5.8]-[8.0]$) CCD  of sources that meet these conditions is plotted in
the top of the right-hand panel of Fig.~\ref{Fig-SAGE}. The bottom panel shows the synthetic photometry based on GS18;
a region delineated by the lines
$[3.6]-[4.5] >  -0.6 \cdot ([5.8]-[8.0]) +0.6$ (for ($[5.8]-[8.0]$) $<$ 1~mag), and
$[3.6]-[4.5] >   1.7 \cdot ([5.8]-[8.0]) -1.7$ (for ($[5.8]-[8.0]$) $>$ 1~mag) is defined.
The initial sample consists of 1958 SAGE sources that meet these constraints.

The next step is the correlation with the VMC database. After inspecting the results using a
search radius of 1\arcsec\ the final correlation was done returning all VMC counterparts
within 0.8\arcsec\ of the SAGE coordinates.
Ten sources had multiple counterparts listed, but in all cases there was no reason not to take the closest match.
In the sample 35 SAGE stars had no counterpart in VMC, and in eight cases the counterpart lacked $K_{\rm s}$-band data,
leaving 1915 objects.

The selection so far was based on SAGE colours, and positions. An additional consistency check between
the IRAC and the $K_{\rm s}$-band VMC magnitude is performed. Based on the synthetic photometry from GS18
we impose a limit of $K_{\rm s} - [3.6] > 1.5$~mag (which is consistent with the limit $(J-K_{\rm s}) > 3$~mag in selecting sample~1).
This removes over 300 sources for a final sample size of 1562. 

%

\subsection{The final sample}

\subsubsection{Sample statistics}

The three samples were combined and together contain 2014 unique sources.
This sample is almost of the same size as that of the known Mira population in the MCs, indicating that the initial
sample selection based on colour criteria has not been very restrictive.

As the main interest in the present paper is in finding new (red, [very] long-period) pulsating stars,
the next step is a correlation with lists of known LPVs.
To be able to compare the properties (in terms of period, amplitude, etc.) of known Miras and LPVs with the new objects, known
LPVs with periods longer than 450 days are kept, while LPVs with shorter periods are removed.
This is an arbitrary choice but a good compromise, because including stars with shorter periods would increase the sample
by including stars that would essentially never become optically obscured and where the period from the relatively poorly
sampled IR LC would never be able to compete in quality with period(s) determined from OGLE data.
At the same time the cut at 450 days allows overlap with OGLE to verify our procedures and compare periods and amplitudes.

A comparison with the OGLE-III catalogue of LPVs in the LMC \citep{SoszynskiLPVLMC} and SMC
\citep{SoszynskiLPVSMC} removes about 450 stars, and keeps about 480 LPVs with periods longer than 450 days.
Similar correlations were done with the list of MACHO detected LPVs in the LMC \citep{Fraser08},
and the catalogues of LPVs detected in the EROS-2 survey \citep{Spano11, Kim14}.

\subsubsection{Light curves}
\label{S-lc}

The final sample of stars for which the LCs were analysed contains 1299 objects and for those the $K_{\rm s}$-band epoch data were retrieved
from the VSA archive.
The information for all these objects is given in Appendix~\ref{App-A}.
Table~\ref{App-Sample} lists general information on the objects: the right ascension and declination
(from VMC, all epochs in this paper are J2000), some common
names and object type (from SIMBAD\footnote{As of January 2019.}), (selected) spectral types (from \citealt{Skiff14}),
the variability-type as listed in the 2nd {\it Gaia} data release (GDR2, \citealt{GDR2Sum,HollGDR2}), and the classification based on the
MIR {\it Spitzer} IRS spectrum; see the footnote to the table for more information.
Table~\ref{App-Periods2} lists known periods and related information from the literature, while
Table~\ref{App-Periods1} lists the results of the period analysis (see Sect.~\ref{S-modelling} for details).

The LCs consist of the VMC $K_{\rm s}$-band data with $K/K_{\rm s}$-band data found in the literature added (as detailed in
columns 7 and 8 of Table~\ref{App-Periods1}).
For some of the brightest stars the VMC photometry is influenced by non-linearity or saturation.
A correction is applied in the VISTA pipeline for this \citep{Irwin09} but for the brightest stars it does not correct sufficiently.
In the LC fitting this was accounted for by shifting the VMC $K_{\rm s}$-band photometry to externally
available photometry (see the remarks in Column~10 of Table~\ref{App-Periods1}).
The shift was determined from the difference in the median of the
VMC and the external photometry.
The shifted VMC photometry was also assigned an error bar of 0.25~mag.

No attempt was made to bring the various $K$-band measurements onto a common system.
The main reason is that colour transformations from one system to another almost always depend on
the $(J-K)$-colour \citep{Carpenter01, Koen07}.
A $J$-band magnitude observed simultaneously with the $K$-band data is not always available, the error on the $J$-band magnitude
is typically large for these red sources, and the $(J-K)$-colour is outside the range considered in the transformation formula.
This would make a colour transformation particularly complicated and uncertain in many cases.
This is a limitation, and in Appendix~\ref{App-Transf} we estimated the possible effect and come to the conclusion that
  it is typically smaller than the error bar quoted for mean magnitudes, amplitudes and periods in Table~\ref{App-Periods1}.
In the following text we refer to the $K$-band for simplicity unless we address a specific survey.

\section{Modelling}
\label{S-modelling}

The automatic analysis of the LCs is done with the Fortran codes available in {\it numerical recipes} \citep{Press1992} as
described in Appendix~A of \citet{Groenewegen04}. The codes were tuned at that time to analyse OGLE-II LCs with hundreds
of data points, while the current LCs have between 6 and 73 data points (92\% of the stars in the sample have 15 or more data points).
This makes the determination of the pulsation period through Fourier analysis (using the subroutine {\sc fasper}) more complicated.
Therefore the LCs of most stars in the sample were analysed manually with the
code {\sc Period04} \citep{Period04} as well.
Once an initial guess of the period was determined (either through the automatic routine, a period found in the literature, or
from the manual fitting of the LC) a function of the form
\begin{equation}
\label{Eq-fit}
K(t) = K_0 +
A \sin (2 \pi \; t \; e^{f}) +
B \cos (2 \pi \; t \; e^{f}) 
\end{equation}
was fitted to the LC using the weighted linear least-squares fitting routine {\sc mrqmin}.
This results in the parameters listed in Table~\ref{App-Periods1}, mean magnitudes, periods ($\exp (-f)$), and
amplitudes ($\sqrt{A^2 + B^2}$) with their associated uncertainties.
Equation~\ref{Eq-fit} implies that the LC can be described by a single period. It is well known that the LCs of LPVs are not
strictly single-periodic. The OGLE team typically fits three periods to the LC. However with the limited number of data points
available in the IR LCs we are not able in general to comment on the presence of more than one period. The exception is
the few stars that have been monitored extensively from the ground.
Table~\ref{App-Periods1} sometimes lists a long secondary period (LSP). Those were derived from the manual fitting of the LC using {\sc Period04}.

When no initial period is known from the literature it is taken from the Fourier analysis.
The reliability of this period determination (also) depends on the number
of data points in the LC. Selecting from Table~\ref{App-Periods1} those stars that have no initial period listed, and a S/N >3
on the final period and amplitude shows that the minimum number of useable data points is 12.

A comparison of the LC with the fit sometimes suggested that alternative periods may be possible as well.
These cases were inspected by manual fitting of the LC using {\sc Period04}, and alternative periods are sometimes
indicated in Col.~9 of  Table~\ref{App-Periods1}.

The table also includes the reduced $\chi^2$, 
defined as $\chi_{\rm r}^2 =  \sum_{\rm i} \left(((m_{\rm i} - o_{\rm i})/\sigma_{\rm o_{\rm i}})^2 \right)/ (N-N_{\rm p}),$ with $m$, $o$ and
$\sigma_{\rm o}$ indicating the model magnitude, the observed magnitude, and the error, $N$ is the number of data points,
and $N_{\rm p}$ = 1 or 4, depending on whether Eq.~\ref{Eq-fit} is fitted without or with the period.
The $\chi_{\rm r}^2$ in Table~\ref{App-Periods1} ranges from 0.2 to 3900, with a median of 70.
These values are very large compared to an expected value of unity in case of a model describing the data correctly and with
correct estimates of the error bars.
This probably indicates that the internal error bars of the $K_{\rm s}$-band photometry in the VSA (typically at the mmag level) are
likely underestimated when compared to other photometry on an absolute external scale, but also that differences in filter curves between the
different sets of photometry, and the simplicity of the model (a single period) play a role.
In the next section a selection is done eliminating stars with a $\chi_{\rm r}^2 > 430$, which represents about 10\% of the sample.

\section{Analysis and selection of new red LPVs}
\label{S-analysis}

%
\begin{figure*}
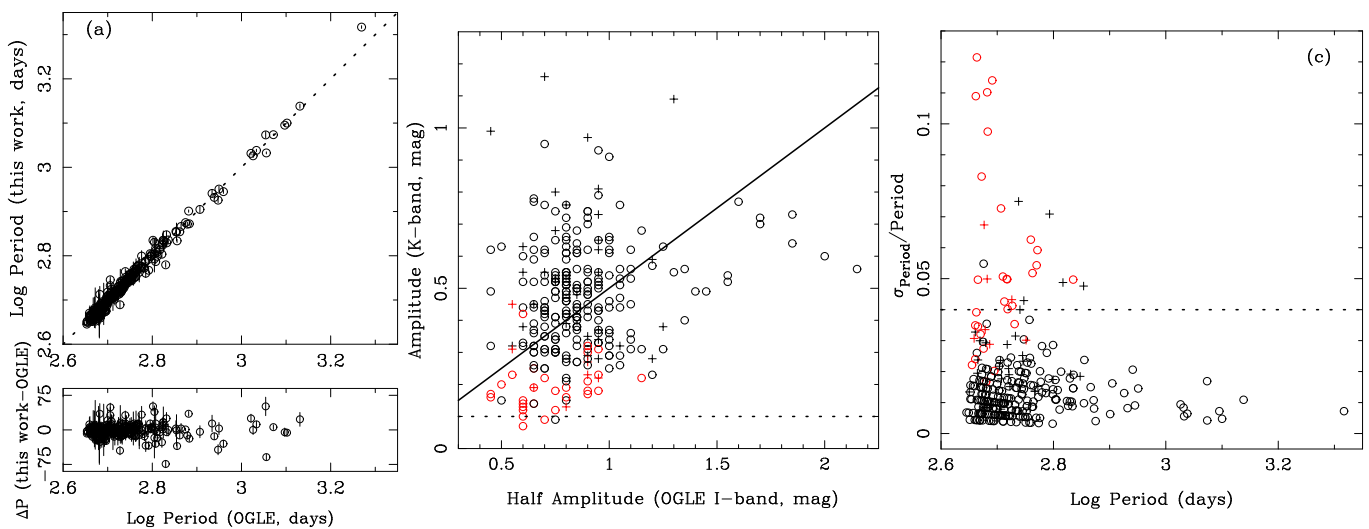

   \centering

\begin{minipage}{0.28\textwidth}
\resizebox{\hsize}{!}{\includegraphics{PerPer.ps}}
\end{minipage}
\begin{minipage}{0.34\textwidth}
\resizebox{\hsize}{!}{\includegraphics{AmpAmp.ps}}
\end{minipage}
\begin{minipage}{0.34\textwidth}
\resizebox{\hsize}{!}{\includegraphics{ePerPer.ps}}
\end{minipage}

\caption[]{
Comparison between the pulsation properties of 294 confirmed Miras by OGLE with the results in the present work.
The left panel (a) compares in the upper part the periods directly, with the dotted line marking the one-to-one relation.
The lower part indicates the difference in periods as a function of the period found by OGLE.
The star at $P= 1858$~d ($\log P= 3.26$) in the top panel is located outside the plot at $\Delta$Period = 216~d in the bottom panel.
The middle panel (b) compares the amplitude in the $K$-band to that in the $I$-band. Note that the OGLE team defines amplitude
as the minimum-to-maximum variation while we use the traditional definition (see Eq.~\ref{Eq-fit}).
Stars with a $\chi_{\rm r}^2 > 430$ are marked by a plus sign, stars where the error on the $K$-band amplitude is larger than the amplitude
are marked in red.
The dotted line indicates the cut-off value of 0.1~mag used to select the maximum number of known Miras from our sample.
The solid line has a slope 0.5 to guide the eye.
The right hand panel (c) plots the relative error of the period determination against the period.
Stars with a $\chi_{\rm r}^2 > 430$ are marked by a plus sign, stars where the error on the $K$-band amplitude is larger
than the amplitude are marked in red.
The dotted line indicates the cut-off value of 4\% error on the period determination used to select Miras from our sample.
}
\label{Fig-Compare}
\end{figure*}

\subsection{Comparison to known Miras}
\label{S-known}

The next step is a comparison of the results of the IR LC fitting to the periods of known Miras from OGLE-III. 
To compare between the OGLE and the IR pulsation properties we selected from Table~\ref{App-Periods2} the stars classified
by OGLE to be Miras and with an amplitude for the second listed period smaller than that listed for the primary period (as given by OGLE).
As mentioned above, many LPVs show a LSP, sometimes with a large amplitude, and this choice makes a more direct comparison of periods
and amplitudes possible.
After removing three stars for which the present analysis reveals no significant period
there remain 294 stars to make the comparison, and the results are shown in Figure~\ref{Fig-Compare}.
The panel on the left compares the primary period found by OGLE to the period derived in the present work.
The dotted line marks the one-to-one relation and shows that the agreement is very good in most cases.
The middle panel compares the amplitudes in the $K$-band as found in our analysis with the amplitude in the $I$-band as found by OGLE.
There is no strong correlation.
One observes the cut-off in $I$-band amplitude that reflects the common definition (as used also by the OGLE team)
that Miras should have an $I$-band amplitude larger than 0.9 mag. The solid line is plotted to guide the eye and indicates
an amplitude in $K$ that is half that in $I$.
The right-hand panel shows the relative error in the period determination plotted against period.
The majority of stars have a period determination better than a few percent.
The middle and right-hand panels identify outliers by the red symbols, stars where the amplitude is ill-defined
(the error on the amplitude being larger than the amplitude itself), and by plus signs where the $\chi_{\rm r}^2$ in the LC fitting is $> 430$.
Using the following selection criteria 291 of the 294 known Miras (based on $I$-band data) with low amplitude secondary periods are selected:
\begin{itemize}
\item error in the period determination better than 4\%.
\item error in the amplitude smaller than the amplitude.
\item amplitude larger than 0.1 mag in the $K$-band.
\item $\chi_{\rm r}^2$ in the LC fitting procedure smaller than 430.
\end{itemize}

When these criteria are applied to the full sample (and also enforcing $P > 100$ days to eliminate a few known Type-II Cepheids)
there are 634 objects of the 1299 that show pulsation properties as derived from the IR LCs consistent with those
of known Miras (as defined based on $I$-band data).
This includes the 291 stars that were used to define the criteria, but also known OGLE LPVs that were not used in the comparison (those
with larger LSP amplitudes). In addition there are objects where the classification of the {\it Spitzer} IRS spectrum
indicates that they are not AGB stars (see Table~\ref{App-Sample}).
The amplitude cut-off of 0.1~mag does not imply that these are Miras (as opposed to SRVs).
  They are (candidate) LPVs and we do not distinguish between the two classes.
  The middle panel in Figure~\ref{Fig-Compare} shows considerable scatter between the amplitude in the $I$ and $K$-band but
  also demonstrates that the majority of stars that have been classified as Miras based on $I$-band data have $K$-band
  amplitudes $>$0.2~mag, in line with expectations.

In a subsequent step all known OGLE objects were removed, with the exception of those stars with periods longer than 1000 days
as they are of prime interest here. This limit for a long period is arbitrary, but a limit of 1000 days was
also used by \citet{Menzies19} recently. Also stars detected by MACHO or EROS (and not by OGLE) were removed if the period listed
by these surveys was in good agreement with the period found here (and not longer than 1000 days).
Although these sources are not retained for further detailed analysis, the pulsation properties derived from the LC fitting
are still available in Table~\ref{App-Periods1}.

In a final step, the {\it Spitzer} IRS spectra of all those stars having a classification in the literature
not being that of an O- or C-rich AGB star (see Table~\ref{App-Sample}) were inspected visually using the
CASSIS tool\footnote{https://cassis.sirtf.com/} (Combined Atlas of Sources with {\it Spitzer} IRS spectra, \citealt{Lebouteiller11}).
All spectra appeared consistent with the literature classification, and thus unlike those of O- or C-rich AGB stars.
Therefore, they were removed from the sample of candidate LPVs.
The exception is SMP LMC 11 ($72.907608 -67.088019$).
Although it is classified as a planetary nebula (PN) in SIMBAD, and the literature classification of its {\it Spitzer} IRS spectrum is that of a
C-rich post-AGB star, its spectrum is, technically speaking, consistent with that of a very red C-rich AGB star and it is thus retained.
At this point the sample of stars that will be investigated further is reduced to 254.

\subsection{Creating and fitting the spectral energy distributions}
\label{S-Template}

As a further step in distinguishing AGB stars from non-AGB stars the SEDs of the 254 selected stars are constructed and
fitted to a large sample of synthetic template SEDs of AGB stars and some non-AGB stars (created from some of the non-AGB stars
previously discarded that have MIR spectra).

The procedure is largely outlined in GS18. Photometry is collected from AllWISE, SAGE, Akari, IRAS, OGLE and GDR2.
In the NIR we use the mean $K$-band magnitude derived from the LC fitting.
The VMC $Y$ and $J$ magnitude were downloaded from the VSA. The $K_{\rm s}$-band listed in the VSA was also downloaded and assigned
a large uncertainty of 0.1~mag so as to give this point less weight in the template fitting.
As mentioned in Sect.~\ref{S-lc}, for some of the brightest stars the VMC photometry is influenced by non-linearity or saturation.
For the SEDs the corresponding data points are plotted with large error bars (arbitrarily set at 0.25~mag) to identify them immediately and
in order not to affect the fitting procedure.
In some cases also the $Y$ and $J$ magnitudes may be affected by saturation as well and these are also assigned large error bars.

The magnitudes are then fitted to the synthetic photometry of several hundred templates. This indicates, with high reliability, whether
the SED of a star is consistent with that of an O-rich, C-rich, or young stellar object (YSO) or other non-AGB objects.
Details are given in Appendix~\ref{App-Template}.
The SEDs of all 254 objects with their best matching template are plotted in Fig.~\ref{Fig-SEDs}.

\begin{figure}
   \centering

\begin{minipage}{0.47\textwidth}

\resizebox{\hsize}{!}{\includegraphics[angle=0]{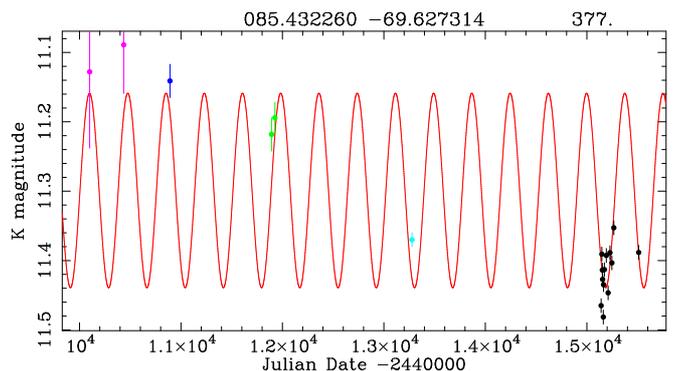}} 
\end{minipage}

\caption[]{
Examples of LCs of objects where the literature classification and the template fitting point to a non-AGB nature.
In the top panel the probable YSO object 2MASS J$04573274-6959032$ (based on a photometric classification; Table~10 in \citealt{Gruendl09})
and in the bottom panel the B1 emission-line star [BE74] 621.
Data points (with error bars) are identified as follows: 
black: VMC, green: 2MASS,  dark blue: 2MASS-6X, light blue: IRSF.
}
\label{Fig-YSO}
\end{figure}

Based on all available information (spectral type, MIR spectral classification, and template fitting) the 254 stars are divided into
217 most likely AGB stars, and 37 most likely non-AGB stars; see Table~\ref{Tab-Selected}.
The latter group is no longer discussed, although it contains sources with variability properties that have similarities
to those of LPVs. Figure~\ref{Fig-YSO} shows the light curves of a YSO and a B1 emission-line star as examples.
Table~\ref{App-Sample} lists a total of about 200 sources that have been associated with YSOs.
An analysis of the VMC-based IR LCs of Spitzer-identified massive YSOs in a $\sim$1.5 $\deg^2$ area in the LMC will be
presented in \citet{Zivkov20}.

\begin{figure}
   \centering

\begin{minipage}{0.45\textwidth}
\resizebox{\hsize}{!}{\includegraphics[angle=0]{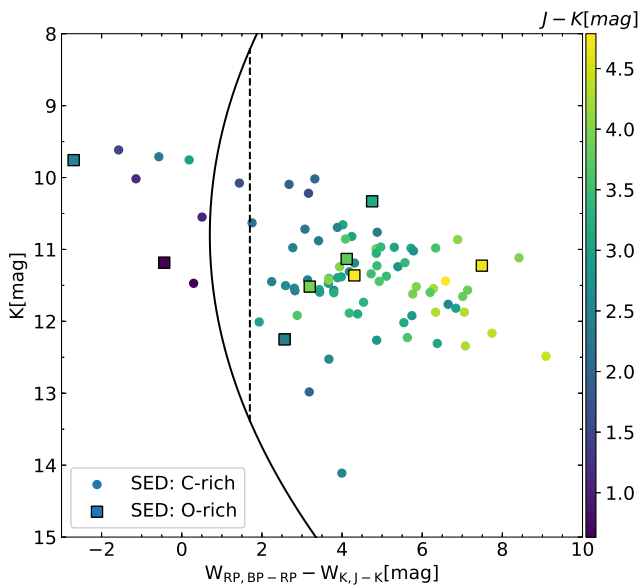}}
\end{minipage}
\caption[]{Classification of the AGB candidates in this work based on the VISTA $K_{\rm s}$-band and {\it GAIA} DR2 photometry \citep{Lebzelter18}.
  The solid line indicates the separation between O-stars (to the left of the curve) and C-stars,
  while the dashed line separates C-stars from the most extreme dust enshrouded stars. All the stars are colour-coded
  according to their $(J-K)$ colours.}
\label{GaiaDR2}
\end{figure}

The classification based on the SED fitting  is compared with that introduced by \citet{Lebzelter18} for
those AGB candidates for which the {\it Gaia} DR2 photometry is available (73 stars), as shown in Fig.~\ref{GaiaDR2}.
In \citet{Lebzelter18} the division between O- and C-rich stars is performed according to their position on a specific diagram
involving the difference between two Wesenheit functions obtained by combining the $J$, $K_{\rm s}$, $G_{BP}$ and
$G_{RP}$ photometric bands, $W_{RP, BP-RP} - W_{K_{\rm s}, J-K_{\rm s}}$
\citep[see the left panel of Figure 1, and Table~A.1 of][]{Lebzelter18}. The Wesenheit functions are defined as:
\begin{equation}
W_{K_{\rm s}, J-K_{\rm s}} = K_{\rm s} - 0.686 \cdot (J-K_{\rm s}),
\end{equation}
where $J$ and $K_{\rm s}$ are the VISTA bands, and:
\begin{equation}
W_{RP, BP-RP} = WRP = G_{RP} - 1.3 \cdot (G_{BP} - G_{RP}).
\end{equation}
Among the selected stars, 65 sources are classified as C-rich and eight as O-rich.
We find that 72 sources are classified as C-stars by both methods, while only six sources classified as extreme stars
  by \citet{Lebzelter18} are classified as O-rich by the SED template fitting  developed here.
  Two stars out of eight that are O-rich for \citet{Lebzelter18} are classified as O-rich on the basis of the templates.
  Therefore, when considering the classification by \citet{Lebzelter18}, the SED classification method yields a comparable
  result for C-stars, but disagree on O-rich stars.
According with the SED fitting classification method, among the 217 AGB candidates, 22 are O-rich, 1 is a PN and the remaining are C-rich stars.

\begin{figure}
   \centering

\begin{minipage}{0.45\textwidth}
\resizebox{\hsize}{!}{\includegraphics[angle=0]{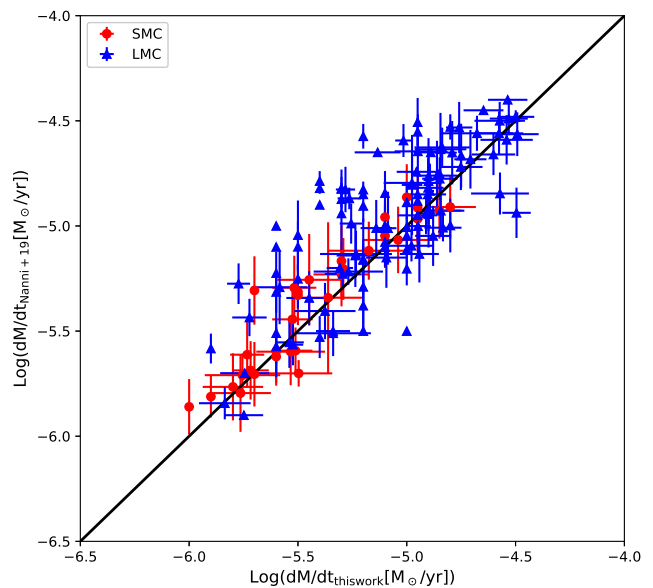}}
\end{minipage}
\caption[]{
Comparison between the MLR obtained by \citet{Nanni19b} against that derived 
in this work for stars in common. The solid black line is the one-to-one relation.
}
\label{P_oldnew}
\end{figure}

The AGB candidates are also fitted using the public database of synthetic spectra and photometry for C-stars presented in \citet{Nanni19b}.
Such spectra are obtained by consistently computing the growth of dust grains of different species (amorphous carbon and
silicon carbide for C-stars) coupled with a stationary dust-driven wind \citep{Nanni13, Nanni14} and a dust radiative
transfer code \citep{Gr_MOD}. The optical constants for amorphous carbon dust are selected to reproduce the photometry in
the IR and in the {\it Gaia} bands \citep{Nanni16, Nanni19a}. We here adopt the data set of \citet{Hanner88} combined with
small grains $<0.04$ $\mu$m. The metallicity of the synthetic models is $Z = 0.004$ and $0.006$ for the SMC
and LMC, respectively. As in \citet{Nanni19b} we select only those models among the ones computed for which the outflow
is accelerated through dust-driven wind when the input MLR is $\log(\dot{M}) >= -5.5$~\msolyr.  The best fit between the observed and
synthetic photometry allows us to estimate the (gas) MLR, the dust production rate and the luminosity of each star,
without the need to assume quantities such as the gas-to-dust ratio and the wind speed of the outflow. The results for the
luminosity and MLRs derived from the fitting procedure are provided in Table~\ref{Tab-Nanni}.  
In Fig.~\ref{P_oldnew} the MLRs derived in this work and from \citet{Nanni19b} for sources in common are compared.
We find that the results are in good agreement with a scatter of 0.23 dex.

Only 44 and eight C-stars analysed in this work are missing from the analysis of \citet{Nanni19b} for the LMC and the SMC,
respectively. Indeed, these sources were not included in the catalogues matched by \citet{Nanni19b}
(which included C-stars from \citep[][GS18]{Riebel12,Ruffle15, Srinivasan16, Jones17}.
The total dust production rate of these missing stars is calculated, and it is found that their total contribution
is negligible with respect to the total dust production rate of the entire AGB population in these galaxies.

\begin{figure}
   \centering

\begin{minipage}{0.4\textwidth}
\resizebox{\hsize}{!}{\includegraphics[angle=0]{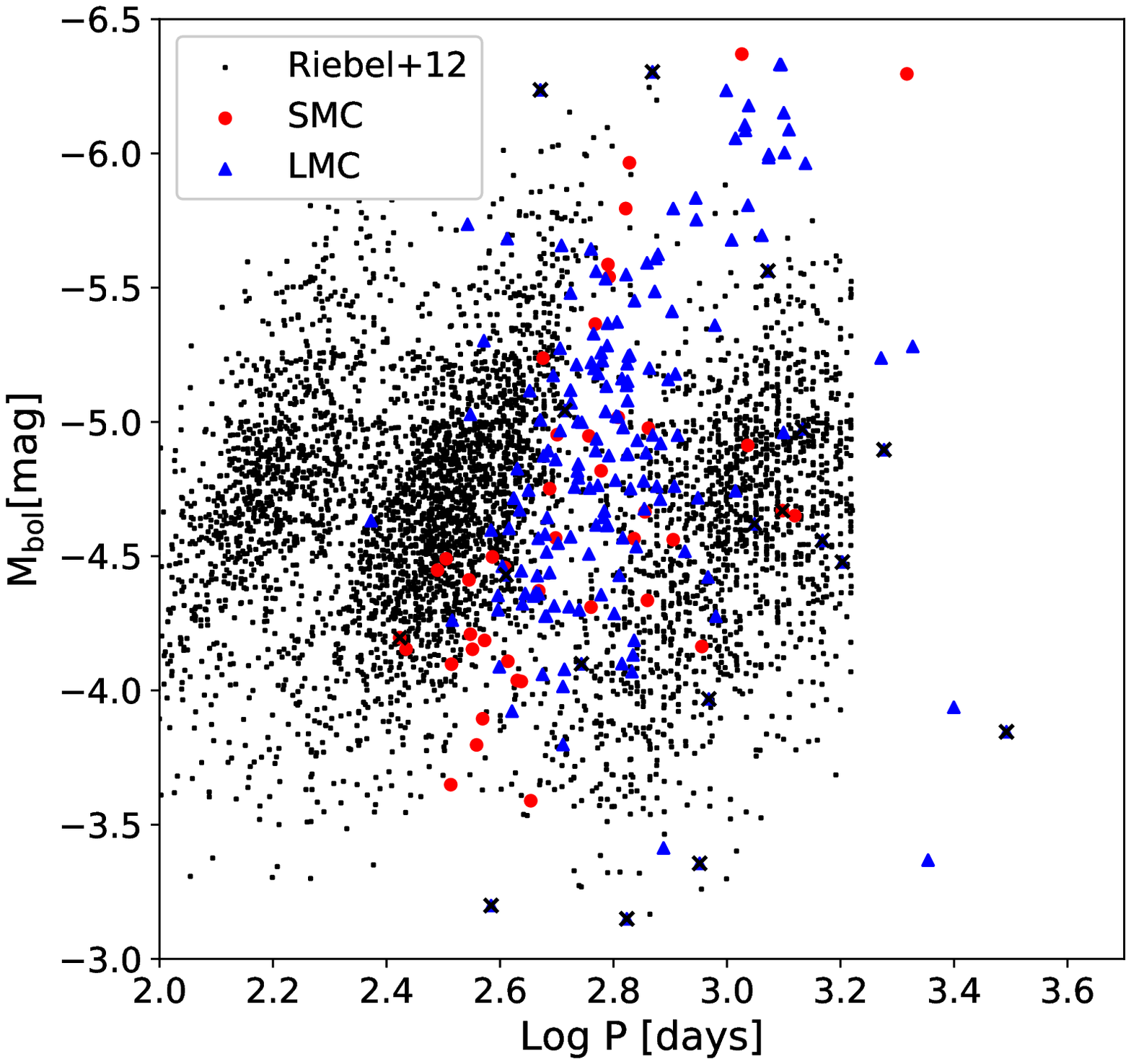}}
\end{minipage}
\begin{minipage}{0.4\textwidth}
\resizebox{\hsize}{!}{\includegraphics[angle=0]{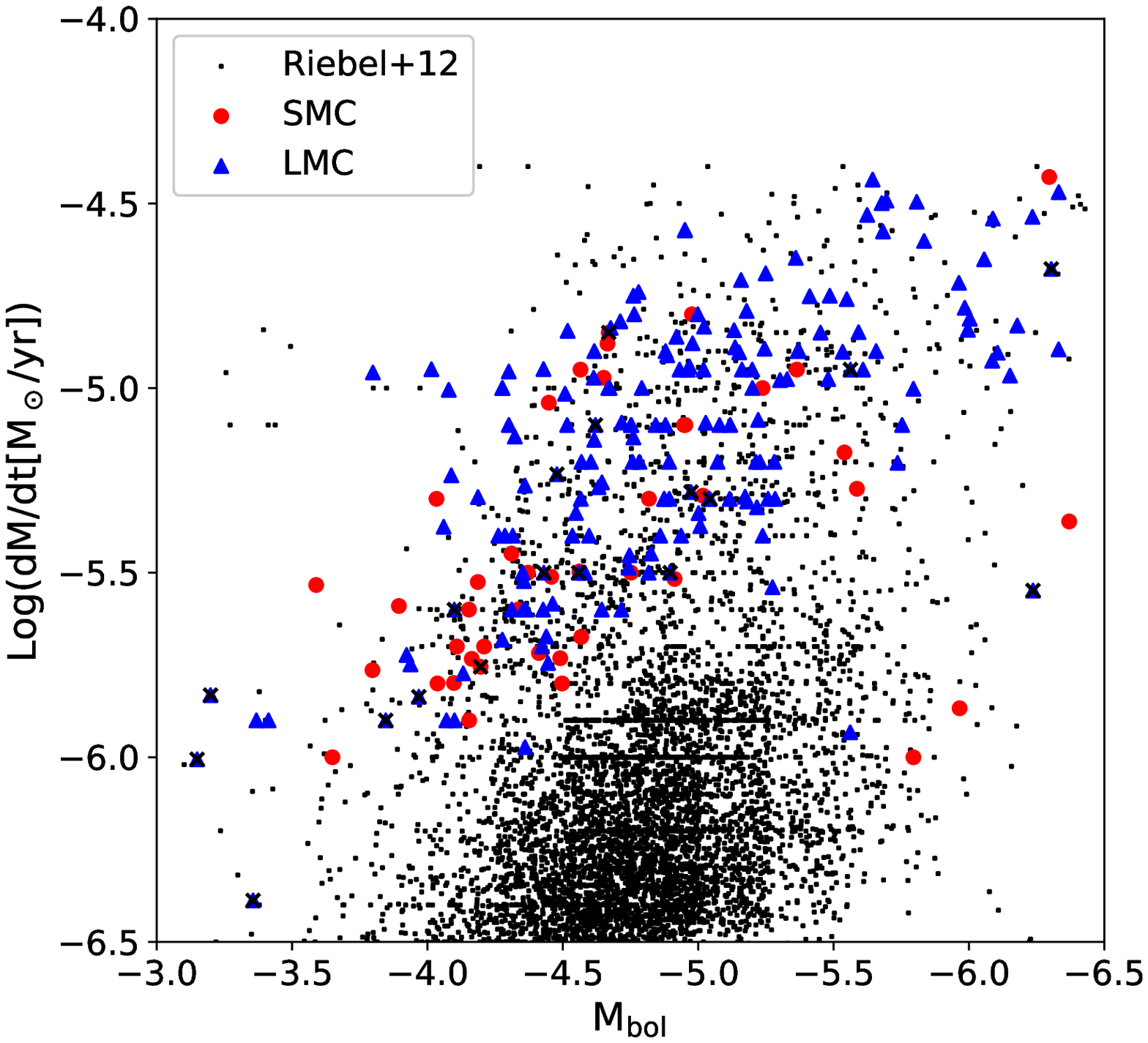}}
\end{minipage}
\begin{minipage}{0.4\textwidth}
\resizebox{\hsize}{!}{\includegraphics[angle=0]{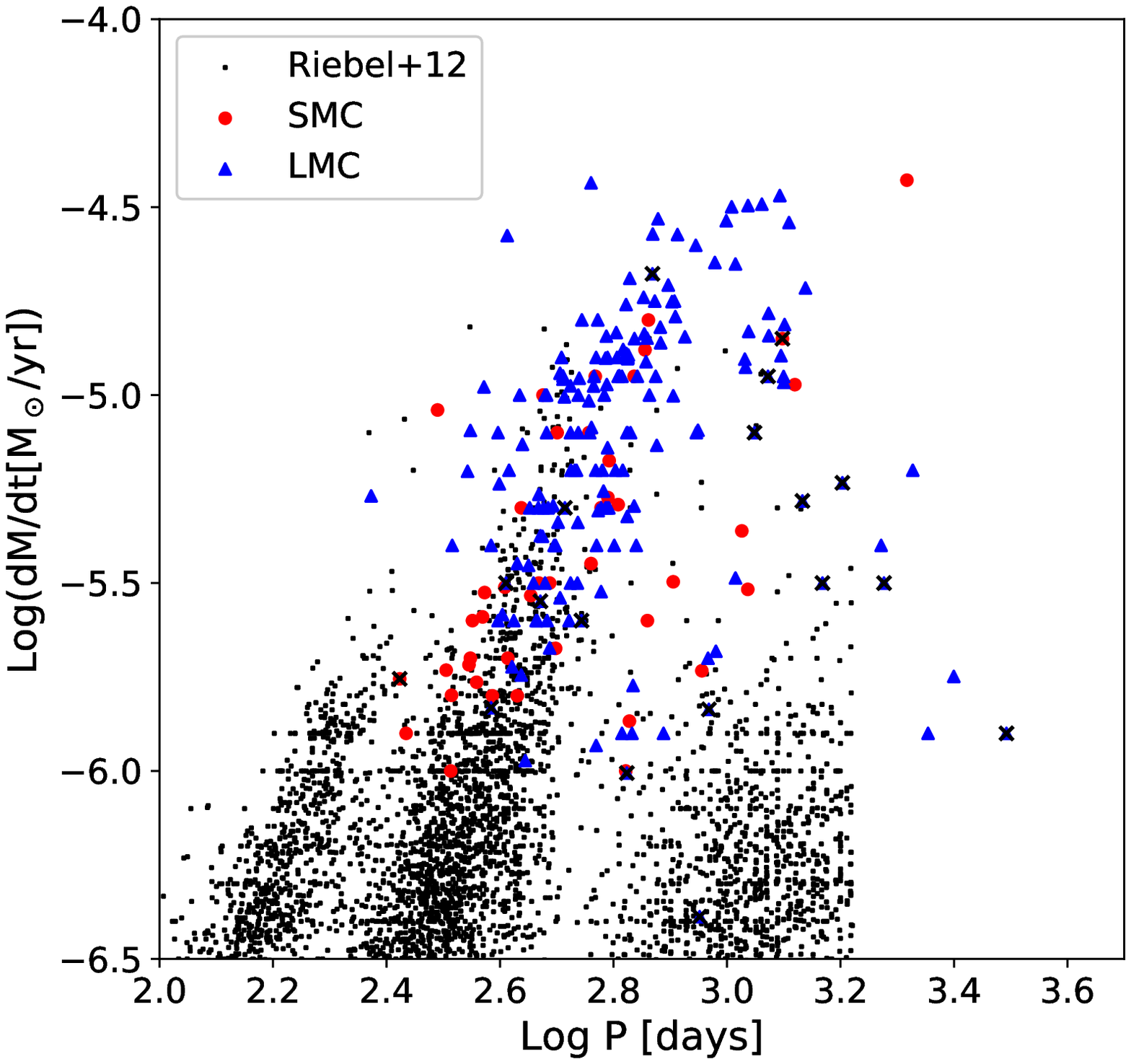}}
\end{minipage}
\caption[]{
Top panel:
bolometric luminosity versus period obtained from the SED fitting of the C-stars selected in this work: C-stars in the SMC are indicated
with red dots, while C-stars in the LMC are plotted with blue triangles. For comparison, we over plot the population of C-stars in the LMC
for which the period has been determined from \citet{Riebel12} which have been fitted by \citet{Nanni19b}.
To avoid superposition in the plot, M$_{\rm bol}$ has been changed by a random number between $-0.25$ and $+0.25$~mag.
The sources for which the period has been flagged with a superscript `a' or `b' in Table~\ref{Tab-Selected} have been indicate with a cross.
The middle and bottom panels show the relations between
MLR versus the bolometric luminosity and
MLR versus period, respectively.
}
\label{P_Mbol}
\end{figure}

The properties derived for the sources analysed in this work (period, bolometric luminosity and MLRs) are compared with those
derived by \citet{Nanni19b} for the entire population of C-stars in the LMC for which the period has been determined \citep{Riebel12}.
The results are shown in Fig.~\ref{P_Mbol}, where we correlate the luminosity with the period, the MLR with the luminosity,
and the MLR with the period. The sources analysed in this work span the entire range of $M_{\rm bol}$ while the periods are
usually between $2.4 <\log P$ (d) $<3.1$. The estimated MLR is in general greater than $\approx 10^{-6}$ M$_\odot$ yr$^{-1}$ and
correlates with the luminosity, even though the scatter is large, and with the period.

\section{Discussion}

\subsection{A comparison with periods from {\it Gaia}}

Table~\ref{Tab-Gaia} contains 32 sources which are listed in GDR2 with a `MIRA\_SR' classification and a period.
The comparison with the period derived from the NIR data in the present paper is excellent;
the value of
$(P_{\rm Gaia} - P_{\rm NIR})/\sigma_{\rm Gaia}$ ranges from $-2.2$ to $+2.4$ with a median value of $+0.1$.
The periods that are presented here are more precise (the median error in the period is 70 days in the {\it Gaia} data
and eight days in the present work), but the error in the period determination by {\it Gaia} is expected
to decrease significantly with further data releases.

\begin{table*}

\caption{Comparison of periods for sources in common between our study and {\it Gaia} DR2.}
\label{Tab-Gaia}
\centering
  \begin{tabular}{ccrr}
  \hline
RA \hfill Dec        & GDR2 Source Id     & Period          & Period           \\ 
(deg) \hfill (deg)   &                    & {\it Gaia} (d)  & this work (d)     \\
\hline
04.876441 $-$72.465755 & 4689594530461666432 &  335 $\pm$ 29 & 265 $\pm$ 1 \\ 
07.572240 $-$72.472302 & 4689184000304539520 &  531 $\pm$ 85 &  508 $\pm$   12 \\ 
08.511784 $-$72.363404 & 4689188428414841216 &  336 $\pm$ 37 &  356 $\pm$ 1 \\ 
09.660904 $-$72.008445 & 4689221379397747968 &  513 $\pm$ 90 &  385 $\pm$    8 \\ 
09.828878 $-$70.131865 & 4702368897059777792 &  343 $\pm$ 27  & 320 $\pm$ 4  \\ 
13.053875 $-$73.148050 & 4685938821845281280 &  821 $\pm$ 284 &  796 $\pm$    5 \\ 
13.292230 $-$72.198509 & 4689061778454380416 &  624 $\pm$ 157 &  491 $\pm$   56 \\ 
21.130633 $-$71.620707 & 4687594377067361280 &  549 $\pm$ 103 &  724 $\pm$ 12 \\
25.642854 $-$71.369942 & 4687832356908117120 &  562 $\pm$ 172 &  545 $\pm$   23  \\
69.714864 $-$68.402870 & 4656274857070901760 &  580 $\pm$ 103 &  513 $\pm$    2  \\
73.110342 $-$68.577069 & 4655516060931435008 &  3937 $\pm$ 3845 &  744 $\pm$    8  \\
73.927019 $-$68.281715 & 4661533550631677184 &  466 $\pm$ 62 &  468 $\pm$    7  \\
74.358023 $-$74.020732 & 4649773380396677376 &  437 $\pm$ 50 &  547 $\pm$   11 \\
74.535830 $-$67.878353 & 4661558083486627328 &  433 $\pm$ 53 &  474 $\pm$   13 \\
74.536373 $-$71.966533 & 4654485474923219968 & 443 $\pm$ 73 &  536 $\pm$    3  \\
74.731098 $-$68.218399 & 4661543033920475392 & 612 $\pm$ 70 &  495 $\pm$   10  \\
76.076144 $-$74.117919 & 4649713079054327424 & 450 $\pm$ 58 &  493 $\pm$   13 \\
77.550727 $-$65.326051 & 4663596882956831104 & 664 $\pm$ 68 &  599 $\pm$   98 \\
78.161438 $-$64.203811 & 4664123102346900864 & 723 $\pm$ 216 &  883 $\pm$   14 \\
79.026012 $-$70.825900 & 4651988037356793600 & 515 $\pm$ 122 &  516 $\pm$   10 \\
79.994330 $-$70.471053 & 4651927113213190272 & 512 $\pm$ 56 &  455 $\pm$    8  \\
80.792265 $-$67.835106 & 4658834421384702208 & 593 $\pm$ 95 &  714 $\pm$   34 \\
80.802424 $-$64.158664 & 4661168096128809216 & 435 $\pm$ 43 &  449 $\pm$    3  \\
81.285698 $-$73.278491 & 4651133854242537728 & 321 $\pm$ 25 &  310 $\pm$    6 \\
82.281667 $-$66.970853 & 4660210735084020608 & 495 $\pm$ 54 &  490 $\pm$    6  \\
82.751418 $-$69.179865 & 4658428873394148352 & 645 $\pm$ 174 &  612 $\pm$   32  \\
83.558539 $-$68.978837 & 4658446499944028800 & 550 $\pm$ 249 &  548 $\pm$    7 \\
83.720398 $-$71.006350 & 4657091970360446464 & 710 $\pm$ 121 &  715 $\pm$    7  \\
83.767073 $-$69.954142 & 4657247619989849472 & 532 $\pm$ 61 &  523 $\pm$   21 \\
85.201902 $-$69.560027 & 4657599188763136896 &  1173 $\pm$ 829 &  576 $\pm$    5 \\
86.488037 $-$64.462946 &  4756251319078635648 &  423 $\pm$ 55 &  388 $\pm$    9  \\
89.223262 $-$67.070124  & 4659600643536501504 & 509 $\pm$ 87 &  414 $\pm$   35 \\
\hline
\end{tabular}
\tablefoot{GDR2 does not list periods but a frequency $\omega$ and its error $\sigma$.
           Period and error are calculated as $1/\omega$ and  $1/\omega - 1/(\omega+\sigma)$ for simplicity.
}
\end{table*}


\subsection{A comparison to the IRSF monitoring data in the SMC}

Three stars in our sample have NIR monitoring data recently published by \citet{ItaSMC} using the
InfraRed Survey Facility (IRSF, \citealt{Kato_IRSF}).
These data were not used in our LC analysis, and can therefore serve as an independent test.
The $K$-band data ($H$-band for 12.627600 $-72.858307$) were analysed in the same way as the VMC data complemented with literature data.
Table~\ref{Tab-IRSF} compares the periods and amplitudes.
For the first two stars the agreement is good.
In the last case the periods do not agree within their mutual error bars, but this is a faint source for the IRSF survey ($K \sim 15.5$~mag)
while our analysis might be hampered by the relatively small number of data points (18 in total).

\begin{table*}

\caption{Comparison with three sources in the IRSF survey.}
\label{Tab-IRSF}
\centering
\begin{tabular}{ccrrrr}
  \hline
RA \hfill Dec        & Filter & Period    & Amplitude    & Period   & Amplitude   \\
(deg) \hfill (deg)   &        &  (d)      &  (mag)       &  (d) & (mag)   \\
  \hline
                     & \multicolumn{3}{c}{IRSF} & \multicolumn{2}{c}{VMC} \\
12.627600 $-$72.858307 &  $H$  &  1035.8 $\pm$ 5.3  & 0.39 $\pm$ 0.04 & 1062 $\pm$ 10 &  0.56 $\pm$ 0.05\tablefootmark{a} \\
14.860309 $-$72.394926 &  $K$  &   421.9 $\pm$ 1.0  & 0.54 $\pm$ 0.03 &  434 $\pm$  3 &  0.52 $\pm$ 0.20 \\
15.173292 $-$72.633497 &  $K$  &   391.6 $\pm$ 2.4  & 0.53 $\pm$ 0.14 &  451 $\pm$  7 &  0.37 $\pm$ 0.15 \\
\hline
\end{tabular}
  \tablefoot{
    \tablefoottext{a}{In the $K$-band.}
    }
\end{table*}

\subsection{LPVs with periods over 1000 days}

Among the likely AGB stars 34 stars have periods longer than 1000 days\footnote{We include one star with a period
  of 997 $\pm$ 12 days.}. The longest period we find is 3108~d, but this object has alternative periods that are
much shorter. 
Six other stars also have shorter possible periods, and therefore the periods beyond 1000 days are only tentative in these cases.
It will be interesting to analyse the IRSF data for the LMC objects when it becomes available (cf. the SMC data
presented in \citealt{ItaSMC}).

Nine stars with periods over 1000~days in the MCs were previously reported by \citet{Wood1992} and \citet{Whitelock2003}
based on NIR observations.
Recently, \citet{Menzies19} presented a compilation of Miras with periods over 1000~days in the Milky Way (17), SMC (3), and LMC (18),
 while \citet{Karambelkar19} published a list of 417 luminous infrared variables, of which 86 have periods over 1000~days, located in
 seven galaxies (the majority in M~83 (49 objects), and M~81 (11 objects)).
 The compilation by   \citet{Menzies19} was based on \citet{Wood1992} and \citet{Whitelock2003}, OGLE data in the SMC and LMC
 \citep{SoszynskiLPVSMC,SoszynskiLPVLMC}, and GS18 (which is partly based on a preliminary analysis of VMC data).
 Of the 21 MC objects, 18 are confirmed with updated periods. Three objects are not in our list.
 One object (IRAS 05506--7053) is too red; the period in GS18 is based on {\it WISE} data. The two others (HV~888 and HV~11417) 
 are too blue to have passed the selection criteria, and both have been classified as RSGs (GS18, \citealt{Kraemer17}).

The longest period in the study of GS18 of almost 400 AGB stars with {\it Spitzer} IRS spectra is 1810~d based on a re-analysis of
OGLE-III data for MSX SMC 055 (OGLE-SMC-LPV-08137, IRAS $00483-7347$).
No error on the period is given in GS18 as they are very hard to reliably estimate for these long periods.
The period in the OGLE-III release is 1859~d which gives some indication of the uncertainty.
In \citet{GS09} such a long-period was already recognised
based on OGLE-II data and it was suggested that the object may be a candidate super-AGB star, based on its long period and very high luminosity.
Updated parameters were given in GS18, where a current pulsation mass of about 9~\msol\ was derived.
It has a large amplitude (0.8 mag in $I$) but its LC is not very regular
in shape (see the middle panel of Fig.~7 in  \citealt{SoszynskiLPVSMC}).
Based on the IR LC a period of 2075 $\pm$ 15~d is derived here, about $4~\sigma$ away from the optically determined
period considering the mutual error bars.

There are two stars in the present sample of likely AGB stars that have very
long periods\footnote{Excluding 70.632433 $-74.796568$ with $P= 3108$~d because of the possible presence of also a shorter period,
  and 81.098476  $-65.536337$ with $P= 2125$~d but a rather poor fit and an uncertain amplitude of 0.38 $\pm$ 0.18~mag.}.
They are
VMC J$052454.84-682958.02$ (81.228480 $-68.499449$) with  $P= 2261 \pm 41$~d and amplitude 0.20 $\pm$ 0.02~mag, and
VMC J$045211.18-701244.83$ (73.046582 $-70.212454$) with  $P= 2510 \pm 67$~d and amplitude 0.15 $\pm$ 0.06~mag.
Little is known about the former source. It is not listed in SIMBAD. Its SED is very well fitted by that of a dusty O-rich star,
but its luminosity based on the template fitting is low ($L \sim 1500$~\lsol), and its amplitude small for a Mira variable.
The latter source is listed as a candidate post-AGB object. In ViZieR there are catalogues associated with a source classified
as a PN, but often located at about 2\arcsec\ away.
VMC $Y$ and $K_{\rm s}$-band images were inspected.
The finding chart in \citet{Morgan94} (object 9) suffers from poor resolution and is not clear.
A slightly better chart was published by \citet{Leisy1997} but it does not allow to unequivocally distinguish between two sources in the VMC images.
The coordinates published there and in \citet{ReidParker06} (RP 1607) refer to a
source located at about  2.2\arcsec\ to the SE of the very red VMC source referred to in this paper.
In any case, the luminosity based on the template fitting is low ($L \sim 3000$~\lsol), and its amplitude small for a Mira variable.

One of the original ideas behind this work was to find stars with very long periods as a tracer of the most massive AGB and super-AGB stars.
The nature of the two stars with 2000+ day periods would be interesting to investigate but they are unlikely to be as
massive as the about 9~\msol\ estimated for MSX SMC 055 (GS18).

\subsection{The bolometric $PL$-relation}

\begin{figure}
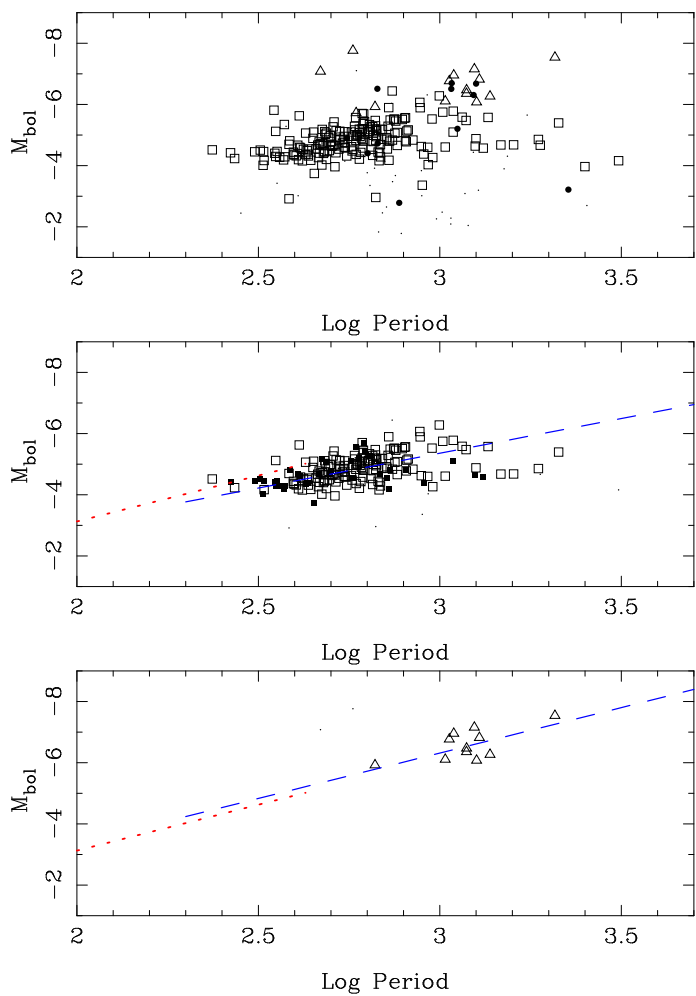

   \centering

\begin{minipage}{0.49\textwidth}
\resizebox{\hsize}{!}{\includegraphics[angle=0]{PMbol_All254.ps}}
\resizebox{\hsize}{!}{\includegraphics[angle=0]{PMbol_C_miras.ps}}
\resizebox{\hsize}{!}{\includegraphics[angle=0]{PMbol_O_miras.ps}}
\end{minipage}

\caption[]{
Bolometric $PL$-relations for all the 254 objects for which the template fitting was performed (top panel), and the subsets of
likely C-rich Miras (middle panel), and O-rich Miras (bottom panel). 
The likely C-rich objects  are plotted as open squares,
the likely O-rich objects are plotted as open triangles,
the other AGB stars as filled circles, see the main text for details.
The 37 likely non-AGB stars are plotted as dots.
The likely C-rich Miras (with an amplitude $> 0.2$~mag) are plotted in the middle panel.
The objects in the LMC are plotted with open symbols, those in the SMC with filled symbols.
The blue dashed line is a fit to the data, excluding outliers (plotted as dots).
The likely O-rich Miras (with an amplitude $> 0.2$~mag) are plotted in the bottom panel.
The blue dashed line is a fit to the data, excluding outliers (plotted as dots).
The red dotted lines in the two panels indicate relations from the literature derived
at shorter periods, see the main text for details.
}
\label{Fig-PL}
\end{figure}

Figure~\ref{Fig-PL} shows the bolometric $PL$-relation.
The top panel shows all 254 objects for which the  template fitting was performed.
The likely C-rich objects (i.e. listed as `C' in col.~10 in Table~\ref{Tab-Selected}) are plotted as open squares,
the likely O-rich objects (i.e. listed as `O' in Table~\ref{Tab-Selected}) are plotted as open triangles,
the other AGB stars (i.e.  listed as O(AGB), O(OTH), C(AGB) in Table~\ref{Tab-Selected}) as filled circles.
The 37 likely non-AGB stars are plotted as dots.
The likely C-rich Miras (with an amplitude $> 0.2$~mag) are plotted in the middle panel.
The blue dashed line is a fit to the data, excluding seven outliers (plotted as dots) based
on iterative 3~$\sigma$ clipping.
The fit is $M_{\rm bol} = (-2.27 \pm 0.20) \cdot \log P +  (1.45 \pm 0.54)$~mag using 182 stars and with an rms of 0.41~mag
using a linear bi-sector fit (using the code SIXLIN from \citealt{Isobe90}).
Restricting the sample to LMC objects only (144 objects), or
eliminating the objects for which an alternative period can not be excluded,  
do not significantly change the coefficients of the fit or the rms value.

The likely O-rich Miras (with an amplitude $> 0.2$~mag) are plotted in the bottom panel.
The blue dashed line is a fit to the data, excluding two outliers (plotted as dots) based
on iterative 3~$\sigma$ clipping.
The fit is $M_{\rm bol} = (-2.97 \pm 0.09) \cdot \log P +  (2.59 \pm 0.28)$~mag based on 11 stars and with an rms of 0.36~mag.
The red dotted lines in the two panels indicate relations from \citet{Feast1989} for C- and O-rich Miras in the LMC respectively.
These relations were derived for $P < 420$~d, and were shifted using an LMC  distance modulus of 18.477~mag from \citet{Pietrzynski19}.

\section{Summary}

Based on CCD and CMD diagrams and the properties of known Mira variables detected by the OGLE survey a sample
of 1299 sources is selected as (candidate) LPVs with periods longer than 450 days. Multiple VMC $K_{\rm s}$-band data are combined with literature
data to construct LCs.
A limitation is that magnitudes from different $K$-type filters are mixed, as it is close to impossible to uniformly and
  consistently convert all data to a uniform system. 
The LCs are analysed for periodic variations. Mean magnitudes are published for all stars, and periods and amplitudes for 949 stars.
Based on the pulsation properties of known OGLE Miras, and removing all stars with a known period (except when longer than 1000 days), a sample
of 254 stars is retained for further study. The spectral energy distributions of these stars are derived and fitted with template distributions of
known objects. A final list of 217 candidate LPVs is obtained. Thirty-four objects have periods longer than 1000 days, but some have a
viable alternative period that is shorter. The longest-period known Mira in the Magellanic Clouds from OGLE data (with $P= 1810$~d) is
derived to have a period of 2075~d based on its IR LC.
Two stars are found to have longer periods, but both have luminosities and pulsation amplitudes unlike those of Miras.
The bolometric magnitude - period relation is derived for both likely O- and C-rich stars in the sample.

\begin{acknowledgements}
We thank the Cambridge Astronomy Survey Unit (CASU) and the Wide Field Astronomy Unit (WFAU) in Edinburgh
for providing calibrated data products under the support of the Science and Technology Facility Council (STFC) in the UK.
M.-R.L.C acknowledges support from the European Research Council (ERC) under the
European Union's Horizon 2020 research and innovation programme (grant agreement No 682115).
A.N. acknowledges support from the Centre National d’Etudes Spatiale (CNES) through a post-doctoral fellowship.
This publication is partly based on the OGLE observations obtained
with the Warsaw Telescope at the Las Campanas Observatory, Chile,
operated by the Carnegie Institution of Washington.
This paper utilises public domain data originally obtained by the MACHO Project,
whose work was performed under the joint auspices of the U.S. Department of Energy,
National Nuclear Security Administration by the University of California, Lawrence Livermore
National Laboratory under contract No. W-7405-Eng-48, the National Science Foundation
through the Center for Particle Astrophysics of the University of California under
cooperative agreement AST-8809616, and the Mount Stromlo and Siding Spring Observatory,
part of the Australian National University.
This publication makes use of data products from the Wide-field
Infrared Survey Explorer, which is a joint project of the University
of California, Los Angeles, and the Jet Propulsion Laboratory/California Institute of Technology,
funded by the National Aeronautics and Space Administration.
This research has made use of the SIMBAD database and the VizieR catalogue access tool, both operated at CDS, Strasbourg, France.
This work has made use of data from the European Space Agency (ESA) mission
{\it Gaia} (\url{https://www.cosmos.esa.int/gaia}), processed by the {\it Gaia}
Data Processing and Analysis Consortium (DPAC,
\url{https://www.cosmos.esa.int/web/gaia/dpac/consortium}). Funding for the DPAC
has been provided by national institutions, in particular the institutions
participating in the {\it Gaia} Multilateral Agreement.
\end{acknowledgements}


\bibliographystyle{aa}
	\bibliography{references}

\begin{thebibliography}{117}
\expandafter\ifx\csname natexlab\endcsname\relax\def\natexlab#1{#1}\fi

\bibitem[{{Aringer} {et~al.}(2009){Aringer}, {Girardi}, {Nowotny}, {Marigo}, \&
  {Lederer}}]{Aringer09}
{Aringer}, B., {Girardi}, L., {Nowotny}, W., {Marigo}, P., \& {Lederer}, M.~T.
  2009, \aap, 503, 913

\bibitem[{{Aringer} {et~al.}(2019){Aringer}, {Marigo}, {Nowotny}, {Girardi},
  {Me{\v{c}}ina}, \& {Nanni}}]{Aringer19}
{Aringer}, B., {Marigo}, P., {Nowotny}, W., {et~al.} 2019, \mnras, 487, 2133

\bibitem[{{Bessell} \& {Brett}(1989)}]{BB89}
{Bessell}, M.~S. \& {Brett}, J.~M. 1989, {JHKLM Photometry: Standard Systems,
  Passbands and Intrinsic Colors}, ed. E.~F. {Milone}, Vol. 341, 61

\bibitem[{{Bolatto} {et~al.}(2007){Bolatto}, {Simon}, {Stanimirovi{\'c}}, {van
  Loon}, {Shah}, {Venn}, {Leroy}, {Sand strom}, {Jackson}, {Israel}, {Li},
  {Staveley-Smith}, {Bot}, {Boulanger}, \& {Rubio}}]{Bolatto07}
{Bolatto}, A.~D., {Simon}, J.~D., {Stanimirovi{\'c}}, S., {et~al.} 2007, \apj,
  655, 212

\bibitem[{{Boyer} {et~al.}(2011){Boyer}, {Srinivasan}, {van Loon}, {McDonald},
  {Meixner}, {Zaritsky}, {Gordon}, {Kemper}, {Babler}, {Block}, {Bracker},
  {Engelbracht}, {Hora}, {Indebetouw}, {Meade}, {Misselt}, {Robitaille},
  {Sewi{\l}o}, {Shiao}, \& {Whitney}}]{Boyer2011}
{Boyer}, M.~L., {Srinivasan}, S., {van Loon}, J.~T., {et~al.} 2011, \aj, 142,
  103

\bibitem[{{Carpenter}(2001)}]{Carpenter01}
{Carpenter}, J.~M. 2001, \aj, 121, 2851

\bibitem[{{Cioni} {et~al.}(2011){Cioni}, {Clementini}, {Girardi}, {Guandalini},
  {Gullieuszik}, {Miszalski}, {Moretti}, {Ripepi}, {Rubele}, {Bagheri},
  {Bekki}, {Cross}, {de Blok}, {de Grijs}, {Emerson}, {Evans}, {Gibson},
  {Gonzales-Solares}, {Groenewegen}, {Irwin}, {Ivanov}, {Lewis}, {Marconi},
  {Marquette}, {Mastropietro}, {Moore}, {Napiwotzki}, {Naylor}, {Oliveira},
  {Read}, {Sutorius}, {van Loon}, {Wilkinson}, \& {Wood}}]{Cioni11}
{Cioni}, M.-R.~L., {Clementini}, G., {Girardi}, L., {et~al.} 2011, \aap, 527,
  A116

\bibitem[{{Cioni} {et~al.}(2017){Cioni}, {Ripepi}, {Clementini}, {Groenewegen},
  {Moretti}, {Muraveva}, \& {Subramanian}}]{Cioni2017}
{Cioni}, M.-R.~L., {Ripepi}, V., {Clementini}, G., {et~al.} 2017, in European
  Physical Journal Web of Conferences, Vol. 152, European Physical Journal Web
  of Conferences, 01008

\bibitem[{{Cross} {et~al.}(2012){Cross}, {Collins}, {Mann}, {Read}, {Sutorius},
  {Blake}, {Holliman}, {Hambly}, {Emerson}, {Lawrence}, \& {Noddle}}]{Cross12}
{Cross}, N.~J.~G., {Collins}, R.~S., {Mann}, R.~G., {et~al.} 2012, \aap, 548,
  A119

\bibitem[{{Cutri} {et~al.}(2003){Cutri}, {Skrutskie}, {van Dyk}, {Beichman},
  {Carpenter}, {Chester}, {Cambresy}, {Evans}, {Fowler}, {Gizis}, {Howard},
  {Huchra}, {Jarrett}, {Kopan}, {Kirkpatrick}, {Light}, {Marsh}, {McCallon},
  {Schneider}, {Stiening}, {Sykes}, {Weinberg}, {Wheaton}, {Wheelock}, \&
  {Zacarias}}]{Cutri_2MASS}
{Cutri}, R.~M., {Skrutskie}, M.~F., {van Dyk}, S., {et~al.} 2003, VizieR Online
  Data Catalog, II/246

\bibitem[{{Cutri} {et~al.}(2012){Cutri}, {Skrutskie}, {van Dyk}, {Beichman},
  {Carpenter}, {Chester}, {Cambresy}, {Evans}, {Fowler}, {Gizis}, {Howard},
  {Huchra}, {Jarrett}, {Kopan}, {Kirkpatrick}, {Light}, {Marsh}, {McCallon},
  {Schneider}, {Stiening}, {Sykes}, {Weinberg}, {Wheaton}, {Wheelock}, \&
  {Zacharias}}]{Cutri_2MASS6X}
{Cutri}, R.~M., {Skrutskie}, M.~F., {van Dyk}, S., {et~al.} 2012, VizieR Online
  Data Catalog, II/281

\bibitem[{{Cutri} {et~al.}(2013){Cutri}, {Wright}, {Conrow}, {Fowler},
  {Eisenhardt}, {Grillmair}, {Kirkpatrick}, {Masci}, {McCallon}, {Wheelock},
  {Fajardo-Acosta}, {Yan}, {Benford}, {Harbut}, {Jarrett}, {Lake}, {Leisawitz},
  {Ressler}, {Stanford}, {Tsai}, {Liu}, {Helou}, {Mainzer}, {Gettings},
  {Gonzalez}, {Hoffman}, {Marsh}, {Padgett}, {Skrutskie}, {Beck}, {Papin}, \&
  {Wittman}}]{Cutri_Allwise}
{Cutri}, R.~M., {Wright}, E.~L., {Conrow}, T., {et~al.} 2013, VizieR Online
  Data Catalog, 2328, 0

\bibitem[{{DENIS Consortium}(2005)}]{DENIS05}
{DENIS Consortium}. 2005, VizieR Online Data Catalog, 2263

\bibitem[{{Emerson} \& {Sutherland}(2010)}]{Emerson10}
{Emerson}, J. \& {Sutherland}, W. 2010, The Messenger, 139, 2

\bibitem[{{Feast} {et~al.}(1989){Feast}, {Glass}, {Whitelock}, \&
  {Catchpole}}]{Feast1989}
{Feast}, M.~W., {Glass}, I.~S., {Whitelock}, P.~A., \& {Catchpole}, R.~M. 1989,
  \mnras, 241, 375

\bibitem[{{Ferraro} {et~al.}(1995){Ferraro}, {Fusi Pecci}, {Testa}, {Greggio},
  {Corsi}, {Buonanno}, {Terndrup}, \& {Zinnecker}}]{Ferraro95}
{Ferraro}, F.~R., {Fusi Pecci}, F., {Testa}, V., {et~al.} 1995, \mnras, 272,
  391

\bibitem[{{Fraser} {et~al.}(2008){Fraser}, {Hawley}, \& {Cook}}]{Fraser08}
{Fraser}, O.~J., {Hawley}, S.~L., \& {Cook}, K.~H. 2008, \aj, 136, 1242

\bibitem[{{Frogel} {et~al.}(1990){Frogel}, {Mould}, \& {Blanco}}]{FMB90}
{Frogel}, J.~A., {Mould}, J., \& {Blanco}, V.~M. 1990, \apj, 352, 96

\bibitem[{{Gaia Collaboration} {et~al.}(2018){Gaia Collaboration}, {Brown},
  {Vallenari}, {Prusti}, {de Bruijne}, {Babusiaux}, {Bailer-Jones}, {Biermann},
  {Evans}, {Eyer}, \& et~al.}]{GDR2Sum}
{Gaia Collaboration}, {Brown}, A.~G.~A., {Vallenari}, A., {et~al.} 2018, \aap,
  616, A1

\bibitem[{{Glass} \& {Evans}(1981)}]{GE81}
{Glass}, I.~S. \& {Evans}, T.~L. 1981, \nat, 291, 303

\bibitem[{{Glass} \& {Reid}(1985)}]{GR1985}
{Glass}, I.~S. \& {Reid}, N. 1985, \mnras, 214, 405

\bibitem[{{Gordon} {et~al.}(2011){Gordon}, {Meixner}, {Meade}, {Whitney},
  {Engelbracht}, {Bot}, {Boyer}, {Lawton}, {Sewi{\l}o}, {Babler}, {Bernard},
  {Bracker}, {Block}, {Blum}, {Bolatto}, {Bonanos}, {Harris}, {Hora},
  {Indebetouw}, {Misselt}, {Reach}, {Shiao}, {Tielens}, {Carlson},
  {Churchwell}, {Clayton}, {Chen}, {Cohen}, {Fukui}, {Gorjian}, {Hony},
  {Israel}, {Kawamura}, {Kemper}, {Leroy}, {Li}, {Madden}, {Marble},
  {McDonald}, {Mizuno}, {Mizuno}, {Muller}, {Oliveira}, {Olsen}, {Onishi},
  {Paladini}, {Paradis}, {Points}, {Robitaille}, {Rubin}, {Sandstrom}, {Sato},
  {Shibai}, {Simon}, {Smith}, {Srinivasan}, {Vijh}, {Van Dyk}, {van Loon}, \&
  {Zaritsky}}]{Gordon2011}
{Gordon}, K.~D., {Meixner}, M., {Meade}, M.~R., {et~al.} 2011, \aj, 142, 102

\bibitem[{{Groenewegen}(2004)}]{Groenewegen04}
{Groenewegen}, M.~A.~T. 2004, \aap, 425, 595

\bibitem[{{Groenewegen}(2012)}]{Gr_MOD}
{Groenewegen}, M.~A.~T. 2012, \aap, 543, A36

\bibitem[{{Groenewegen} \& {Blommaert}(1998)}]{GB98}
{Groenewegen}, M.~A.~T. \& {Blommaert}, J.~A.~D.~L. 1998, \aap, 332, 25

\bibitem[{{Groenewegen} {et~al.}(2002){Groenewegen}, {Sevenster}, {Spoon}, \&
  {P{\'e}rez}}]{Gr02}
{Groenewegen}, M.~A.~T., {Sevenster}, M., {Spoon}, H.~W.~W., \& {P{\'e}rez}, I.
  2002, \aap, 390, 511

\bibitem[{{Groenewegen} \& {Sloan}(2018)}]{GS18}
{Groenewegen}, M.~A.~T. \& {Sloan}, G.~C. 2018, \aap, 609, A114

\bibitem[{{Groenewegen} {et~al.}(2009){Groenewegen}, {Sloan}, {Soszy{\'n}ski},
  \& {Petersen}}]{GS09}
{Groenewegen}, M.~A.~T., {Sloan}, G.~C., {Soszy{\'n}ski}, I., \& {Petersen},
  E.~A. 2009, \aap, 506, 1277

\bibitem[{{Groenewegen} {et~al.}(2016){Groenewegen}, {Vlemmings}, {Marigo},
  {Sloan}, {Decin}, {Feast}, {Goldman}, {Justtanont}, {Kerschbaum}, {Matsuura},
  {McDonald}, {Olofsson}, {Sahai}, {van Loon}, {Wood}, {Zijlstra},
  {Bernard-Salas}, {Boyer}, {Guzman-Ramirez}, {Jones}, {Lagadec}, {Meixner},
  {Rawlings}, \& {Srinivasan}}]{Groenewegen2016}
{Groenewegen}, M.~A.~T., {Vlemmings}, W.~H.~T., {Marigo}, P., {et~al.} 2016,
  \aap, 596, A50

\bibitem[{{Gruendl} \& {Chu}(2009)}]{Gruendl09}
{Gruendl}, R.~A. \& {Chu}, Y.-H. 2009, \apjs, 184, 172

\bibitem[{{Gullieuszik} {et~al.}(2012){Gullieuszik}, {Groenewegen}, {Cioni},
  {de Grijs}, {van Loon}, {Girardi}, {Ivanov}, {Oliveira}, {Emerson}, \&
  {Guandalini}}]{2012A&A...537A.105G}
{Gullieuszik}, M., {Groenewegen}, M.~A.~T., {Cioni}, M.-R.~L., {et~al.} 2012,
  \aap, 537, A105

\bibitem[{{Hanner}(1988)}]{Hanner88}
{Hanner}, M. 1988, {Grain optical properties}, Tech. rep.

\bibitem[{{Hauschildt} {et~al.}(1999){Hauschildt}, {Allard}, \&
  {Baron}}]{Hauschildt1999}
{Hauschildt}, P.~H., {Allard}, F., \& {Baron}, E. 1999, \apj, 512, 377

\bibitem[{{H{\"o}fner} \& {Olofsson}(2018)}]{HO18}
{H{\"o}fner}, S. \& {Olofsson}, H. 2018, \aapr, 26, 1

\bibitem[{{Holl} {et~al.}(2018){Holl}, {Audard}, {Nienartowicz}, {Jevardat de
  Fombelle}, {Marchal}, {Mowlavi}, {Clementini}, {De Ridder}, {Evans}, {Guy},
  {Lanzafame}, {Lebzelter}, {Rimoldini}, {Roelens}, {Zucker}, {Distefano},
  {Garofalo}, {Lecoeur-Ta{\"i}bi}, {Lopez}, {Molinaro}, {Muraveva}, {Panahi},
  {Regibo}, {Ripepi}, {Sarro}, {Aerts}, {Anderson}, {Charnas}, {Barblan},
  {Blanco-Cuaresma}, {Busso}, {Cuypers}, {De Angeli}, {Glass}, {Grenon},
  {Juh{\'a}sz}, {Kochoska}, {Koubsky}, {Lanza}, {Leccia}, {Lorenz}, {Marconi},
  {Marschalk{\'o}}, {Mazeh}, {Messina}, {Mignard}, {Moitinho}, {Moln{\'a}r},
  {Morgenthaler}, {Musella}, {Ordenovic}, {Ord{\'o}{\~n}ez}, {Pagano},
  {Palaversa}, {Pawlak}, {Plachy}, {Pr{\v s}a}, {Riello}, {S{\"u}veges},
  {Szabados}, {Szegedi-Elek}, {Votruba}, \& {Eyer}}]{HollGDR2}
{Holl}, B., {Audard}, M., {Nienartowicz}, K., {et~al.} 2018, \aap, 618, A30

\bibitem[{{Houck} {et~al.}(2004){Houck}, {Roellig}, {van Cleve}, {Forrest},
  {Herter}, {Lawrence}, {Matthews}, {Reitsema}, {Soifer}, {Watson}, {Weedman},
  {Huisjen}, {Troeltzsch}, {Barry}, {Bernard-Salas}, {Blacken}, {Brandl},
  {Charmandaris}, {Devost}, {Gull}, {Hall}, {Henderson}, {Higdon}, {Pirger},
  {Schoenwald}, {Sloan}, {Uchida}, {Appleton}, {Armus}, {Burgdorf},
  {Fajardo-Acosta}, {Grillmair}, {Ingalls}, {Morris}, \& {Teplitz}}]{Houck2004}
{Houck}, J.~R., {Roellig}, T.~L., {van Cleve}, J., {et~al.} 2004, \apjs, 154,
  18

\bibitem[{{Huang} {et~al.}(2018){Huang}, {Riess}, {Hoffmann}, {Klein}, {Bloom},
  {Yuan}, {Macri}, {Jones}, {Whitelock}, {Casertano}, \& {Anderson}}]{Huang18}
{Huang}, C.~D., {Riess}, A.~G., {Hoffmann}, S.~L., {et~al.} 2018, \apj, 857, 67

\bibitem[{{Huang} {et~al.}(2020){Huang}, {Riess}, {Yuan}, {Macri}, {Zakamska},
  {Casertano}, {Whitelock}, {Hoffmann}, {Filippenko}, \& {Scolnic}}]{Huang19}
{Huang}, C.~D., {Riess}, A.~G., {Yuan}, W., {et~al.} 2020, \apj, 889, 5

\bibitem[{{Hughes}(1989)}]{Hughes1989}
{Hughes}, S.~M.~G. 1989, \aj, 97, 1634

\bibitem[{{Hughes} \& {Wood}(1990)}]{HW1990}
{Hughes}, S.~M.~G. \& {Wood}, P.~R. 1990, \aj, 99, 784

\bibitem[{{Irwin}(2009)}]{Irwin09}
{Irwin}, M.~J. 2009, {UKIRT Newsletter, 25, 15}

\bibitem[{{Isobe} {et~al.}(1990){Isobe}, {Feigelson}, {Akritas}, \&
  {Babu}}]{Isobe90}
{Isobe}, T., {Feigelson}, E.~D., {Akritas}, M.~G., \& {Babu}, G.~J. 1990, \apj,
  364, 104

\bibitem[{{Ita} {et~al.}(2018){Ita}, {Matsunaga}, {Tanab{\'e}}, {Nakada},
  {Kato}, {Nagayama}, {Nagashima}, {Kurita}, {Nakajima}, {Whitelock},
  {Menzies}, {Feast}, {Nagata}, {Tamura}, \& {Nakaya}}]{ItaSMC}
{Ita}, Y., {Matsunaga}, N., {Tanab{\'e}}, T., {et~al.} 2018, \mnras, 481, 4206

\bibitem[{{Ita} {et~al.}(2004){Ita}, {Tanab{\'e}}, {Matsunaga}, {Nakajima},
  {Nagashima}, {Nagayama}, {Kato}, {Kurita}, {Nagata}, {Sato}, {Tamura},
  {Nakaya}, \& {Nakada}}]{Ita04}
{Ita}, Y., {Tanab{\'e}}, T., {Matsunaga}, N., {et~al.} 2004, \mnras, 347, 720

\bibitem[{{Ivezi{\'c}} {et~al.}(1999){Ivezi{\'c}}, {Nenkova}, \&
  {Elitzur}}]{Ivezic_D}
{Ivezi{\'c}}, {\v{Z}}., {Nenkova}, M., \& {Elitzur}, M. 1999, {DUSTY: Radiation
  transport in a dusty environment}, Astrophysics Source Code Library

\bibitem[{{Jones} {et~al.}(2017){Jones}, {Woods}, {Kemper}, {Kraemer}, {Sloan},
  {Srinivasan}, {Oliveira}, {van Loon}, {Boyer}, {Sargent}, {McDonald},
  {Meixner}, {Zijlstra}, {Ruffle}, {Lagadec}, {Pauly}, {Sewi{\l}o}, {Clayton},
  \& {Volk}}]{Jones17}
{Jones}, O.~C., {Woods}, P.~M., {Kemper}, F., {et~al.} 2017, \mnras, 470, 3250

\bibitem[{{Kamath} {et~al.}(2010){Kamath}, {Wood}, {Soszy{\'n}ski}, \&
  {Lebzelter}}]{Kamath10}
{Kamath}, D., {Wood}, P.~R., {Soszy{\'n}ski}, I., \& {Lebzelter}, T. 2010,
  \mnras, 408, 522

\bibitem[{{Karambelkar} {et~al.}(2019){Karambelkar}, {Adams}, {Whitelock},
  {Kasliwal}, {Jencson}, {Boyer}, {Goldman}, {Masci}, {Cody}, {Bally}, {Bond},
  {Gehrz}, {Parthasarathy}, {Lau}, \& {SPIRITS Collaboration}}]{Karambelkar19}
{Karambelkar}, V.~R., {Adams}, S.~M., {Whitelock}, P.~A., {et~al.} 2019, \apj,
  877, 110

\bibitem[{{Kastner} {et~al.}(2008){Kastner}, {Thorndike}, {Romanczyk},
  {Buchanan}, {Hrivnak}, {Sahai}, \& {Egan}}]{Kastner08}
{Kastner}, J.~H., {Thorndike}, S.~L., {Romanczyk}, P.~A., {et~al.} 2008, \aj,
  136, 1221

\bibitem[{{Kato} {et~al.}(2007){Kato}, {Nagashima}, {Nagayama}, {Kurita},
  {Koerwer}, {Kawai}, {Yamamuro}, {Zenno}, {Nishiyama}, {Baba}, {Kadowaki},
  {Haba}, {Hatano}, {Shimizu}, {Nishimura}, {Nagata}, {Sato}, {Murai},
  {Kawazu}, {Nakajima}, {Nakaya}, {Kandori}, {Kusakabe}, {Ishihara},
  {Kaneyasu}, {Hashimoto}, {Tamura}, {Tanab{\'e}}, {Ita}, {Matsunaga},
  {Nakada}, {Sugitani}, {Wakamatsu}, {Glass}, {Feast}, {Menzies}, {Whitelock},
  {Fourie}, {Stoffels}, {Evans}, \& {Hasegawa}}]{Kato_IRSF}
{Kato}, D., {Nagashima}, C., {Nagayama}, T., {et~al.} 2007, \pasj, 59, 615

\bibitem[{{Kim} {et~al.}(2014){Kim}, {Protopapas}, {Bailer-Jones}, {Byun},
  {Chang}, {Marquette}, \& {Shin}}]{Kim14}
{Kim}, D.-W., {Protopapas}, P., {Bailer-Jones}, C.~A.~L., {et~al.} 2014, \aap,
  566, A43

\bibitem[{{Koen} {et~al.}(2007){Koen}, {Marang}, {Kilkenny}, \&
  {Jacobs}}]{Koen07}
{Koen}, C., {Marang}, F., {Kilkenny}, D., \& {Jacobs}, C. 2007, \mnras, 380,
  1433

\bibitem[{{Kraemer} {et~al.}(2017){Kraemer}, {Sloan}, {Wood}, {Jones}, \&
  {Egan}}]{Kraemer17}
{Kraemer}, K.~E., {Sloan}, G.~C., {Wood}, P.~R., {Jones}, O.~C., \& {Egan},
  M.~P. 2017, \apj, 834, 185

\bibitem[{{Lebouteiller} {et~al.}(2011){Lebouteiller}, {Barry}, {Spoon},
  {Bernard-Salas}, {Sloan}, {Houck}, \& {Weedman}}]{Lebouteiller11}
{Lebouteiller}, V., {Barry}, D.~J., {Spoon}, H.~W.~W., {et~al.} 2011, \apjs,
  196, 8

\bibitem[{{Lebzelter} {et~al.}(2018){Lebzelter}, {Mowlavi}, {Marigo},
  {Pastorelli}, {Trabucchi}, {Wood}, \& {Lecoeur-Ta{\"i}bi}}]{Lebzelter18}
{Lebzelter}, T., {Mowlavi}, N., {Marigo}, P., {et~al.} 2018, \aap, 616, L13

\bibitem[{{Leisy} {et~al.}(1997){Leisy}, {Dennefeld}, {Alard}, \&
  {Guibert}}]{Leisy1997}
{Leisy}, P., {Dennefeld}, M., {Alard}, C., \& {Guibert}, J. 1997, \aaps, 121,
  407

\bibitem[{{Lenz} \& {Breger}(2005)}]{Period04}
{Lenz}, P. \& {Breger}, M. 2005, Communications in Asteroseismology, 146, 53

\bibitem[{{Meixner} {et~al.}(2006){Meixner}, {Gordon}, {Indebetouw}, {Hora},
  {Whitney}, {Blum}, {Reach}, {Bernard}, {Meade}, {Babler}, {Engelbracht},
  {For}, {Misselt}, {Vijh}, {Leitherer}, {Cohen}, {Churchwell}, {Boulanger},
  {Frogel}, {Fukui}, {Gallagher}, {Gorjian}, {Harris}, {Kelly}, {Kawamura},
  {Kim}, {Latter}, {Madden}, {Markwick-Kemper}, {Mizuno}, {Mizuno}, {Mould},
  {Nota}, {Oey}, {Olsen}, {Onishi}, {Paladini}, {Panagia}, {Perez-Gonzalez},
  {Shibai}, {Sato}, {Smith}, {Staveley-Smith}, {Tielens}, {Ueta}, {van Dyk},
  {Volk}, {Werner}, \& {Zaritsky}}]{Meixner2006}
{Meixner}, M., {Gordon}, K.~D., {Indebetouw}, R., {et~al.} 2006, \aj, 132, 2268

\bibitem[{{Menzies} {et~al.}(2019){Menzies}, {Whitelock}, {Feast}, \&
  {Matsunaga}}]{Menzies19}
{Menzies}, J.~W., {Whitelock}, P.~A., {Feast}, M.~W., \& {Matsunaga}, N. 2019,
  \mnras, 483, 5150

\bibitem[{{Morgan}(1994)}]{Morgan94}
{Morgan}, D.~H. 1994, \aaps, 103, 235

\bibitem[{{Muraveva} {et~al.}(2018){Muraveva}, {Subramanian}, {Clementini},
  {Cioni}, {Palmer}, {van Loon}, {Moretti}, {de Grijs}, {Molinaro}, {Ripepi},
  {Marconi}, {Emerson}, \& {Ivanov}}]{Muraveva2017}
{Muraveva}, T., {Subramanian}, S., {Clementini}, G., {et~al.} 2018, \mnras,
  473, 3131

\bibitem[{{Nanni}(2019)}]{Nanni19a}
{Nanni}, A. 2019, \mnras, 482, 4726

\bibitem[{{Nanni} {et~al.}(2013){Nanni}, {Bressan}, {Marigo}, \&
  {Girardi}}]{Nanni13}
{Nanni}, A., {Bressan}, A., {Marigo}, P., \& {Girardi}, L. 2013, \mnras, 434,
  2390

\bibitem[{{Nanni} {et~al.}(2014){Nanni}, {Bressan}, {Marigo}, \&
  {Girardi}}]{Nanni14}
{Nanni}, A., {Bressan}, A., {Marigo}, P., \& {Girardi}, L. 2014, \mnras, 438,
  2328

\bibitem[{{Nanni} {et~al.}(2019){Nanni}, {Groenewegen}, {Aringer}, {Rubele},
  {Bressan}, {van Loon}, {Goldman}, \& {Boyer}}]{Nanni19b}
{Nanni}, A., {Groenewegen}, M. A.~T., {Aringer}, B., {et~al.} 2019, \mnras,
  487, 502

\bibitem[{{Nanni} {et~al.}(2018){Nanni}, {Marigo}, {Girardi}, {Rubele},
  {Bressan}, {Groenewegen}, {Pastorelli}, \& {Aringer}}]{Nanni18}
{Nanni}, A., {Marigo}, P., {Girardi}, L., {et~al.} 2018, \mnras, 473, 5492

\bibitem[{{Nanni} {et~al.}(2016){Nanni}, {Marigo}, {Groenewegen}, {Aringer},
  {Girardi}, {Pastorelli}, {Bressan}, \& {Bladh}}]{Nanni16}
{Nanni}, A., {Marigo}, P., {Groenewegen}, M.~A.~T., {et~al.} 2016, \mnras, 462,
  1215

\bibitem[{{Nishida} {et~al.}(2000){Nishida}, {Tanab{\'e}}, {Nakada},
  {Matsumoto}, {Sekiguchi}, \& {Glass}}]{Nishida2000}
{Nishida}, S., {Tanab{\'e}}, T., {Nakada}, Y., {et~al.} 2000, \mnras, 313, 136

\bibitem[{{Oliveira} {et~al.}(2009){Oliveira}, {van Loon}, {Chen}, {Tielens},
  {Sloan}, {Woods}, {Kemper}, {Indebetouw}, {Gordon}, {Boyer}, {Shiao},
  {Madden}, {Speck}, {Meixner}, \& {Marengo}}]{Oliveira09}
{Oliveira}, J.~M., {van Loon}, J.~T., {Chen}, C. H.~R., {et~al.} 2009, \apj,
  707, 1269

\bibitem[{{Payne-Gaposchkin}(1971)}]{PG1971}
{Payne-Gaposchkin}, C.~H. 1971, Smithsonian Contributions to Astrophysics, 13

\bibitem[{{Pietrzy{\'n}ski} {et~al.}(2019){Pietrzy{\'n}ski}, {Graczyk},
  {Gallenne}, {Gieren}, {Thompson}, {Pilecki}, {Karczmarek}, {G{\'o}rski},
  {Suchomska}, {Taormina}, {Zgirski}, {Wielg{\'o}rski}, {Ko{\l}aczkowski},
  {Konorski}, {Villanova}, {Nardetto}, {Kervella}, {Bresolin}, {Kudritzki},
  {Storm}, {Smolec}, \& {Narloch}}]{Pietrzynski19}
{Pietrzy{\'n}ski}, G., {Graczyk}, D., {Gallenne}, A., {et~al.} 2019, \nat, 567,
  200

\bibitem[{{Pojmanski}(2002)}]{Pojmanski02}
{Pojmanski}, G. 2002, \actaa, 52, 397

\bibitem[{{Press} {et~al.}(1992){Press}, {Teukolsky}, {Vetterling}, \&
  {Flannery}}]{Press1992}
{Press}, W.~H., {Teukolsky}, S.~A., {Vetterling}, W.~T., \& {Flannery}, B.~P.
  1992, {Numerical recipes in FORTRAN. The art of scientific computing}

\bibitem[{{Ramstedt} \& {Olofsson}(2014)}]{Ramstedt14}
{Ramstedt}, S. \& {Olofsson}, H. 2014, \aap, 566, A145

\bibitem[{{Reid} {et~al.}(1995){Reid}, {Hughes}, \& {Glass}}]{Reid1995}
{Reid}, I.~N., {Hughes}, S.~M.~G., \& {Glass}, I.~S. 1995, \mnras, 275, 331

\bibitem[{{Reid}(1991)}]{Reid1991}
{Reid}, N. 1991, \apj, 382, 143

\bibitem[{{Reid} {et~al.}(1990){Reid}, {Tinney}, \& {Mould}}]{Reid1990}
{Reid}, N., {Tinney}, C., \& {Mould}, J. 1990, \apj, 348, 98

\bibitem[{{Reid}(2014)}]{Reid14}
{Reid}, W.~A. 2014, \mnras, 438, 2642

\bibitem[{{Reid} \& {Parker}(2006)}]{ReidParker06}
{Reid}, W.~A. \& {Parker}, Q.~A. 2006, \mnras, 373, 521

\bibitem[{{Riebel} {et~al.}(2012){Riebel}, {Srinivasan}, {Sargent}, \&
  {Meixner}}]{Riebel12}
{Riebel}, D., {Srinivasan}, S., {Sargent}, B., \& {Meixner}, M. 2012, \apj,
  753, 71

\bibitem[{{Ripepi} {et~al.}(2017){Ripepi}, {Cioni}, {Moretti}, {Marconi},
  {Bekki}, {Clementini}, {de Grijs}, {Emerson}, {Groenewegen}, {Ivanov},
  {Molinaro}, {Muraveva}, {Oliveira}, {Piatti}, {Subramanian}, \& {van
  Loon}}]{Ripepi17}
{Ripepi}, V., {Cioni}, M.-R.~L., {Moretti}, M.~I., {et~al.} 2017, \mnras, 472,
  808

\bibitem[{{Ripepi} {et~al.}(2016){Ripepi}, {Marconi}, {Moretti}, {Clementini},
  {Cioni}, {de Grijs}, {Emerson}, {Groenewegen}, {Ivanov}, \&
  {Piatti}}]{Ripepi16}
{Ripepi}, V., {Marconi}, M., {Moretti}, M.~I., {et~al.} 2016, \apjs, 224, 21

\bibitem[{{Ripepi} {et~al.}(2015){Ripepi}, {Moretti}, {Marconi}, {Clementini},
  {Cioni}, {de Grijs}, {Emerson}, {Groenewegen}, {Ivanov}, {Muraveva},
  {Piatti}, \& {Subramanian}}]{Ripepi15}
{Ripepi}, V., {Moretti}, M.~I., {Marconi}, M., {et~al.} 2015, \mnras, 446, 3034

\bibitem[{{Ripepi} {et~al.}(2012){Ripepi}, {Moretti}, {Marconi}, {Clementini},
  {Cioni}, {Marquette}, {Girardi}, {Rubele}, {Groenewegen}, {de Grijs},
  {Gibson}, {Oliveira}, {van Loon}, \& {Emerson}}]{Ripepi12}
{Ripepi}, V., {Moretti}, M.~I., {Marconi}, M., {et~al.} 2012, \mnras, 424, 1807

\bibitem[{{Ruffle} {et~al.}(2015){Ruffle}, {Kemper}, {Jones}, {Sloan},
  {Kraemer}, {Woods}, {Boyer}, {Srinivasan}, {Antoniou}, {Lagadec}, {Matsuura},
  {McDonald}, {Oliveira}, {Sargent}, {Sewi{\l}o}, {Szczerba}, {van Loon},
  {Volk}, \& {Zijlstra}}]{Ruffle15}
{Ruffle}, P.~M.~E., {Kemper}, F., {Jones}, O.~C., {et~al.} 2015, \mnras, 451,
  3504

\bibitem[{{Sanduleak} {et~al.}(1978){Sanduleak}, {MacConnell}, \&
  {Philip}}]{SMP78}
{Sanduleak}, N., {MacConnell}, D.~J., \& {Philip}, A.~G.~D. 1978, \pasp, 90,
  621

\bibitem[{{Seale} {et~al.}(2009){Seale}, {Looney}, {Chu}, {Gruendl}, {Brandl},
  {Chen}, {Brandner}, \& {Blake}}]{Seale09}
{Seale}, J.~P., {Looney}, L.~W., {Chu}, Y.-H., {et~al.} 2009, \apj, 699, 150

\bibitem[{{Skiff}(2014)}]{Skiff14}
{Skiff}, B.~A. 2014, VizieR Online Data Catalog, 1

\bibitem[{{Sloan} {et~al.}(2016){Sloan}, {Kraemer}, {McDonald}, {Groenewegen},
  {Wood}, {Zijlstra}, {Lagadec}, {Boyer}, {Kemper}, {Matsuura}, {Sahai},
  {Sargent}, {Srinivasan}, {van Loon}, \& {Volk}}]{Sloan16}
{Sloan}, G.~C., {Kraemer}, K.~E., {McDonald}, I., {et~al.} 2016, \apj, 826, 44

\bibitem[{{Soszy{\'n}ski} {et~al.}(2007){Soszy{\'n}ski}, {Dziembowski},
  {Udalski}, {Kubiak}, {Szyma{\'n}ski}, {Pietrzy{\'n}ski}, {Wyrzykowski},
  {Szewczyk}, \& {Ulaczyk}}]{Soszynski07}
{Soszy{\'n}ski}, I., {Dziembowski}, W.~A., {Udalski}, A., {et~al.} 2007,
  \actaa, 57, 201

\bibitem[{{Soszy{\'n}ski} {et~al.}(2008{\natexlab{a}}){Soszy{\'n}ski},
  {Poleski}, {Udalski}, {Szyma{\'n}ski}, {Kubiak}, {Pietrzy{\'n}ski},
  {Wyrzykowski}, {Szewczyk}, \& {Ulaczyk}}]{Soszynski2008_DCEP}
{Soszy{\'n}ski}, I., {Poleski}, R., {Udalski}, A., {et~al.} 2008{\natexlab{a}},
  \actaa, 58, 163

\bibitem[{{Soszy{\'n}ski} {et~al.}(2012){Soszy{\'n}ski}, {Udalski}, {Poleski},
  {Koz{\l}owski}, {Wyrzykowski}, {Pietrukowicz}, {Szyma{\'n}ski}, {Kubiak},
  {Pietrzy{\'n}ski}, {Ulaczyk}, \& {Skowron}}]{Soszynski12}
{Soszy{\'n}ski}, I., {Udalski}, A., {Poleski}, R., {et~al.} 2012, \actaa, 62,
  219

\bibitem[{{Soszy{\'n}ski} {et~al.}(2010){Soszy{\'n}ski}, {Udalski},
  {Szyma{\'n}ski}, {Kubiak}, {Pietrzy{\~n}ski}, {Wyrzykowski}, {Ulaczyk}, \&
  {Poleski}}]{Soszynski2010}
{Soszy{\'n}ski}, I., {Udalski}, A., {Szyma{\'n}ski}, M.~K., {et~al.} 2010,
  \actaa, 60, 91

\bibitem[{{Soszy{\'n}ski} {et~al.}(2008{\natexlab{b}}){Soszy{\'n}ski},
  {Udalski}, {Szyma{\'n}ski}, {Kubiak}, {Pietrzy{\'n}ski}, {Wyrzykowski},
  {Szewczyk}, {Ulaczyk}, \& {Poleski}}]{Soszynski2008}
{Soszy{\'n}ski}, I., {Udalski}, A., {Szyma{\'n}ski}, M.~K., {et~al.}
  2008{\natexlab{b}}, \actaa, 58, 293

\bibitem[{{Soszy{\'n}ski} {et~al.}(2009{\natexlab{a}}){Soszy{\'n}ski},
  {Udalski}, {Szyma{\'n}ski}, {Kubiak}, {Pietrzy{\'n}ski}, {Wyrzykowski},
  {Szewczyk}, {Ulaczyk}, \& {Poleski}}]{SoszynskiLPVLMC}
{Soszy{\'n}ski}, I., {Udalski}, A., {Szyma{\'n}ski}, M.~K., {et~al.}
  2009{\natexlab{a}}, \actaa, 59, 239

\bibitem[{{Soszy{\'n}ski} {et~al.}(2009{\natexlab{b}}){Soszy{\'n}ski},
  {Udalski}, {Szyma{\'n}ski}, {Kubiak}, {Pietrzy{\'n}ski}, {Wyrzykowski},
  {Szewczyk}, {Ulaczyk}, \& {Poleski}}]{Soszynski2009}
{Soszy{\'n}ski}, I., {Udalski}, A., {Szyma{\'n}ski}, M.~K., {et~al.}
  2009{\natexlab{b}}, \actaa, 59, 335

\bibitem[{{Soszy{\'n}ski} {et~al.}(2011){Soszy{\'n}ski}, {Udalski},
  {Szyma{\'n}ski}, {Kubiak}, {Pietrzy{\'n}ski}, {Wyrzykowski}, {Ulaczyk},
  {Poleski}, {Koz{\l}owski}, \& {Pietrukowicz}}]{SoszynskiLPVSMC}
{Soszy{\'n}ski}, I., {Udalski}, A., {Szyma{\'n}ski}, M.~K., {et~al.} 2011,
  \actaa, 61, 217

\bibitem[{{Spano} {et~al.}(2011){Spano}, {Mowlavi}, {Eyer}, {Burki},
  {Marquette}, {Lecoeur-Ta{\"i}bi}, \& {Tisserand}}]{Spano11}
{Spano}, M., {Mowlavi}, N., {Eyer}, L., {et~al.} 2011, \aap, 536, A60

\bibitem[{{Srinivasan} {et~al.}(2016){Srinivasan}, {Boyer}, {Kemper},
  {Meixner}, {Sargent}, \& {Riebel}}]{Srinivasan16}
{Srinivasan}, S., {Boyer}, M.~L., {Kemper}, F., {et~al.} 2016, \mnras, 457,
  2814

\bibitem[{{Tanab{\'e}} {et~al.}(1997){Tanab{\'e}}, {Nishida}, {Matsumoto},
  {Onaka}, {Nakada}, {Soyano}, {Ono}, {Sekiguchi}, \& {Glass}}]{Tanabe1997}
{Tanab{\'e}}, T., {Nishida}, S., {Matsumoto}, S., {et~al.} 1997, \nat, 385, 509

\bibitem[{{Trabucchi} {et~al.}(2017){Trabucchi}, {Wood}, {Montalb{\'a}n},
  {Marigo}, {Pastorelli}, \& {Girardi}}]{Trabucchi2017}
{Trabucchi}, M., {Wood}, P.~R., {Montalb{\'a}n}, J., {et~al.} 2017, \apj, 847,
  139

\bibitem[{{Ulaczyk} {et~al.}(2013){Ulaczyk}, {Szyma{\'n}ski}, {Udalski},
  {Kubiak}, {Pietrzy{\'n}ski}, {Soszy{\'n}ski}, {Wyrzykowski}, {Poleski},
  {Gieren}, {Walker}, \& {Garcia-Varela}}]{Ulaczyk13}
{Ulaczyk}, K., {Szyma{\'n}ski}, M.~K., {Udalski}, A., {et~al.} 2013, \actaa,
  63, 159

\bibitem[{{van Loon} {et~al.}(1997){van Loon}, {Zijlstra}, {Whitelock},
  {Waters}, {Loup}, \& {Trams}}]{vLoon97}
{van Loon}, J.~T., {Zijlstra}, A.~A., {Whitelock}, P.~A., {et~al.} 1997, \aap,
  325, 585

\bibitem[{{Westerlund} \& {Smith}(1964)}]{WS64}
{Westerlund}, B.~E. \& {Smith}, L.~F. 1964, \mnras, 128, 311

\bibitem[{{Whitelock} {et~al.}(1989){Whitelock}, {Feast}, {Menzies}, \&
  {Catchpole}}]{Whitelock1989}
{Whitelock}, P.~A., {Feast}, M.~W., {Menzies}, J.~W., \& {Catchpole}, R.~M.
  1989, \mnras, 238, 769

\bibitem[{{Whitelock} {et~al.}(2003){Whitelock}, {Feast}, {van Loon}, \&
  {Zijlstra}}]{Whitelock2003}
{Whitelock}, P.~A., {Feast}, M.~W., {van Loon}, J.~T., \& {Zijlstra}, A.~A.
  2003, \mnras, 342, 86

\bibitem[{{Whitney} {et~al.}(2008){Whitney}, {Sewilo}, {Indebetouw},
  {Robitaille}, {Meixner}, {Gordon}, {Meade}, {Babler}, {Harris}, {Hora},
  {Bracker}, {Povich}, {Churchwell}, {Engelbracht}, {For}, {Block}, {Misselt},
  {Vijh}, {Leitherer}, {Kawamura}, {Blum}, {Cohen}, {Fukui}, {Mizuno},
  {Mizuno}, {Srinivasan}, {Tielens}, {Volk}, {Bernard}, {Boulanger}, {Frogel},
  {Gallagher}, {Gorjian}, {Kelly}, {Latter}, {Madden}, {Kemper}, {Mould},
  {Nota}, {Oey}, {Olsen}, {Onishi}, {Paladini}, {Panagia}, {Perez-Gonzalez},
  {Reach}, {Shibai}, {Sato}, {Smith}, {Staveley-Smith}, {Ueta}, {Van Dyk},
  {Werner}, {Wolff}, \& {Zaritsky}}]{Whitney08}
{Whitney}, B.~A., {Sewilo}, M., {Indebetouw}, R., {et~al.} 2008, \aj, 136, 18

\bibitem[{{Wood}(1998)}]{Wood1998}
{Wood}, P.~R. 1998, \aap, 338, 592

\bibitem[{{Wood}(2000)}]{Wood2000}
{Wood}, P.~R. 2000, \pasa, 17, 18

\bibitem[{{Wood} {et~al.}(1999){Wood}, {Alcock}, {Allsman}, {Alves}, {Axelrod},
  {Becker}, {Bennett}, {Cook}, {Drake}, {Freeman}, {Griest}, {King}, {Lehner},
  {Marshall}, {Minniti}, {Peterson}, {Pratt}, {Quinn}, {Stubbs}, {Sutherland},
  {Tomaney}, {Vandehei}, \& {Welch}}]{Wood1999}
{Wood}, P.~R., {Alcock}, C., {Allsman}, R.~A., {et~al.} 1999, in IAU Symposium,
  Vol. 191, Asymptotic Giant Branch Stars, ed. T.~{Le Bertre}, A.~{L\`ebre}, \&
  C.~{Waelkens}, 151

\bibitem[{{Wood} {et~al.}(1983){Wood}, {Bessell}, \& {Fox}}]{WBF83}
{Wood}, P.~R., {Bessell}, M.~S., \& {Fox}, M.~W. 1983, \apj, 272, 99

\bibitem[{{Wood} {et~al.}(1992){Wood}, {Whiteoak}, {Hughes}, {Bessell},
  {Gardner}, \& {Hyland}}]{Wood1992}
{Wood}, P.~R., {Whiteoak}, J.~B., {Hughes}, S.~M.~G., {et~al.} 1992, \apj, 397,
  552

\bibitem[{{Woods} {et~al.}(2011){Woods}, {Oliveira}, {Kemper}, {van Loon},
  {Sargent}, {Matsuura}, {Szczerba}, {Volk}, {Zijlstra}, {Sloan}, {Lagadec},
  {McDonald}, {Jones}, {Gorjian}, {Kraemer}, {Gielen}, {Meixner}, {Blum},
  {Sewi{\l}o}, {Riebel}, {Shiao}, {Chen}, {Boyer}, {Indebetouw}, {Antoniou},
  {Bernard}, {Cohen}, {Dijkstra}, {Galametz}, {Galliano}, {Gordon}, {Harris},
  {Hony}, {Hora}, {Kawamura}, {Lawton}, {Leisenring}, {Madden}, {Marengo},
  {McGuire}, {Mulia}, {O'Halloran}, {Olsen}, {Paladini}, {Paradis}, {Reach},
  {Rubin}, {Sandstrom}, {Soszy{\'n}ski}, {Speck}, {Srinivasan}, {Tielens}, {van
  Aarle}, {van Dyk}, {van Winckel}, {Vijh}, {Whitney}, \& {Wilkins}}]{Woods11}
{Woods}, P.~M., {Oliveira}, J.~M., {Kemper}, F., {et~al.} 2011, \mnras, 411,
  1597

\bibitem[{{Zickgraf} {et~al.}(1989){Zickgraf}, {Wolf}, {Stahl}, \&
  {Humphreys}}]{Zickgraf89}
{Zickgraf}, F.~J., {Wolf}, B., {Stahl}, O., \& {Humphreys}, R.~M. 1989, \aap,
  220, 206

\bibitem[{{Zickgraf} {et~al.}(1986){Zickgraf}, {Wolf}, {Stahl}, {Leitherer}, \&
  {Appenzeller}}]{Zickgraf86}
{Zickgraf}, F.~J., {Wolf}, B., {Stahl}, O., {Leitherer}, C., \& {Appenzeller},
  I. 1986, \aap, 163, 119

\bibitem[{{Zijlstra} {et~al.}(1996){Zijlstra}, {Loup}, {Waters}, {Whitelock},
  {van Loon}, \& {Guglielmo}}]{Zijlstra96}
{Zijlstra}, A.~A., {Loup}, C., {Waters}, L.~B.~F.~M., {et~al.} 1996, \mnras,
  279, 32

\bibitem[{{Zivkov} {et~al.}(2020){Zivkov}, {Oliveira}, {Petr-Gotzens},
  {Rubele}, {Cioni}, {van Loon}, {de Grijs}, {Emerson}, {Ivanov}, {Marconi},
  {Moretti}, {Ripepi}, {Niederhofer}, \& {Sun}}]{Zivkov20}
{Zivkov}, V., {Oliveira}, J.~M., {Petr-Gotzens}, M.~G., {et~al.} 2020, arXiv
  e-prints, arXiv:2002.12291

\end{thebibliography}

\begin{appendix}

\section{The sample, literature and period analysis}
\label{App-A}

Tables~\ref{App-Sample}, \ref{App-Periods2} and \ref {App-Periods1} provide the first entries
of the tables available at the CDS. They provide information about the 1299 sources for which
the $K$-band LCs were analysed.
Table~\ref{App-Sample} provides basic information about the sample,
Table~\ref{App-Periods2} provides information on known periodicity from the literature, while
Table~\ref{App-Periods1} provides the results of the LC analysis.
%
The meaning of the columns is explained in the footnotes to the Tables.

\begin{figure*}
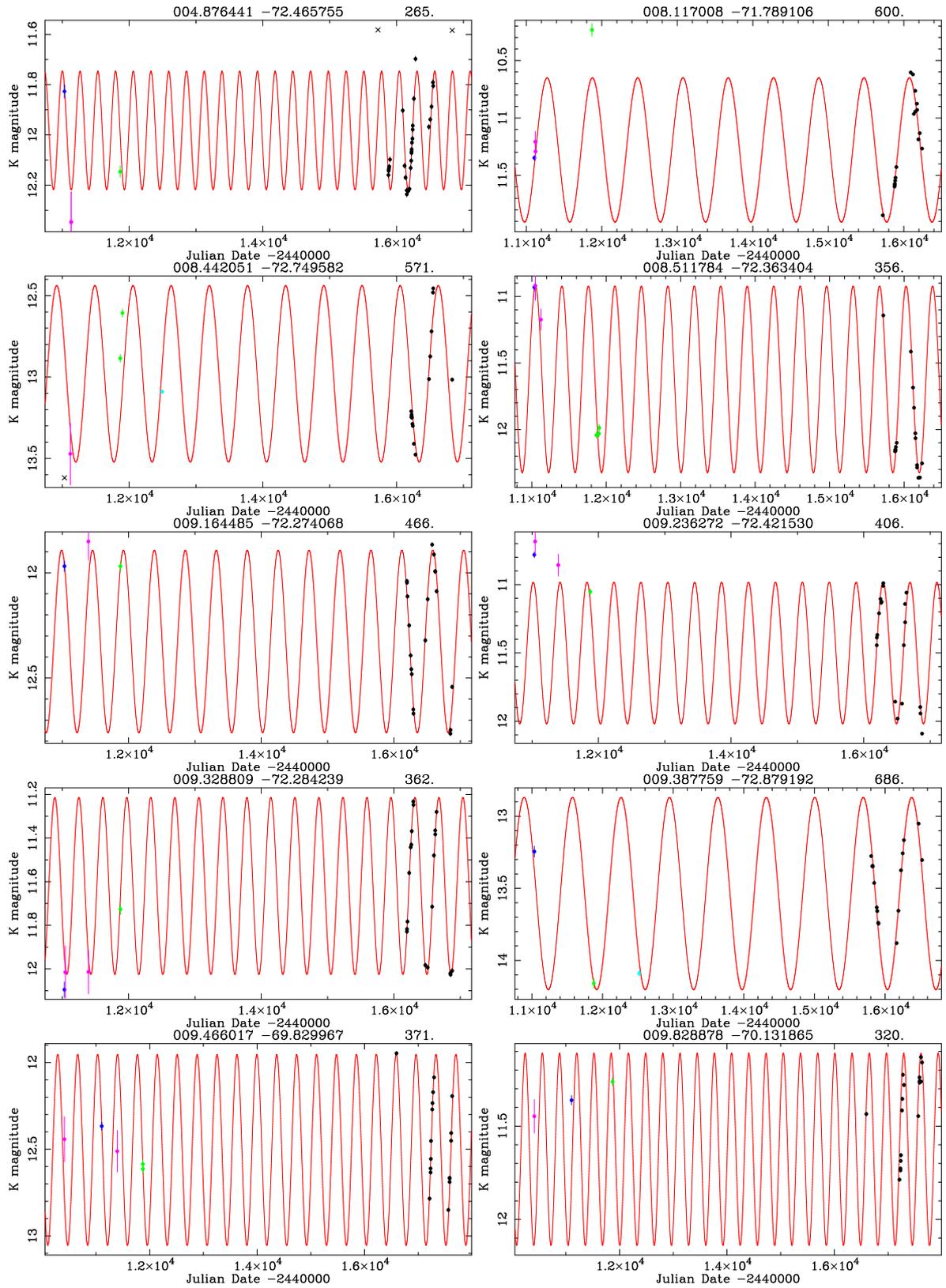

   \centering

\begin{minipage}{0.42\textwidth}
\resizebox{\hsize}{!}{\includegraphics[angle=0]{004.876441_-72.465755.ps}}
\end{minipage}
 \begin{minipage}{0.42\textwidth}
\resizebox{\hsize}{!}{\includegraphics[angle=0]{008.117008_-71.789106.ps}}
\end{minipage}

\begin{minipage}{0.42\textwidth}
\resizebox{\hsize}{!}{\includegraphics[angle=0]{008.442051_-72.749582.ps}}
\end{minipage}
\begin{minipage}{0.42\textwidth}
\resizebox{\hsize}{!}{\includegraphics[angle=0]{008.511784_-72.363404.ps}}
\end{minipage}

\begin{minipage}{0.42\textwidth}
\resizebox{\hsize}{!}{\includegraphics[angle=0]{009.164485_-72.274068.ps}}
\end{minipage}
\begin{minipage}{0.42\textwidth}
\resizebox{\hsize}{!}{\includegraphics[angle=0]{009.236272_-72.421530.ps}}
\end{minipage}

\begin{minipage}{0.42\textwidth}
\resizebox{\hsize}{!}{\includegraphics[angle=0]{009.328809_-72.284239.ps}}
\end{minipage}
\begin{minipage}{0.42\textwidth}
\resizebox{\hsize}{!}{\includegraphics[angle=0]{009.387759_-72.879192.ps}}
\end{minipage}

\begin{minipage}{0.42\textwidth}
\resizebox{\hsize}{!}{\includegraphics[angle=0]{009.466017_-69.829967.ps}}
\end{minipage}
\begin{minipage}{0.42\textwidth}
\resizebox{\hsize}{!}{\includegraphics[angle=0]{009.828878_-70.131865.ps}}
\end{minipage}

\caption[]{
Examples of the LC fitting (red solid lines) to the observed time photometry
On top of the plot the identifier and the period (in days) are listed. 
Data points (with error bars) are identified as follows:
black: VMC, green: 2MASS, dark blue: 2MASS-6X, light blue: IRSF,
magenta: DENIS, orange: other sources (see references in the footnote to Table~\ref{App-Periods1}).
Black crosses indicate points excluded from the fits.
}
\label{Fig-LCs}
\end{figure*}

Figure~\ref{Fig-LCs} shows examples of the LC fitting.
The complete set for the 1299 objects for which this was performed is available in the on-line edition.
The model is represented by the (red) solid line.
On top of the plot  the identifier and the period (in days) are listed.

\begin{sidewaystable*}
\caption{Sample for which a period analysis and LC fitting was performed.}
\label{App-Sample}
\centering
\footnotesize
\begin{tabular}{rlllllcc}
\hline \hline
RA    \hfill Dec           & Names & Object Type  & Spectral Type & {\it Gaia} Type & Spitzer Type \\
(deg) \hfill (deg)         &       &              &               &           &         \\
\hline
  4.876441  $-$72.465755  &  2MASS J00193036-7227567                       &       Can. AGB   &  & MIRA\_SR &           \\
  5.993721  $-$73.631895  &  OGLE SMC-LPV-190, 2MASS J00235849-7337548     &              Mira     &  &       &           \\
  6.088240  $-$72.107443  &                                 &                       &  &       &           \\
  6.498306  $-$73.895774  &  OGLE SMC-LPV-367 2MASS J00255959-7353448   &              Mira     &  &       &           \\
  6.795538  $-$73.408413  &  OGLE SMC-LPV-486 2MASS J00271093-7324303   &              Mira     &  &       &           \\
  7.329711  $-$71.063852  &  IRAS F00271-7120 2MASS J00291916-7103499   &                   C   & C &       &           \\
  7.572240  $-$72.472302  &  OGLE SMC-LPV-881 2MASS J00301737-7228202   &              Mira     &  & MIRA\_SR &           \\
  7.669748  $-$73.712530  &  OGLE SMC-LPV-929 2MASS J00304075-7342450   &              Mira     &  &       &           \\
  7.917060  $-$73.798217  &                                 &                       &  &       &           \\
  7.987056  $-$73.520435  &                                 &                       &  &       &           \\
  8.051002  $-$74.801395  &  OGLE SMC-LPV-1168 2MASS J00321227-7448049   &              Mira     &  &       &           \\
  8.117008  $-$71.789106  &      2MASS J00322809-7147207    &       Can. AGB   &  &       &           \\
  8.442051  $-$72.749582  &      2MASS J00334612-7244584    &       Can. AGB   &  &       &           \\
  8.511784  $-$72.363404  &      2MASS J00340283-7221482    &       Can. AGB   &  & MIRA\_SR &           \\
  8.816105  $-$73.424329  &  OGLE SMC-LPV-1690 2MASS J00351587-7325275   &              Mira     &  &       &           \\
  8.931123  $-$73.352948  &  LIN  19 2MASS J00354347-7321106   &                  Em   &  &       &           \\
  8.955451  $-$74.127210  &      2MASS J00354932-7407381    &              Star     &  &       &           \\
  9.108186  $-$73.432101  &                                 &                       &  &       &           \\
  9.164485  $-$72.274068  &      2MASS J00363946-7216266    &       Can. AGB   &  &       &           \\
  9.193053  $-$73.526474  &  MSX SMC 029 2MASS J00364631-7331351     &            PAGB   & C &       & C (1)  CPAGB (2) CPAGB (3)    \\
  9.230702  $-$74.741285  &                                 &                       &  &       &           \\
  9.236272  $-$72.421530  &  MSX SMC 091 2MASS J00365671-7225175   &                   C   &  &       & C (1)  CAGB (2) CAGB (3)     \\
  9.246025  $-$71.636344  &     [FBR2002] J003659-713813    &             Radio     &  &       &           \\
  9.328809  $-$72.284239  &      2MASS J00371893-7217031    &       Can. AGB   &  &       &           \\
  9.384601  $-$73.506030  &  OGLE J003732.32-733021.3 2MASS J00373232-7330216   &                 LPV   &  &       &           \\
  9.387759  $-$72.879192  &       2MASS J00373306-7252451   &       Can. AGB   &  &       &           \\
  9.446274  $-$73.650656  &  RAW   21 2MASS J00374710-7339022   &                   C   & C &       &           \\
  9.466017  $-$69.829967  &                                 &                       &  &       &           \\
  9.491884  $-$75.208617  &  [KID97] C0036-7529 2MASS J00375807-7512309   &                   C   &  &       &           \\
  9.507699  $-$73.790561  &  OGLE J003801.87-734725.7 2MASS J00380186-7347259   &              Mira     &  &       &           \\
  9.614211  $-$74.260202  &  OGLE SMC-LPV-2488 2MASS J00382740-7415366   &              Mira     &  &       &           \\
  9.660904  $-$72.008445  &       2MASS J00383862-7200304   &       Can. AGB   &  & MIRA\_SR &           \\
  9.794922  $-$71.569425  &                                 &                       &  &       &           \\
  9.828878  $-$70.131865  &  [MH95]  414 2MASS J00391894-7007546   &                   C   & C & MIRA\_SR &           \\
  9.887294  $-$70.295891  &                                 &                       &  &       &           \\
 10.067795  $-$73.020987  &  OGLE J004016.25-730115.1 2MASS J00401627-7301156   &              Mira     &  &       &           \\
 10.137255  $-$73.477785  &  OGLE J004032.93-732839.7 2MASS J00403293-7328399   &              Mira     &  &       &           \\
 10.147153  $-$73.324740  &  OGLE J004035.31-731928.7 2MASS J00403531-7319291   &              Mira     &  &       &           \\
\hline
\end{tabular}
\tablefoot{
Column 1 lists the coorfinates (right ascension and declination) of the source in decimal degrees.
Column 2 gives some names, as listed by the SIMBAD database. A `?' indicates the name of a source that is located close to the nominal position.
Column 3 gives the object type, as listed by the SIMBAD database (As of January 2019. As the SIMBAD database is continuously updated this may change in time).
Column 4 gives the spectral type(s), as listed in \citet{Skiff14}. In case of many entries only a selection is listed.
Column 5 gives the variable star classification as listed in the GDR2 \citet{GDR2Sum,HollGDR2}.
Column 6 gives the classification based on the MIR {\it Spitzer} IRS spectra.
References: (1) \citet{GS18}; (2) \citet{Kraemer17}; (3) \citet{Ruffle15}; (4) \citet{Jones17}; (5) \citet{Seale09};
            (6) \citet{Sloan16} (all stars listed as `C' in \citet{GS18}, also listed in \citealt{Sloan16}); (7) \citet{Woods11}.
The classifications are taken directly from those papers. For the exact meaning of the abbreviations see the original papers.
}
\end{sidewaystable*}

\begin{sidewaystable*}
\caption{Literature data.}
\label{App-Periods2}
\centering
\footnotesize
\setlength{\tabcolsep}{1.1mm}
\begin{tabular}{rcccccrlrlccrlrlcccccrcll}
\hline \hline
RA \hfill Dec       &  Type  &  Evo  &  SpT  &  Vm  &  Im  &  P1  &  A1  &  P2  &  A2   &  VC  &  RC  &  PR  &  AR  &  PB  &  AB  &  erBm  &  erBR &   erP    &  erRm  &  erBR  &  erP1  & erAR & Pother & Ref. \\
(deg) \hfill  (deg) & & & & (mag) & (mag) & (d) & (mag) & (d) & (mag) & (mag) & (mag) & (d) & (mag) & (d) & (mag) & (mag) & (mag) & (d) & (mag) &  (mag) & (d) & (mag) & (d) & \\
\hline
  4.876441  $-$72.465755   &       &     &   &       &       &      &     &      &     &       &       &      &     &      &     &       &       &                &       &       &      &     & 335 $\pm$ 29 & 80 \\
  5.993721  $-$73.631895   & Mira  &  -  & - &   -   & 17.24 &  479 & 1.6 &  175 & 0.1 &       &       &      &     &      &     &       &       &                &       &       &      &     &       &     \\
  6.088240  $-$72.107443   &       &     &   &       &       &      &     &      &     &       &       &      &     &      &     &       &       &                &       &       &      &     &       &     \\
  6.498306  $-$73.895774   & Mira  &  -  & - & 21.06 & 17.27 &  463 & 1.6 &  306 & 0.4 &       &       &      &     &      &     &       &       &                &       &       &      &     &       &     \\
  6.795538  $-$73.408413   & Mira  &  -  & - & 20.04 & 16.17 &  457 & 1.3 &  245 & 0.2 &       &       &      &     &      &     &       &       &                &       &       &      &     &       &     \\
  7.329711  $-$71.063852   &       &     &   &       &       &      &     &      &     &       &       &      &     &      &     &       &       &                &       &       &      &     &       &     \\
  7.572240  $-$72.472302   & Mira  &  -  & - & 18.23 & 15.84 &  508 & 1.6 &  202 & 0.1 &       &       &      &     &      &     &       &       &                &       &       &      &     & 531 $\pm$ 85 & 80 \\
  7.669748  $-$73.712530   & Mira  &  -  & - & 20.77 & 17.80 &  456 & 1.9 &  211 & 0.3 &       &       &      &     &      &     &       &       &                &       &       &      &     &       &     \\
  7.917060  $-$73.798217   &       &     &   &       &       &      &     &      &     &       &       &      &     &      &     &       &       &                &       &       &      &     &       &     \\
  7.987056  $-$73.520435   &       &     &   &       &       &      &     &      &     &       &       &      &     &      &     &       &       &                &       &       &      &     &       &     \\
  8.051002  $-$74.801395   & Mira  &  -  & - & 20.46 & 17.79 &  495 & 2.0 &  259 & 0.4 &       &       &      &     &      &     &       &       &                &       &       &      &     &       &     \\
  8.117008  $-$71.789106   &       &     &   &       &       &      &     &      &     &       &       &      &     &      &     &       &       &                &       &       &      &     &       &     \\
  8.442051  $-$72.749582   &       &     &   &       &       &      &     &      &     &       &       &      &     &      &     &       &       &                &       &       &      &     &       &     \\
  8.511784  $-$72.363404   &       &     &   &       &       &      &     &      &     &       &       &      &     &      &     &       &       &                &       &       &      &     & 336 $\pm$ 37 & 80 \\
  8.816105  $-$73.424329   & Mira  &  -  & - & 21.97 & 17.82 &  478 & 1.6 &  263 & 0.2 &       &       &      &     &      &     &       &       &                &       &       &      &     &       &     \\
  8.931123  $-$73.352948   &       &     &   &       &       &      &     &      &     &       &       &      &     &      &     &       &       &                &       &       &      &     &       &     \\
  8.955451  $-$74.127210   &       &     &   &       &       &      &     &      &     &       &       &      &     &      &     &       &       &                &       &       &      &     &       &     \\
  9.108186  $-$73.432101   &       &     &   &       &       &      &     &      &     &       &       &      &     &      &     &       &       &                &       &       &      &     &       &     \\
  9.164485  $-$72.274068   &       &     &   &       &       &      &     &      &     &       &       &      &     &      &     &       &       &                &       &       &      &     &       &     \\
  9.193053  $-$73.526474   &       &     &   &       &       &      &     &      &     &       &       &      &     &      &     &       &       &                &       &       &      &     &       &     \\
  9.230702  $-$74.741285   &       &     &   &       &       &      &     &      &     &       &       &      &     &      &     &       &       &                &       &       &      &     &       &     \\
  9.236272  $-$72.421530   &       &     &   &       &       &      &     &      &     &       &       &      &     &      &     &       &       &                &       &       &      &     & 405 & 99 \\
  9.246025  $-$71.636344   &       &     &   &       &       &      &     &      &     &       &       &      &     &      &     &       &       &                &       &       &      &     &       &     \\
  9.328809  $-$72.284239   &       &     &   &       &       &      &     &      &     &       &       &      &     &      &     &       &       &                &       &       &      &     &       &     \\
  9.384601  $-$73.506030   &  SRV  &  -  & - & 21.32 & 18.72 &  573 & 0.5 & 1007 & 0.4 &       &       &      &     &      &     &       &       &                &       &       &      &     &       &     \\
  9.387759  $-$72.879192   &       &     &   &       &       &      &     &      &     &       &       &      &     &      &     &       &       &                &       &       &      &     &       &     \\
  9.446274  $-$73.650656   &       &     &   &       &       &      &     &      &     &       &       &      &     &      &     &       &       &                &       &       &      &     &       &     \\
  9.466017  $-$69.829967   &       &     &   &       &       &      &     &      &     &       &       &      &     &      &     &       &       &                &       &       &      &     &       &     \\
  9.491884  $-$75.208617   &       &     &   &       &       &      &     &      &     &       &       &      &     &      &     &       &       &                &       &       &      &     &       &     \\
  9.507699  $-$73.790561   & Mira  &  -  & - & 18.87 & 15.63 &  511 & 1.4 &  466 & 0.2 &       &       &      &     &      &     &       &       &                &       &       &      &     &       &     \\
  9.614211  $-$74.260202   & Mira  &  -  & - & 20.47 & 16.28 &  463 & 1.5 & 2137 & 0.3 &       &       &      &     &      &     &       &       &                &       &       &      &     &       &     \\
  9.660904  $-$72.008445   &       &     &   &       &       &      &     &      &     &       &       &      &     &      &     &       &       &                &       &       &      &     & 513 $\pm$ 90 & 80 \\
  9.794922  $-$71.569425   &       &     &   &       &       &      &     &      &     &       &       &      &     &      &     &       &       &                &       &       &      &     &       &     \\
  9.828878  $-$70.131865   &       &     &   &       &       &      &     &      &     &       &       &      &     &      &     &       &       &                &       &       &      &     & 343 $\pm$ 27 & 80 \\
  9.887294  $-$70.295891   &       &     &   &       &       &      &     &      &     &       &       &      &     &      &     &       &       &                &       &       &      &     &       &     \\
 10.067795  $-$73.020987   & Mira  &  -  & - &   -   & 17.72 &  489 & 1.6 &  305 & 0.1 &       &       &      &     &      &     &       &       &                &       &       &      &     &       &     \\
 10.137255  $-$73.477785   & Mira  &  -  & - &   -   & 18.27 &  566 & 2.2 &  897 & 0.2 &       &       &      &     &      &     &       &       &                &       &       &      &     &       &     \\
 10.147153  $-$73.324740   & Mira  &  -  & - &   -   & 18.28 &  515 & 1.6 &  612 & 0.2 &       &       &      &     &      &     &       &       &                &       &       &      &     &       &     \\
\hline
\end{tabular}
\tablefoot{
Column 1 lists the right ascension and declination of the source in decimal degrees.
Columns 2--10 are taken from the OGLE-III database of LPVs in the LMC and SMC \citep{SoszynskiLPVLMC, SoszynskiLPVSMC}.
They list the pulsation type (Mira, SRV, or OSARG) in Col.~2, and for the LMC only, the evolutionary status (AGB or RGB) and spectral type (C or O) in Cols.~3 and 4.
Columns 5 and 6 list the mean magnitudes in the $V$ and $I$ bands.
Columns 7--10 list the period and amplitude of the primary and secondary period (rounded values are listed, and the OGLE catalogue also lists a tertiary period).
Columns 11--16 are taken from the analysis of MACHO data for LPVs in the LMC by \citep{Fraser08}.
Mean magnitudes in the $V$ and $R$ bands (in the Cousins system) are listed in Cols.~11 and 12.
Period and amplitude in the red and blue band are listed in Cols.~13--16.
%
Columns 17--23 are taken from the analysis of  EROS-2 data by \citet{Kim14} (Cols.~17--19), and \citet{Spano11} (Cols.~20--23).
Listed are the EROS $B$ magnitude, the $B-R$ colour index, and the period and its error (The error is derived from the period and the
SNR of the period listed by \citealt{Kim14}), respectively, the
EROS $R$ magnitude, the $B-R$ colour index, the first fitted period (out of up to five periods listed in \citealt{Spano11}), and
the amplitude in the $R$-band.
Column 24 lists other periods from the literature, with references listed in Col. 25 as follows:
(34) \citet{Wood1992};
(35) \citet{Whitelock2003};
(36) \citet{Whitelock1989};
(38) \citet{WBF83};
(43) \citet{Nishida2000};
(45) \citet{Kamath10};
(80) \citet{HollGDR2};
(89) present work: re-analysis of the data in ref. (90);
(90) \citet{Soszynski12};   
(91) \citet{Ulaczyk13};     
(92) \citet{Hughes1989};
(93) \citet{PG1971};
(94) present work: re-analysis of ASAS \citep{Pojmanski02} data;
(95) \citet{Soszynski2009};  
(96) \citet{Soszynski2008_DCEP}; 
(97) \citet{Soszynski2008};      
(98) \citet{Soszynski2010};      
(99) \citet{GS18}.
}
\end{sidewaystable*}

\begin{sidewaystable*}
\caption{Results from the period analysis.}
\label{App-Periods1}
\centering
\small
\begin{tabular}{rrlllllllll}
\hline \hline
RA \hfill Dec              & $\chi_{\rm r}^2$ &  $K$     &  Period       & Amplitude      & Data  &       References     & P$_{\rm ini}$   & Remarks \\
(deg) \hfill (deg)         &                &  (mag)   &  (d)          & (mag)          &        &                     &     (d)        &  \\
\hline
  4.876441  -72.465755  &   19.8 & 11.982 $\pm$  0.011 &  265 $\pm$    1 & 0.24 $\pm$ 0.05 & 32 36 &               1 10 11 13 &      &   Palt=333 \\
  5.993721  -73.631895  &  308.8 & 11.473 $\pm$  0.077 &  479 $\pm$    6 & 0.42 $\pm$ 0.08 & 14 18 &               1 10 11 13 &  479 &    \\
  6.088240  -72.107443  &    6.7 & 16.452 $\pm$  0.054 &                 &                 & 10 10 &                        1 &      &    \\
  6.498306  -73.895774  &  268.6 & 11.427 $\pm$  0.081 &  468 $\pm$    5 & 0.55 $\pm$ 0.18 & 14 18 &            1 10 11 12 13 &  463 &    \\
  6.795538  -73.408413  &   42.2 & 11.455 $\pm$  0.025 &  451 $\pm$    5 & 0.32 $\pm$ 0.12 & 15 18 &                  1 11 13 &  457 &    \\
  7.329711  -71.063852  &  863.7 & 13.422 $\pm$  0.227 &  343 $\pm$    4 & 0.73 $\pm$ 0.38 & 18 24 &               1 10 11 13 &      &   Palt=953 \\
  7.572240  -72.472302  &  695.6 & 10.792 $\pm$  0.062 &  508 $\pm$   12 & 0.39 $\pm$ 0.18 & 31 33 &                  1 10 11 &  508 &    \\
  7.669748  -73.712530  & 1317.8 & 11.524 $\pm$  0.154 &  458 $\pm$   15 & 0.62 $\pm$ 0.52 & 14 18 &            1 10 11 12 13 &  456 &    \\
  7.917060  -73.798217  &    4.8 & 14.382 $\pm$  0.008 &  740 $\pm$   14 & 0.09 $\pm$ 0.05 & 14 18 &               1 10 11 12 &      &   \\
  7.987056  -73.520435  &    1.3 & 14.786 $\pm$  0.003 &                 &                 & 15 19 &               1 10 11 12 &      &    \\
  8.051002  -74.801395  &   48.8 & 11.475 $\pm$  0.038 &  491 $\pm$    5 & 0.29 $\pm$ 0.15 & 13 18 &               1 10 11 13 &  495 &   \\
  8.117008  -71.789106  &   70.3 & 11.278 $\pm$  0.023 &  600 $\pm$    4 & 0.63 $\pm$ 0.07 & 17 21 &               1 10 11 13 &      &    \\
  8.442051  -72.749582  &  175.6 & 12.980 $\pm$  0.037 &  571 $\pm$   10 & 0.54 $\pm$ 0.30 & 15 20 &            1 10 11 12 13 &      &    \\
  8.511784  -72.363404  &   30.2 & 11.622 $\pm$  0.025 &  356 $\pm$    1 & 0.70 $\pm$ 0.04 & 17 23 &               1 10 11 13 &      &    \\
  8.816105  -73.424329  &  119.3 & 11.139 $\pm$  0.039 &  482 $\pm$    5 & 0.46 $\pm$ 0.07 & 16 22 &            1 10 11 12 13 &  478 &    \\
  8.931123  -73.352948  &   37.8 & 13.720 $\pm$  0.016 &                 &                 & 16 23 &            1 10 11 12 13 &      &    \\
  8.955451  -74.127210  &    1.6 & 14.680 $\pm$  0.003 &                 &                 & 18 22 &                  1 10 11 &      &    \\
  9.108186  -73.432101  &    7.5 & 14.453 $\pm$  0.008 &                 &                 & 16 19 &               1 10 11 12 &      &    \\
  9.164485  -72.274068  &   19.7 & 12.326 $\pm$  0.011 &  466 $\pm$    2 & 0.43 $\pm$ 0.05 & 19 22 &               1 10 11 13 &      &    \\
  9.193053  -73.526474  &   24.7 & 13.372 $\pm$  0.013 &                 &                 & 16 20 &            1 10 11 12 13 &      &    \\
  9.230702  -74.741285  &   10.9 & 14.856 $\pm$  0.016 &  392 $\pm$   11 & 0.05 $\pm$ 0.04 & 33 36 &                  1 10 11 &      &    \\
  9.236272  -72.421530  &   78.7 & 11.501 $\pm$  0.025 &  406 $\pm$    2 & 0.52 $\pm$ 0.05 & 19 23 &               1 10 11 13 &  405 &    \\
  9.246025  -71.636344  &    1.7 & 14.671 $\pm$  0.004 &  960 $\pm$   39 & 0.05 $\pm$ 0.03 & 19 21 &                  1 10 11 &      &    \\
  9.328809  -72.284239  &    5.1 & 11.619 $\pm$  0.010 &  362 $\pm$    1 & 0.41 $\pm$ 0.01 & 19 23 &               1 10 11 13 &      &    \\
  9.384601  -73.506030  &  180.6 & 13.187 $\pm$  0.046 &  571 $\pm$   16 & 0.42 $\pm$ 0.33 & 16 20 &            1 10 11 12 13 &  573 &    \\
  9.387759  -72.879192  &   28.0 & 13.536 $\pm$  0.016 &  686 $\pm$    5 & 0.67 $\pm$ 0.06 & 16 19 &               1 10 11 12 &      &    \\
  9.446274  -73.650656  &  605.8 & 13.146 $\pm$  0.071 &  992 $\pm$   12 & 1.19 $\pm$ 0.19 & 34 39 &            1 10 11 12 13 &      &    \\
  9.466017  -69.829967  &   20.3 & 12.504 $\pm$  0.027 &  371 $\pm$    2 & 0.55 $\pm$ 0.05 & 17 22 &               1 10 11 13 &      &    \\
  9.491884  -75.208617  &  432.8 & 11.599 $\pm$  0.098 &  305 $\pm$    3 & 0.41 $\pm$ 0.20 & 15 19 &               1 10 11 13 &      &    \\
  9.507699  -73.790561  &  226.4 & 10.818 $\pm$  0.048 &  505 $\pm$   11 & 0.31 $\pm$ 0.19 & 18 22 &               1 10 11 13 &  511 &    \\
  9.614211  -74.260202  &   70.8 & 11.272 $\pm$  0.033 &  448 $\pm$    2 & 0.49 $\pm$ 0.12 & 18 21 &                  1 10 13 &  463 &    \\
  9.660904  -72.008445  &  111.4 & 11.029 $\pm$  0.036 &  385 $\pm$    8 & 0.25 $\pm$ 0.31 & 19 23 &               1 10 11 13 &      &    \\
  9.794922  -71.569425  &   42.1 & 14.554 $\pm$  0.017 &                 &                 & 19 23 &               1 10 11 13 &      &   long term drop \\
  9.828878  -70.131865  &  356.8 & 11.624 $\pm$  0.384 &  320 $\pm$    4 & 0.51 $\pm$ 0.38 & 17 20 &               1 10 11 13 &      &    \\
  9.887294  -70.295891  &  428.8 & 11.502 $\pm$  0.132 &  386 $\pm$    6 & 0.35 $\pm$ 0.21 & 17 21 &               1 10 11 13 &      &    \\
 10.067795  -73.020987  &  291.6 & 11.568 $\pm$  0.056 &  482 $\pm$    7 & 0.43 $\pm$ 0.23 & 16 19 &                  1 10 11 &  489 &    \\
 10.137255  -73.477785  &  100.0 & 11.423 $\pm$  0.032 &  566 $\pm$    5 & 0.34 $\pm$ 0.04 & 16 22 &            1 10 11 12 13 &  566 &    \\
 10.147153  -73.324740  &   75.1 & 12.085 $\pm$  0.028 &  507 $\pm$    3 & 0.55 $\pm$ 0.07 & 16 21 &               1 10 11 13 &  515 &    \\
\hline
\end{tabular}
\tablefoot{
Column 1 lists the coordinates (right ascension and declination) of the source in decimal degrees.
Column 2 Lists the reduced $\chi^2$ statistics.
Column 3 lists the mean $K$-band magnitude.
Column 4 lists the pulsation period with error.
Column 5 lists the amplitude in the $K$-band.
Column 6 lists the number of available data points. The first number indicates the number from the VMC survey, the second the total number.
Column 7 lists the references for the available data points as follows,
(1) present work from the VMC survey;
(10) 2MASS \citep{Cutri_2MASS}
(11) 2MASS 6X \citep{Cutri_2MASS6X}
(12) IRSF \citep{Kato_IRSF}
(13) \citet{DENIS05}
(31) \citet{Zijlstra96};
(32) \citet{vLoon97};
(33)  \citet{Reid1991};
(34) \citet{Wood1992};
(35) \citet{Whitelock2003};
(36) \citet{Whitelock1989};
(37) Peter Wood and Greg Sloan (private communication 2016);
(38) \citet{WBF83};
(39) \citet{HW1990};
(40) \citet{Reid1995};
(41) \citet{Reid1990};
(42) \citet{GR1985};
(43) \citet{Nishida2000};
(44) \citet{FMB90};
(45) \citet{Kamath10};
(46) \citet{GB98};
(47) \citet{Tanabe1997};
(48) \citet{Ferraro95}.
Column 8 gives the initial period used in the LC fitting (overwriting any period found from the Fourier analysis).
This value comes from the literature, see Table~\ref{App-Periods2}.
Column 9 lists any remarks on the source or the LC fitting. Sometimes possible alternative periods are given (`Palt'),
or the long secondary period (`Plsp').
}
\end{sidewaystable*}

\section{The effect of mixing different $K$-band filters in the analysis}
\label{App-Transf}

No attempt was made to bring the various $K$-band measurements onto a common system.
Various reasons (see Sect.~\ref{S-lc}) would make a colour transformation particularly complicated and uncertain in many cases.
This is a limitation, and we try to estimate the possible effect now.

Examination of the bolometric corrections computed by\citet{Aringer09, Aringer19}, limited to the coolest of
their spectra ($T_{\rm eff}$ = 2600~K), reveals that the expected magnitude differences between
a $K$ passband (from \citealt{BB89}) and a 2MASS $K_{\rm s}$ filter are typically smaller than 0.04~mag
for O-rich stars, and smaller than 0.015~mag for C-rich stars. Such differences are smaller for stars with
hotter photospheres.

We also investigated the synthetic photometry of the best fits to the SEDs and IRS spectra of almost 400 AGB and RSG stars (GS18) that
span a very wide range in colour.
We compared $K$-band magnitudes in the VISTA, DENIS, IRSF, 2MASS and SAAO systems.
For stars with moderate circumstellar reddening ($J-K \less 2$) the differences between VISTA and DENIS, and VISTA and 2MASS are of the
order quoted above, i.e. $\less 0.02$~mag. For redder stars, the differences are of order $0.03-0.04$~mag at $J-K= 5$.
The effect in the IRSF filter is less, and in the SAAO system greater, up to 0.2~mag for the redder stars.
It is noted that these colours are much redder than the transformation formula given in \citet{Koen07} which are derived in the
range $-0.087 \leq (J-K) \leq 1.390$.

The effect of a 0.2~mag change in the externally available photometry (independent of the filter system) is tested on eight stars
monitored by \citet{Whitelock2003} without detection in the $J$-band, which, based on their $(H-K)$ colour, have an estimated $(J-K) \more 4$~mag.
Their periods range from 550 to 1375~days and their amplitudes from 0.4 to 0.9~mag.
The corresponding offset is applied to each of the stars considered, and the period analysis repeated. 
The absolute differences in the mean magnitudes are 0.01-0.07~mag, and this corresponds to, at most, a 1.2$\sigma$ difference with respect to the quoted
error bar (for six stars the difference is $< 0.5~\sigma$).
The absolute differences in periods are 1-8 days, and this corresponds to, at most, a 1.8$\sigma$ difference with respect to the quoted
error bar (for six stars the difference is $< 0.3~\sigma$).
The absolute differences in amplitudes are 0.01-0.06~mag, and this corresponds to, at most, a 1.3$\sigma$ difference with respect to the quoted
error bar (for five stars the difference is $< 0.3~\sigma$).

We conclude therefore that the differences between different filter systems is in most cases smaller than the quoted error bars and do
not influence the results of this paper.


 \section{Spectral energy distributions and template fitting}
\label{App-Template}

To obtain insights into the nature of the objects the SEDs are constructed and compared in a
quantitative way to the synthetic SEDs of sources of known composition.
For the AGB stars we use as templates the synthetic photometry of the best-fits to the SEDs and {\it Spitzer} spectra
of O-rich AGB stars, RSGs, and C-rich AGB stars from GS18, respectively, 82, 76 and 204 objects.
These stars cover a large range in effective temperatures, MLRs and dust compositions.
The distinction made in GS18 between O-rich AGB stars and RSGs is based on a number of properties
  including luminosity and pulsation period following the discussion in Sect.~5.1 in \citet{GS09}.

From Table~\ref{App-Sample} and the discussion in Sect.~\ref{S-known} it is clear that the main interlopers among possible
LPVs on the AGB are YSOs, but also objects classified as H{\sc ii} regions, hot stars (Be stars or blue supergiants (BSG)),
Wolf-Rayet (WR) stars or post-AGB/PN stars could be among the sample of 254 candidate LPVs. To have these types
of sources represented among the templates a few sources were picked and analysed in detail, as outlined in the next section.

\subsection{The fitting of some non-AGB templates}
\label{App-Template1}

\begin{table*}

\caption{SED and spectra fitted to some non-AGB stars.}
\label{Tab-nonAGB}
\centering
\footnotesize
\setlength{\tabcolsep}{1.4mm}
  \begin{tabular}{clllll}
  \hline
RA \hfill Dec        & Name   & SIMBAD       & IRS               & Reference & Adopted    \\
(deg) \hfill (deg)  &        & Object type  & Classification    &           &   classification         \\
\hline
72.90 $-$67.08 & SMP LMC 11        & PN  & CPAGB (4) & \citet{SMP78} & PN  \\ 
81.51 $-$67.49 & HD  36402         & WR  & WR (4)    & \citet{WS64}  & WR \\ 
81.05 $-$68.49 & IRAS $05244-6832$ & YSO & HII (4), HII (7) & \citet{Kastner08} & H\,{\sc ii} \\ 
13.53 $-$72.69 & LHA 115-S 18      & Em  & Be (2), B[e] star (3) & \citet{Zickgraf89} & Be \\ 
74.40 $-$67.79 & LHA 120-S 12      & BSG & B[e] (4)  & \citet{Zickgraf86} & Be \\ 
74.19 $-$69.84 & HD 268835         & BSG & B[e] (4)  & \citet{Zickgraf86} & Be \\ 
78.70 $-$67.20 & IRAS $05148-6715$ & YSO & YSO1 (4), YSO1 (7) & \citet{Oliveira09} & YSO \\ 
83.16 $-$69.51 & [RP2006] 774      & YSO & YSO4 (4), YSO4 (7) & \citet{Gruendl09,Reid14} & YSO  \\ 
84.00 $-$67.75 & 2MASS J$05360241-6745171$ & YSO candidate & YSO4 (4), YSO4 (7)   & \citet{Reid14} & YSO \\ 
85.14 $-$69.41 & 2MASS J$05403400-6925099$ & YSO           & YSO4 (4), O Group (5) & \citet{Seale09} & YSO \\ 
73.19 $-$69.19 & IRAS $04530-6916$         & YSO candidate & B[e] (4), F Group (5) & \citet{Seale09} & YSO \\ 
80.37 $-$67.85 & IRAS $05216-6753$         & PN            & H\,{\sc ii} (4) & \citet{Whitney08} & YSO (?) \\ 
\hline
\end{tabular}
  \tablefoot{
Column~1 gives the right ascension and declination.    
Column~2 gives the name of the object.
Column~3 gives the object type listed in SIMBAD.
Column~4 gives the classification based on the IRS spectrum, copied from Table~\ref{App-Sample}.
References are: 
(2) \citet{Kraemer17}; (3) \citet{Ruffle15}; (4) \citet{Jones17}; (5) \citet{Seale09}; 
(7) \citet{Woods11}.
Column~5 gives a reference supporting the adopted classification listed in Col.~6.
    }
\end{table*}

One PN, one Wolf-Rayet star, one H\,{\sc ii} region, three blue BSGs and six (candidate) YSOs with IRS spectra are selected
for a detailed study from the objects in Table~\ref{App-Sample}. The exception is iras05216 (80.37\_$-67.85$) that was initially
considered by GS18 as an AGB star but eliminated in the process as its SED and IRS spectrum were not consistent with that of an AGB star.
Information on these stars is provided in Table~\ref{Tab-nonAGB}.

Like in GS18 the SEDs and the {\it Spitzer} IRS spectra are fitted with More of DUSTY (MoD, \citealt{Gr_MOD}), an extension of the
radiative transfer code DUSTY \citep{Ivezic_D}. For a given set of photometry and spectra as input data
the programme determines the best fitting luminosity, 
dust optical depth, 
dust temperature at the inner radius, 
and slope of the density profile. 

The stars are hotter than AGB stars and the PHOENIX model atmospheres \citep{Hauschildt1999} are
used to represent the central star\footnote{https://phoenix.ens-lyon.fr/Grids/BT-NextGen/SPECTRA/}.
The dust is a combination of amorphous silicates, corundum, and metallic iron, or amorphous carbon and silicon carbide
similar to that in GS18. The proportions between the species are chosen as to reasonably fit the dust continuum and
broad dust features of the {\it Spitzer} spectrum. The actual dust composition may be more complicated, but it is not our aim
to study this in detail in this paper.

The fits are shown in Figure~\ref{Fig-Temps}.
Overall the fits are reasonable representations of the photometric data and that is the main purpose of this exercise.
The synthetic photometry corresponding to these fits can serve as representative templates.

There are also clear shortcomings, for example the spectra show the presence of polycyclic aromatic hydrocarbon and other features
that are not included in the model.
In addition DUSTY and MoD are one-dimensional codes while the geometry around the YSOs is often clearly non-spherical.
For this reason the fit parameters are not discussed in detail as they are sometimes unphysical, such as dust
temperatures at the inner radius above 2000~K.

\begin{figure*}
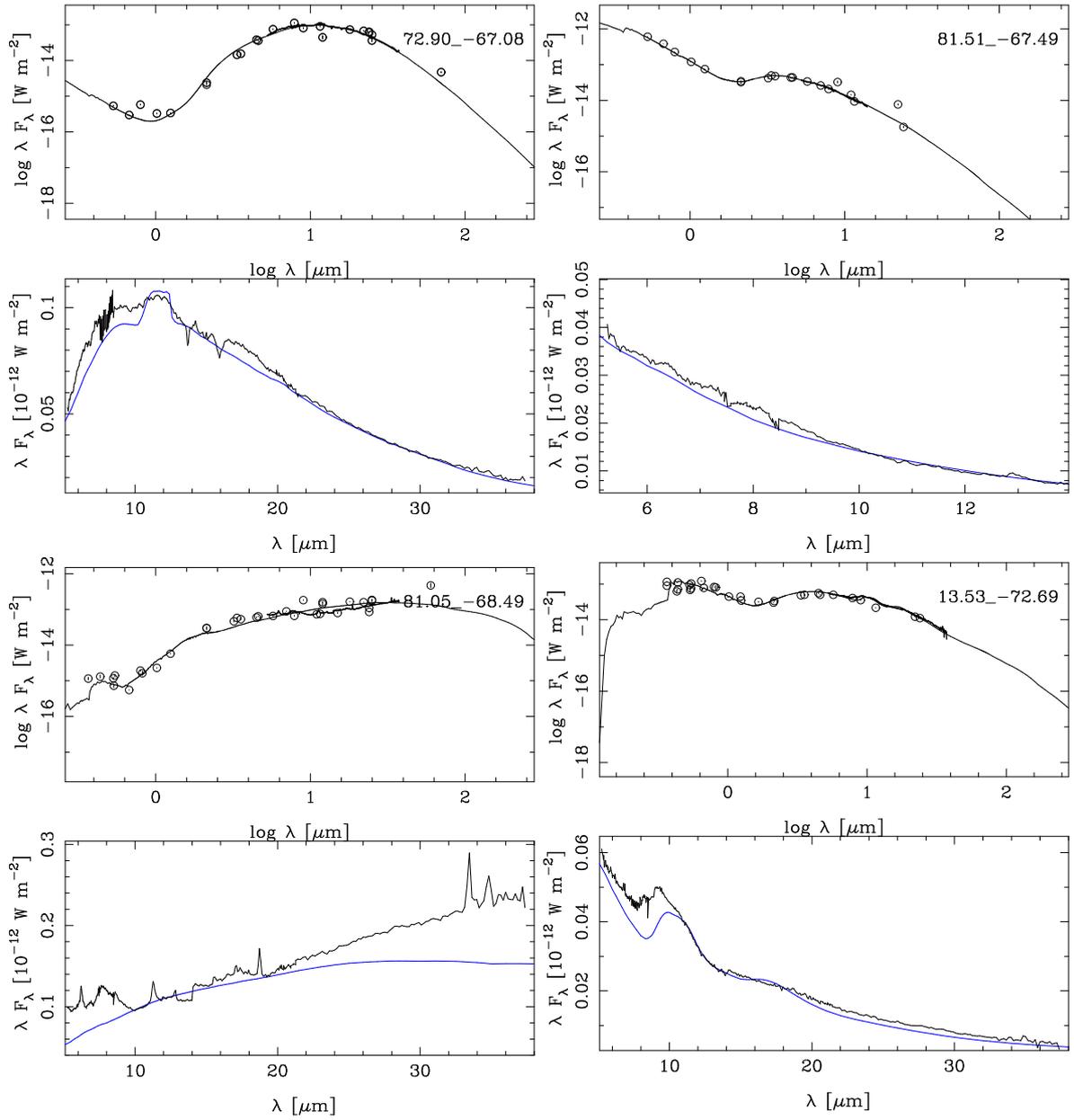

   \centering

\begin{minipage}{0.42\textwidth}
\resizebox{\hsize}{!}{\includegraphics[angle=0]{72.90_-67.08_sed.ps}}
\end{minipage}
\begin{minipage}{0.42\textwidth}
\resizebox{\hsize}{!}{\includegraphics[angle=0]{81.51_-67.49_sed.ps}}
\end{minipage}

\begin{minipage}{0.42\textwidth}
\resizebox{\hsize}{!}{\includegraphics[angle=0]{81.05_-68.49_sed.ps}}
\end{minipage}
\begin{minipage}{0.42\textwidth}
\resizebox{\hsize}{!}{\includegraphics[angle=0]{13.53_-72.69_sed.ps}}
\end{minipage}

\caption[]{
SEDs and {\it Spitzer} spectra of a C-rich post-AGB/PN star (top left), a Wolf-Rayet star (top right),
an H{\sc ii} region (bottom left) and a Be star (bottom right). See the notes to Table~\ref{Tab-nonAGB} for the spectroscopic classification.
In the panels with the {\it Spitzer} IRS spectra, the models (blue lines) are scaled to the observed spectra
based on the average flux in the 12-13~$\mu$m region.
For simplicity, the identifiers are the RA and Dec from Table~\ref{App-Sample} truncated to two decimal figures.
}
\label{Fig-Temps}
\end{figure*}
\setcounter{figure}{0}

\begin{figure*}
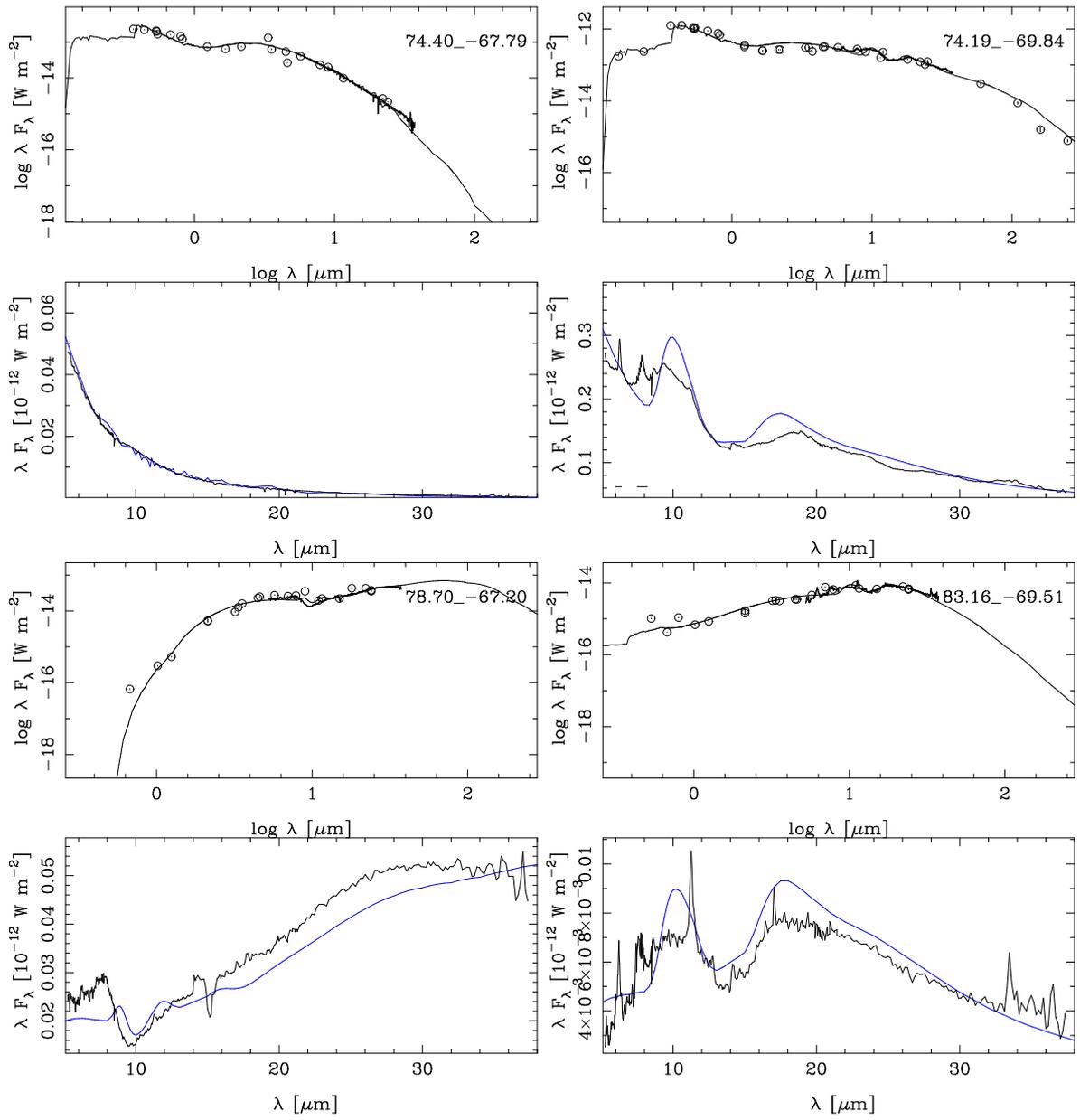

   \centering

\begin{minipage}{0.42\textwidth}
\resizebox{\hsize}{!}{\includegraphics[angle=0]{74.40_-67.79_sed.ps}}
\end{minipage}
\begin{minipage}{0.42\textwidth}
\resizebox{\hsize}{!}{\includegraphics[angle=0]{74.19_-69.84_sed.ps}}
\end{minipage}

\begin{minipage}{0.42\textwidth}
\resizebox{\hsize}{!}{\includegraphics[angle=0]{78.70_-67.20_sed.ps}}
\end{minipage}
\begin{minipage}{0.42\textwidth}
\resizebox{\hsize}{!}{\includegraphics[angle=0]{83.16_-69.51_sed.ps}}
\end{minipage}

\caption[]{Continued. The SEDs and {\it Spitzer} spectra of two Be stars (top panels) and two YSOs (bottom panels).
  See the notes to Table~\ref{Tab-nonAGB} for the spectroscopic classification. }
\end{figure*}

\begin{figure*}
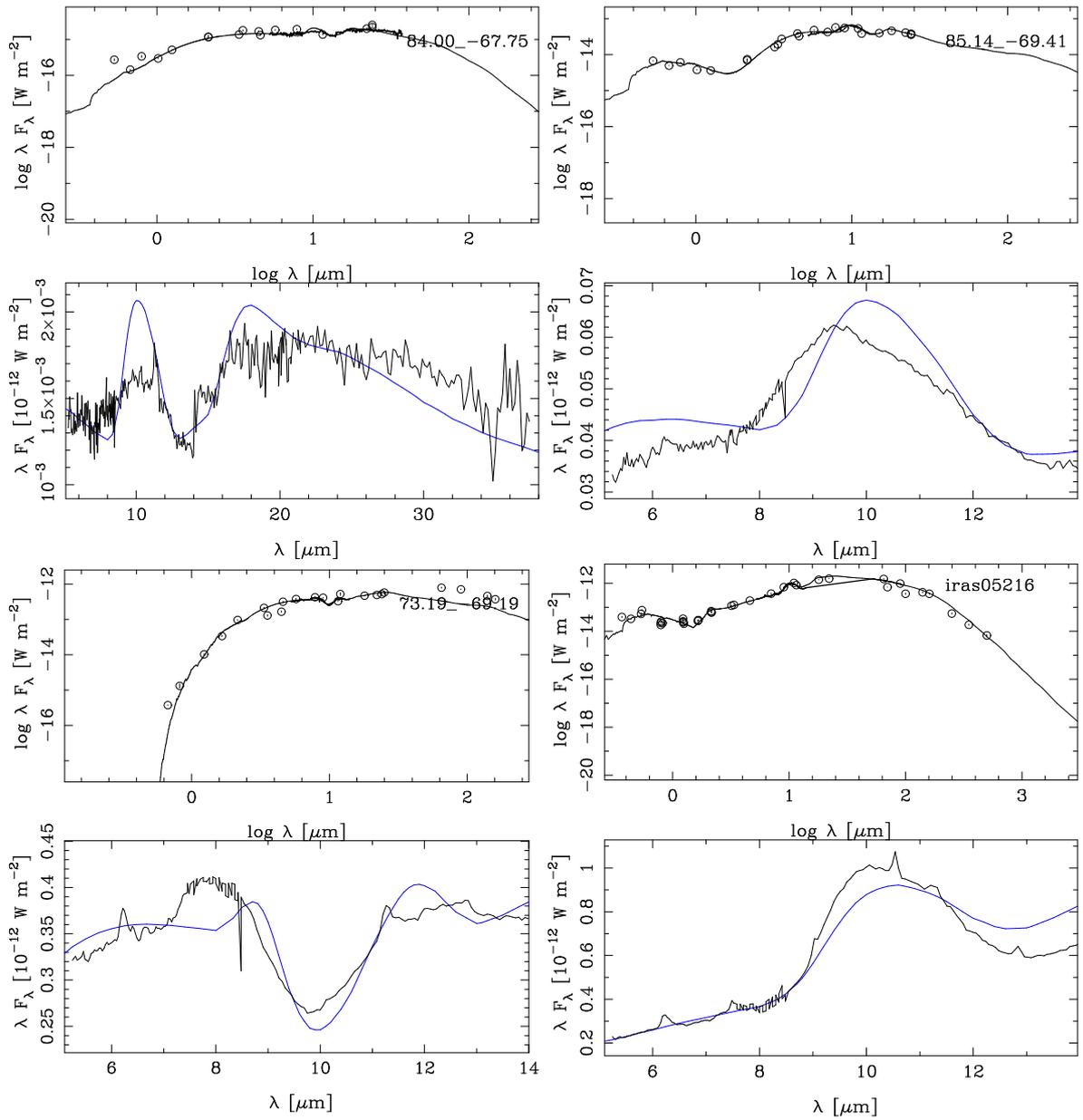

   \centering
\setcounter{figure}{0}

\begin{minipage}{0.42\textwidth}
\resizebox{\hsize}{!}{\includegraphics[angle=0]{84.00_-67.75_sed.ps}}
\end{minipage}
\begin{minipage}{0.42\textwidth}
\resizebox{\hsize}{!}{\includegraphics[angle=0]{85.14_-69.41_sed.ps}}
\end{minipage}

\begin{minipage}{0.42\textwidth}
\resizebox{\hsize}{!}{\includegraphics[angle=0]{73.19_-69.19_sed.ps}}
\end{minipage}
\begin{minipage}{0.42\textwidth}
\resizebox{\hsize}{!}{\includegraphics[angle=0]{iras05216_sed.ps}}
\end{minipage}

\caption[]{Continued. The SEDs and {\it Spitzer} spectra of four YSOs.
  See the notes to Table~\ref{Tab-nonAGB} for the spectroscopic classification. }
\end{figure*}

\subsection{Fitting the SEDs to the templates}
\label{App-Template2}

Given a template with luminosity ($L_{\rm t}$), distance ($d_{\rm t}$) and synthetic absolute magnitudes in many filters ($M_{\rm t}$) and
an object with observed absolute magnitudes with errors ($M_{\rm o}, \sigma$) for an assumed distance ($d_{\rm o}$),
the comparison between template and object is
\begin{equation}
\label{Eq-chi}
\chi^2 =  \sum ((M_{\rm t} + o - M_{\rm o})/\sigma)^2,
\end{equation}
where the sum runs over the observed magnitudes in different filters.
As the template was originally fitted to the photometric and spectroscopic data of a star for a certain distance resulting
in a best-fitting luminosity there is a degeneracy as a template could fit any observed SED by a simple offset ($o$) in magnitude.

Setting the derivative of $\chi^2$ to zero, the offset that will provide the best fit is
\begin{equation}
o =  \sum (m_{\rm t} - m_{\rm o})/\sigma^2,
\end{equation}
that is the weighted mean difference between observed and template magnitudes.

The SEDs of the 254 targets are compared in this way to a total of 374 templates.
As additional constraint it was imposed that in order to be a valid template a C-star had to have a predicted
luminosity above 1~000~\lsol\ and below 70~000~\lsol, an O-rich AGB star to be above 500~\lsol\
and for an O-rich RSG to be above 5~000~\lsol.
This results in an ordered list of $\chi^2$s with the type of the template, and the luminosity of the target
(for an assumed distance of 50~kpc to the LMC and 61~kpc for the objects in the SMC).

There are 50 objects among the 254 that were fitted by GS18. This allows to check the level of `contamination':
if a star is a known C-rich object, is the best-fitting template that of a C-star, and at what level of $\chi^2$
is a non-C star considered a possible match.
This is not a trivial exercise as the SEDs used in GS18 and constructed here are not the same.
For example, GDR2 photometry and VMC magnitudes were not used by GS18.
Instead NIR photometry was taken from the available 2MASS, 2MASS6X, IRSF, and DENIS data.
In addition the fits in GS18 were made to the photometry and {\it Spitzer} spectra, which leads to different synthetic
magnitudes than if only the photometry were fitted as is done here.
Among the 35 known C-stars the best fitting template is always that of a C-star and the best-fitting O-rich template
has a $\chi^2$ that is (when ordered) 1.36, 1.99, 2.05, $\ldots$ times larger than the $\chi^2$ of the best matching template.
Among the 15 known O-rich stars the best fitting template is always that of an O-rich star and the best-fitting C-rich template
has a $\chi^2$ that is 1.28, 1.31, 1.32, 1.51, 1.65,  $\ldots$ times larger than the $\chi^2$ of the best matching template.

Based on this result, the list of templates with a $\chi^2$ less than 1.5 times that of the best-fitting template
is retained and used in the classification.
In this case, 34 stars are classified as
`C'       (all templates are of C-stars), one as
`C (AGB)' (the best fitting template is C, and the alternative(s) is (are) (an) O-rich AGB stars or RSGs), 12 as
`O'       (all templates are O-rich AGB stars or RSGs), and 3 as
`O (AGB)' (the best fitting template is that of an O-rich AGB star or RSG, and the alternative(s) is (are) C-stars).

The finding that the classification of AGB stars based on a fit to the SED gives reliable results, and that the success rate of classification
is better for C-stars than for O-stars was also found by \citet{2012A&A...537A.105G}.
They fitted the SEDs of 374 stars in one VMC field selected to be AGB stars based on CCDs and CMDs
both with a C-star atmosphere model and carbonaceous dust grains, and an O-star model and silicate dust grains and
selected the best model based on a $\chi^2$ comparison. Eighty-seven stars in the field were spectroscopically identified as C-stars
and the SED fitting classified 87\% of them correctly, even 100\% for stars with $(J-K_{\rm s}) > 1.5$~mag.
They also fitted the SEDs of stars with a known classification from {\it Spitzer} spectra in the MCs.
The classification was correct in more than 90\% of the cases for C-stars and about 75\% for the O-rich stars.

If among the list of templates there is a non-AGB star this is marked by OTHER, for example. `C(OTH)'.
The best-fitting template could be a YSO, and in that case one could have the classification
`YSO'        (all templates are those of YSOs),
`YSO(AGB)'  (the best fitting template is that of a YSO, and the alternatives are only O-rich AGB stars, RSGs, or C-rich stars), or
`YSO(OTH)'  (the best fitting template is that of a YSO, and the alternatives could be (an) AGB star(s), RSG(s), WR, Be star, H{\sc ii}, or PN).

\begin{figure*}
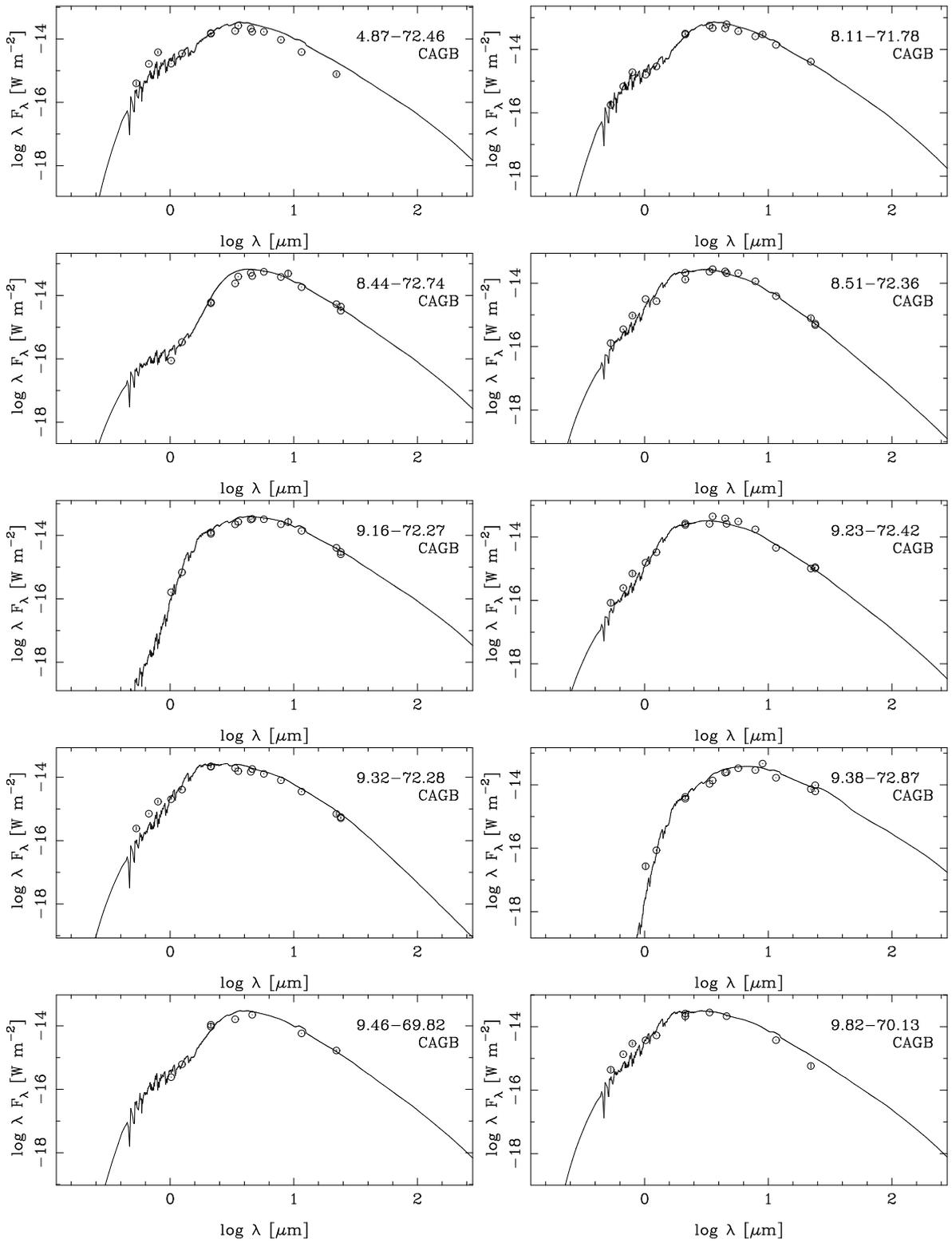

   \centering

\begin{minipage}{0.42\textwidth}
\resizebox{\hsize}{!}{\includegraphics[angle=0]{4.876441_-72.465755_sed.ps}}
\end{minipage}
 \begin{minipage}{0.42\textwidth}
\resizebox{\hsize}{!}{\includegraphics[angle=0]{8.117008_-71.789106_sed.ps}}
\end{minipage}

\begin{minipage}{0.42\textwidth}
\resizebox{\hsize}{!}{\includegraphics[angle=0]{8.442051_-72.749582_sed.ps}}
\end{minipage}
\begin{minipage}{0.42\textwidth}
\resizebox{\hsize}{!}{\includegraphics[angle=0]{8.511784_-72.363404_sed.ps}}
\end{minipage}

\begin{minipage}{0.42\textwidth}
\resizebox{\hsize}{!}{\includegraphics[angle=0]{9.164485_-72.274068_sed.ps}}
\end{minipage}
\begin{minipage}{0.42\textwidth}
\resizebox{\hsize}{!}{\includegraphics[angle=0]{9.236272_-72.421530_sed.ps}}
\end{minipage}

\begin{minipage}{0.42\textwidth}
\resizebox{\hsize}{!}{\includegraphics[angle=0]{9.328809_-72.284239_sed.ps}}
\end{minipage}
\begin{minipage}{0.42\textwidth}
\resizebox{\hsize}{!}{\includegraphics[angle=0]{9.387759_-72.879192_sed.ps}}
\end{minipage}

\begin{minipage}{0.42\textwidth}
\resizebox{\hsize}{!}{\includegraphics[angle=0]{9.466017_-69.829967_sed.ps}}
\end{minipage}
\begin{minipage}{0.42\textwidth}
\resizebox{\hsize}{!}{\includegraphics[angle=0]{9.828878_-70.131865_sed.ps}}
\end{minipage}

\caption[]{
Examples of the fitting of a template (the solid line) to the observed photometry.
The top right corner lists the identifier of the star (RA and Dec truncated to two decimal figures), and
below it the type of the template.
}
\label{Fig-SEDs}
\end{figure*}

The results of the template fitting are listed in Col.~10 in Table~\ref{Tab-Selected}
with the predicted luminosity in Col.~11 (the error in $L$ is estimated to be 15\% based
on the typical scatter in luminosity among the best-fitting templates).
The SEDs of all 254 objects with their best matching templates are plotted in Fig.~\ref{Fig-SEDs}.
In total 197 stars are classified as C-rich (in 2 cases the matching templates include O-rich stars, and in 3 cases the matching templates
include `other' objects). The luminosities are between 1000 and 29~000~\lsol, in agreement with GS18 who derived luminosities for 225 C-stars between
1125 and 56~700~\lsol.
All three objects marked as `C(OTH)' are at relatively low luminosities  between 1250 and 2850~\lsol\ and based on a visual inspection
of their SED and additional information (one object is classified as an emission-line object) the three are classified as likely non-AGB stars.
In total 22 stars are classified as O-rich (in 5 cases the matching templates include C-rich stars, and in 2 cases the matching templates
include `other' objects). The luminosities are between 7000 and 98~000~\lsol\ with one exception (1500~\lsol) in agreement with GS18.
For 82 stars classified as O-rich AGB stars they found a range between 1500 and 107~000~\lsol\ and for 76 stars they classified as RSG a range
from  24~500 to $\sim$ 250~000~\lsol.
That means that 217 stars are very likely AGB stars and the period analysis has revealed LPV like properties.

Twenty stars are classified as YSOs, in 5 cases AGB stars are among the best-fitting templates.
Fifteen have luminosities between 400 and 2800~\lsol\ while five have luminosities between 4500 and 20~000~\lsol.
Two stars are classified as PNe, six as H{\sc ii} regions and seven as Be-stars.

\longtab[2]{
\footnotesize
\begin{landscape}
\vfill
\setlength{\tabcolsep}{1.0mm}
\begin{longtable}{rllllrrcrcr}
\caption{\label{Tab-Selected}
Sample of stars with fitted energy distributions.}\\
\hline
\hline
\centering
\footnotesize
RA        \hfill  Dec        & Name & Obj. Type & Sp. Type  & IRS  & $K$     & Period &  Amp  &  $\chi_{\rm r}^2$ & SEDT  & SEDL   \\
(deg)     \hfill (deg)       &      &           &           &               & (mag) & (d) & (mag) &       &      & (\lsol) \\
\hline
\endfirsthead
\caption{continued.}\\
\hline\hline
\footnotesize
RA        \hfill  Dec        & Name & Obj. Type & Sp. Type  & IRS & $K$     & Period &  Amp  & $\chi_{\rm r}^2$  & SEDT  & SEDL  \\
(deg) \hfill (deg)            &      &         &           &     & (mag) & (d) & (mag) &       &      & (\lsol) \\
\hline
\endhead
                                            \multicolumn{11}{c}{\underline{Most likely AGB stars and LPVs} } \\
  4.876441  $-$72.465755   &   2MASS J00193036-7227567                               &        Can. AGB    &            & &  11.982 $\pm$  0.011  &   265 $\pm$    1$^a$  &  0.24 $\pm$ 0.05  &    19.8  & C & 4530  \\
  8.117008  $-$71.789106   &       2MASS J00322809-7147207                           &        Can. AGB    &            & &  11.278 $\pm$  0.023  &   600 $\pm$    4  &  0.63 $\pm$ 0.07  &    70.3  & C & 9231 \\
  8.442051  $-$72.749582   &       2MASS J00334612-7244584                           &        Can. AGB    &            & &  12.980 $\pm$  0.037  &   571 $\pm$   10  &  0.54 $\pm$ 0.30  &   175.6  & C & 8462 \\
  8.511784  $-$72.363404   &       2MASS J00340283-7221482                           &        Can. AGB    &            & &  11.622 $\pm$  0.025  &   356 $\pm$    1  &  0.70 $\pm$ 0.04  &    30.2  & C & 4719 \\
  9.164485  $-$72.274068   &       2MASS J00363946-7216266                           &        Can. AGB    &            & &  12.326 $\pm$  0.011  &   466 $\pm$    2  &  0.43 $\pm$ 0.05  &    19.7  & C & 5802 \\
  9.236272  $-$72.421530   &   MSX SMC 091 2MASS J00365671-7225175                   &                    C    &            & C(1) &  11.501 $\pm$  0.025  &   406 $\pm$    2  &  0.52 $\pm$ 0.05  &    78.7  & C & 5698 \\
  9.328809  $-$72.284239   &       2MASS J00371893-7217031                           &        Can. AGB    &            & &  11.619 $\pm$  0.010  &   362 $\pm$    1  &  0.41 $\pm$ 0.01  &     5.1  & C & 3959 \\
\hline
\end{longtable}
\tablefoot{
Column 1 lists the right ascension and declination of the source in decimal degrees.
Column 2 gives some names, copied from Table~\ref{App-Sample}.
Column 3 gives the object type, copied from Table~\ref{App-Sample}.
Column 4 gives the main spectral type, copied from Table~\ref{App-Sample}.
Column 5 gives the main {\it Spitzer} IRS classification and the reference, copied from Table~\ref{App-Sample}.
Column 6--9 give the mean magnitude, period and amplitude, and $\chi^2$ statistics copied  from Table~\ref{App-Periods1}.
When a superscript $^a$ is listed after the period it implies that an alternative period is given in Table~\ref{App-Periods1}.
When a superscript $^b$ is listed after the classification the star is an outlier in the bolometric $PL$-relation (see Fig.~\ref{Fig-PL}).
Column 10 lists the classification based on the SED template fitting, see Appendix~\ref{App-Template}.
Column 11 lists the luminosity of the star based on the template fitting.
}
\end{landscape}
}

\section{Results of the fits to synthetic spectra from \citet{Nanni19b} }
\label{App-Nanni}

Table~\ref{Tab-Nanni} lists the first entries of the table available at the CDS.
It provides the results of the SED fitting performed to synthetic spectra
from \citet{Nanni19b} for the 217 objects classified as AGB stars in this work.

The calculation of the uncertainty on the MLR and luminosity is the same as that described by \citet{Nanni18, Nanni19b}.
The synthetic photometry that best fits the observations provides $\chi^2=\chi^2_{\rm best}$.
Each of the synthetic fluxes are randomly modified within the observed flux errors, and a new value of
$\chi^2$ between the observed and modified flux is calculated. Such a procedure is reiterated $100$ times and an equal number of $\chi^2$
values are computed. From the $100$ $\chi^2$ we extract the minimum and the 34$^{\rm th}$ values (corresponding to 1$\sigma$).
The difference between the minimum and the 1$\sigma$ value provides the statistical variation of $\Delta\chi^2$ and thus the maximum
acceptable value of $\chi^2$, $\chi^2_{\rm max}=\chi^2_{\rm best}+\Delta\chi^2$. We then select from the grid of models those
that yield a value of $\chi^2$: $\chi^2\le\chi^2_{\rm max}$, and we compute their average and standard deviation,
which provides the corresponding uncertainty of the quantity calculated. If fewer than 4 models
satisfy the condition $\chi^2\le\chi^2_{\rm max}$, the stellar quantities are assumed to be represented by the best-fitting value and
zero uncertainty.

\begin{table*}

\caption{
Results from the SED fitting performed to synthetic spectra from \citet{Nanni19b} for all AGB candidates
selected in this work.} 
\label{Tab-Nanni}
\centering
\begin{tabular}{rllll}
\hline
  RA    & Dec   & Name  & $\dot{M}$          & Luminosity \\
  (deg) & (deg) &       & ($10^{-6}$~\msolyr) & (\lsol)    \\
\hline 
  4.876441 & $-$72.465755 &   2MASS J00193036-7227567 & 1.76 $\pm$ 0.67 & 3162  \\
  8.117008 & $-$71.789106 &   2MASS J00322809-7147207 & 5.0 $\pm$  1.5  & 7943  \\
  8.442051 & $-$72.749582 &   2MASS J00334612-7244584 & 7.94            & 7943  \\
  8.511784 & $-$72.363404 &   2MASS J00340283-7221482 & 2.51            & 3981  \\
  9.164485 & $-$72.274068 &   2MASS J00363946-7216266 & 3.16            & 5012  \\
  9.236272 & $-$72.421530 &   MSX SMC 091  2MASS J00365671-7225175 & 3.08  $\pm$  0.54  & 5012  \\
  9.328809 & $-$72.284239 &   2MASS J00371893-7217031 & 1.72  $\pm$ 0.54  & 3162 \\
  9.387759 & $-$72.879192 &   2MASS J00373306-7252451 & 11.22 & 5012  \\
  9.466017 & $-$69.829967 &   & 2.57  $\pm$ 0.58 & 3162  \\
  9.828878 & $-$70.131865 & [MH95] 414  2MASS J00391894-7007546 & 1.85 $\pm$ 0.33 & 3981\\
  9.887294 & $-$70.295891 &                   & 1.58    & 5012 \\
  10.466052 & $-$71.648376 &   2MASS J00415184-7138541 & 2.98  $\pm$ 0.63 & 3981  \\
  10.590777 & $-$72.401692 &   2MASS J00422179-7224060 & 1.59  $\pm$ 0.49 & 3981  \\
  10.803761 & $-$72.249720 &   2MASS J00431291-7214590 & 2.0   & 3981  \\
  11.433544 & $-$72.137625 &   2MASS J00454404-7208154 & 1.92 $\pm$ 0.36 & 3981  \\
  11.844981 & $-$74.952560 &   2MASS J00472280-7457092 & 1.00   & 2512  \\
  12.148646 & $-$74.177498 & IRAS F00468-7427  2MASS J00483568-7410391 & 15.8 $\pm$ 4.2  & 6310  \\
  12.529927 & $-$73.523596 & IRAS 00483-7347  2MASS J00500719-7331251 & 37.3 $\pm$ 2.6  & 25120 \\
  12.627578 & $-$72.858313 & IRAS F00486-7308  2MASS J00503062-7251298 & 4.35  $\pm$ 1.55  & 25120  \\
  12.707210 & $-$73.815508 &   2MASS J00504974-7348557 & 14.13 $\pm$ 3.23  & 5012  \\
  13.332679 & $-$75.138802 & OGLE J005337.29-723434.9 & 2.12  $\pm$ 0.56 & 5012 \\
  13.595669 & $-$74.832455 &                           & 13.19 $\pm$ 3.48  & 5012 \\
  13.791431 & $-$73.712404 &   2MASS J00550995-7342447 & 9.1 $\pm$ 3.2  & 3981  \\
  13.923324 & $-$70.409980 &   2MASS J00554160-7024359 & 3.2  $\pm$ 1.1  & 5012  \\
  14.860307 & $-$72.394923 &   2MASS J00592646-7223417 & 5.01  & 3981  \\
  15.058298 & $-$74.958569 &                    & 1.58   & 3981  \\
  15.173262 & $-$72.633511 &   2MASS J01004152-7238003 & 2.9 $\pm$ 1.0  & 2081 $\pm$ 211\\
  15.194747 & $-$74.420436 &   2MASS J01004675-7425136 & 1.85 $\pm$ 0.33 & 3162  \\
  15.636536 & $-$72.320172 &   2MASS J01023277-7219125 & 3.56  $\pm$ 1.16  & 3981\\
  15.851503 & $-$72.634413 & SAGEMA J010324.38-723803.8 & 1.36 $\pm$ 0.28  & 20560 $\pm$ 1220\\
  15.891398 & $-$72.734837 &   2MASS J01033393-7244053 & 1.00   & 14130\\
  15.893094 & $-$70.597998 &   2MASS J01033435-7035528 & 3.04 $\pm$ 0.70 & 6310  \\
  15.941578 & $-$75.266262 &   & 5.34 $\pm$ 0.65 & 12590  \\
  16.062660 & $-$70.405257 & SAGEMA J010415.07-702418.9 & 10.7 $\pm$ 1.7  & 5012 \\
  17.634352 & $-$73.084464 & IRAS F01091-7320  2MASS J01103224-7305040 & 7.94  & 6310  \\
  19.320585 & $-$74.804827 &   2MASS J01171696-7448173 & 2.0   & 3162  \\
  19.322041 & $-$74.604738 & IRAS F01160-7451  2MASS J01171728-7436170 & 11.2  & 11220  \\
  20.617749 & $-$71.157234 & IRAS F01210-7125  2MASS J01222827-7109260 & 6.7 $\pm$ 2.7  & 11220  \\
  21.130633 & $-$71.620707 &   2MASS J01243139-7137145 & 2.51   & 3981  \\
  22.765869 & $-$71.318003 &   2MASS J01310381-7119048 & 5.11 $\pm$ 0.36 & 7943  \\
  23.417313 & $-$73.823852 &   2MASS J01334016-7349257 & 10.0  & 7943 \\
  24.195532 & $-$72.641560 &   2MASS J01364695-7238295 & 3.16   & 6310  \\
  25.517064 & $-$75.171631 &   2MASS J01420412-7510177 & 1.26  & 3162  \\
  66.778920 & $-$73.458705 &   & 4.23  $\pm$ 0.98  & 6310 \\
  68.402239 & $-$67.659170 &   2MASS J04333653-6739329 & 7.94  & 5012  \\
  68.629927 & $-$70.830315 &   & 3.52 $\pm$ 0.67 & 6310  \\
  68.850394 & $-$66.947047 &   2MASS J04352409-6656493 & 3.00 $\pm$ 0.33 & 3981  \\
  69.049993 & $-$71.370090 &   2MASS J04361200-7122123 & 10.0 $\pm$ 1.2  & 13668 $\pm$ 744\\
  69.178078 & $-$68.536224 & MSX LMC 1045 & 2.51   & 5012  \\
  69.781620 & $-$74.922266 &   & 3.16   & 6310  \\
  70.088825 & $-$75.332959 &   & 15.2 $\pm$ 2.5  & 5012  \\
  70.104244 & $-$74.743169 &   & 3.26 $\pm$ 0.89 & 5012  \\
  70.118694 & $-$69.920432 & IRAS 04407-7000  2MASS J04402848-6955135 & 14.4 $\pm$ 3.0  & 22390  \\
  70.438656 & $-$72.645357 &   2MASS J04414529-7238431 & 5.1 $\pm$ 1.1  & 3981  \\
  70.632433 & $-$74.796568 &   & 1.26   & 3162  \\
  71.369404 & $-$72.260494 &   2MASS J04452864-7215379 & 10.7 $\pm$ 2.8  & 5012  \\
\hline
\end{tabular}
\tablefoot{
Column 1 lists the right ascension and declination of the source in decimal degrees.
Column 2 gives some names, copied from Table~\ref{App-Sample}.
Column 3 lists the MLR.
Column 4 lists the luminosity.
}
\end{table*}


\end{appendix}

\end{document}